\documentclass{aa}

\usepackage{graphicx}
\usepackage{txfonts}
\usepackage{natbib}
\usepackage[dvipsnames]{xcolor}
\usepackage{xspace}
\usepackage{hvfloat}
\usepackage{multirow}
\usepackage{array,booktabs,ragged2e}
\usepackage[colorlinks=true,linkcolor=blue,citecolor=blue]{hyperref}
\usepackage{etoolbox}

\graphicspath{{./}}

\makeatletter

\patchcmd{\NAT@citex}
  {\@citea\NAT@hyper@{%
     \NAT@nmfmt{\NAT@nm}%
     \hyper@natlinkbreak{\NAT@aysep\NAT@spacechar}{\@citeb\@extra@b@citeb}%
     \NAT@date}}
  {\@citea\NAT@nmfmt{\NAT@nm}%
   \NAT@aysep\NAT@spacechar\NAT@hyper@{\NAT@date}}{}{}

\patchcmd{\NAT@citex}
  {\@citea\NAT@hyper@{%
     \NAT@nmfmt{\NAT@nm}%
     \hyper@natlinkbreak{\NAT@spacechar\NAT@@open\if*#1*\else#1\NAT@spacechar\fi}%
       {\@citeb\@extra@b@citeb}%
     \NAT@date}}
  {\@citea\NAT@nmfmt{\NAT@nm}%
   \NAT@spacechar\NAT@@open\if*#1*\else#1\NAT@spacechar\fi\NAT@hyper@{\NAT@date}}
  {}{}

\makeatother

\newcolumntype{R}[1]{>{\RaggedLeft\arraybackslash}p{#1}}
\newcolumntype{C}[1]{>{\centering\arraybackslash}p{#1}}

\newcommand{\hlink}[1]{\url{http://#1}\xspace}

\newcommand{\rfig}[1]{Fig.~\ref{#1}}
\newcommand{\rfigs}[1]{Figs.~\ref{#1}}
\newcommand{\req}[1]{Eq.~\ref{#1}}
\newcommand{\reqs}[1]{Eqs.~\ref{#1}}
\newcommand{\rtab}[1]{Table \ref{#1}}

\newcommand{\rapp}[1]{Appendix \ref{#1}}

\newcommand{\rsec}[1]{section \ref{#1}}
\newcommand{\rsecs}[1]{sections \ref{#1}}

\newcommand{\herschel}{{\it Herschel}\xspace}
\newcommand{\spitzer}{{\it Spitzer}\xspace}
\newcommand{\hubble}{{\it Hubble}\xspace}
\newcommand{\hst}{{\it HST}\xspace}
\newcommand{\jwst}{{\it JWST}\xspace}

\newcommand{\um}{\mu{\rm m}}

\newcommand{\uJy}{\mu{\rm Jy}}
\newcommand{\mJy}{{\rm mJy}}

\newcommand{\sfr}{{\rm SFR}}

\newcommand{\ssfr}{{\rm sSFR}}
\newcommand{\lir}{L_{\rm IR}}

\newcommand{\lsun}{L_\odot}
\newcommand{\msun}{M_\odot}
\newcommand{\Mpc}{{\rm Mpc}}

\newcommand{\Gyr}{{\rm Gyr}}
\newcommand{\Myr}{{\rm Myr}}
\newcommand{\yr}{{\rm yr}}
\newcommand{\dex}{{\rm dex}}
\newcommand{\mstar}{M_\ast}

\newcommand{\kms}{{\rm km}/{\rm s}}

\newcommand{\tdust}{T_{\rm dust}}

\newcommand{\restew}{{\rm EW}_{\rm rest}}
\newcommand{\obsew}{{\rm EW}_{\rm obs}}
\newcommand{\uvj}{$UVJ$\xspace}

\newcommand{\mean}[1]{\left<#1\right>}

\newcommand{\kelvin}{{\rm K}}
\newcommand{\zphot}{z_{\rm phot}}
\newcommand{\zspec}{z_{\rm spec}}
\newcommand{\zform}{z_{\rm form}}
\newcommand{\oiii}{[\ion{O}{III}]}
\newcommand{\oiiifour}{[\ion{O}{III}]_{4363}}
\newcommand{\oii}{[\ion{O}{II}]}
\newcommand{\oone}{[\ion{O}{I}]_{6300}}
\newcommand{\nii}{[\ion{N}{II}]}
\newcommand{\sii}{[\ion{S}{II}]}
\newcommand{\neiii}{[\ion{Ne}{III}]_{3689}}
\newcommand{\neiv}{[\ion{Ne}{IV}]_{2422}}
\newcommand{\nev}{[\ion{Ne}{V}]_{3426}}
\newcommand{\heii}{\ion{He}{ii}_{4686}}
\newcommand{\hei}{\ion{He}{i}_{5876}}
\newcommand{\mgii}{\ion{Mg}{ii}_{2799}}

\newcommand{\halpha}{{\rm H}\alpha}
\newcommand{\hbeta}{{\rm H}\beta}
\newcommand{\hgamma}{{\rm H}\gamma}
\newcommand{\hdelta}{{\rm H}\delta}
\newcommand{\hepsilon}{{\rm H}\varepsilon}
\newcommand{\hzeta}{{\rm H}\zeta}
\newcommand{\heta}{{\rm H}\eta}

\newcommand{\mgi}{\ion{Mg}{I}}

\newcommand{\Ks}{$K_{\rm s}$\xspace}

\defcitealias{straatman2014}{S14}
\defcitealias{glazebrook2017}{G17}

\let\oldAA\AA
\renewcommand{\AA}{\text{\normalfont\oldAA}}

\newcommand*\dd{\ensuremath{{\rm d}}}

\begin{document}

\title{Near infrared spectroscopy and star-formation histories \\ of $3 \le z \le 4$ quiescent galaxies \thanks{Tables \ref{TAB:final_props} and \ref{TAB:galaxies_zspecs} are available in electronic form at the CDS via anonymous ftp to \url{cdsarc.u-strasbg.fr} (130.79.128.5) or via \url{http://cdsweb.u-strasbg.fr/cgi-bin/qcat?J/A+A/}}}
\titlerunning{Spectroscopy and star-formation histories of $3 \le z \le 4$ quiescent galaxies}
\authorrunning{C.~Schreiber et al.}

\author{C.~Schreiber\inst{1}
  \and K.~Glazebrook\inst{2}
  \and T.~Nanayakkara\inst{1,2}
  \and G.~G.~Kacprzak\inst{2}
  \and I.~Labb\'e\inst{1,2}
  \and P.~Oesch\inst{3}
  \and T.~Yuan\inst{4}
  \and K.-V.~Tran\inst{5,6}
  \and C.~Papovich\inst{7}
  \and L.~Spitler\inst{8,9,10}
  \and C.~Straatman\inst{11}
}

\institute{
    Leiden Observatory, Leiden University, NL-2300 RA Leiden, The Netherlands \\
    \email{cschreib@strw.leidenuniv.nl} 
    \and Centre for Astrophysics and Supercomputing, Swinburne University of Technology, Hawthorn, VIC 3122, Australia 
    \and Observatoire de Gen\`eve, 1290 Versoix, Switzerland 
    \and Research School of Astronomy and Astrophysics, The Australian National University, Cotter Road, Weston Creek, ACT 2611, Australia 
    \and Australia Telescope National Facility, CSIRO Astronomy and Space Science, PO Box 76, Epping, NSW 1710, Australia 
    \and School of Physics, University of New South Wales, Sydney, NSW 2052, Australia 
    \and George P.~and Cynthia W.~Mitchell Institute for Fundamental Physics and Astronomy, Department of Physics and Astronomy, Texas A\&M University, College Station, TX 77843, USA 
    \and Research Centre for Astronomy, Astrophysics \& Astrophotonics, Macquarie University, Sydney, NSW 2109, Australia 
    \and Department of Physics \& Astronomy, Macquarie University, Sydney, NSW 2109, Australia 
    \and Australian Astronomical Observatory, 105 Delhi Rd., Sydney NSW 2113, Australia 
    \and Max-Planck Institut f\"ur Astronomie, K\"onigstuhl 17, D-69117, Heidelberg, Germany 
}

\date{Received 22 March 2018; accepted 4 July 2018}

\abstract{We present Keck--MOSFIRE $H$ and $K$ spectra for a sample of $24$ candidate quiescent galaxies at $3<z<4$, identified from their rest-frame \uvj colors and photometric redshifts in the ZFOURGE and 3DHST surveys. With median integration times of one hour in $H$ and five in $K$, we obtain spectroscopic redshifts for half of the sample, using either Balmer absorption lines or nebular emission lines. We confirm the high accuracy of the photometric redshifts for this spectroscopically-confirmed sample, with a median $|\zphot - \zspec|/(1+\zspec)$ of $1.2\%$. Two galaxies turn out to be dusty $\halpha$ emitters at lower redshifts ($z<2.5$), and these are the only two detected in the sub-mm with ALMA. High equivalent-width $\oiii$ emission is observed in two galaxies, contributing up to $30\%$ of the $K$-band flux and mimicking the \uvj colors of an old stellar population. This implies a failure rate of only $20\%$ for the \uvj selection at these redshifts. Lastly, Balmer absorption features are identified in four galaxies, among the brightest of the sample, confirming the absence of OB stars. We then modeled the spectra and photometry of all quiescent galaxies with a wide range of star-formation histories. We find specific star-formation rates ($\ssfr$) lower than $0.15\,\Gyr^{-1}$ (a factor of ten below the main sequence) for all but one galaxy, and lower than $0.01\,\Gyr^{-1}$ for half of the sample. These values are consistent with the observed $\hbeta$ and $\oii$ luminosities, and the ALMA non-detections. The implied formation histories reveal that these galaxies have quenched on average $300\,\Myr$ prior to being observed, between $z=3.5$ and $5$, and that half of their stars were formed by $z\sim5.5$ with a mean $\sfr\sim300\,\msun/\yr$. We finally compared the \uvj selection to a selection based instead on the $\ssfr$, as measured from the photometry. We find that galaxies a factor of ten below the main sequence are $40\%$ more numerous than \uvj-selected quiescent galaxies, implying that the \uvj selection is pure but incomplete. Current models fail at reproducing our observations, and underestimate either the number density of quiescent galaxies by more than an order of magnitude, or the duration of their quiescence by a factor two. Overall, these results confirm the existence of an unexpected population of quiescent galaxies at $z>3$, and offer the first insights on their formation histories.}


\maketitle

\section{Introduction}

In the present-day Universe, clear links have been observed between the stellar mass of a galaxy, the effective age of its stellar population, its optical colors, its morphology, and its immediate environment. The most massive galaxies, in particular, tend to be located in galaxy over-densities (e.g., clusters or groups), have old stellar populations and little on-going star formation, and display red, featureless spheroidal light profiles with compact cores \citep[e.g.,][]{baldry2004}. These different observables have been used broadly to identify galaxies belonging to this population, sometimes interchangeably, and be it from their morphology [``early-type galaxies'' (ETGs), ``spheroids'', ``ellipticals``], their colors [``red'' or ``red sequence galaxies'', ``extremely red objects'' (EROs), ``luminous red galaxies'' (LRGs)], their star formation history [``old'', ``quiescent'', ``evolved'', ``passive'', or ``passively evolving galaxies'' (PEGs)], their mass [``massive galaxies''], their environment [``bright cluster galaxy'' (BCGs), ``central galaxy''], or any combination thereof.

However, these links tend to dissolve at earlier epochs. While massive galaxies always seem to have red optical colors, at higher redshifts this is increasingly caused by dust obscuration rather than old stellar populations \citep[e.g.,][]{cimatti2002,dunlop2007,spitler2014,martis2016}. Similarly, the proportion of star-forming objects among massive galaxies, compact galaxies, or within over-dense structures was larger in the past \citep[e.g.,][]{butcher1978,elbaz2007,vandokkum2010,brammer2011,barro2013,barro2016-a,wang2016-a,elbaz2017}. When exploring the evolution of galaxies through cosmic time, it is therefore crucial not to assume that the aforementioned observables independently map to the same population of objects, and to precisely define which population is under study.

In the present paper, we aim to constrain and understand the emergence of massive galaxies with low levels of on-going star formation, which we will dub hereafter ``quiescent'' galaxies (QGs), in opposition to ``star-forming'' galaxies (SFGs). In our view, for a galaxy to qualify as quiescent its star formation rate ($\sfr$) needs not be strictly zero, but remain significantly lower than the average for SFGs of similar masses at the same epoch. In other words, these galaxies must reside ``below'' the so-called star-forming main sequence (MS, \citealt{elbaz2007,noeske2007}). If found with $\sfr$s sufficiently lower than that expected for an MS galaxy, say below the MS by an order of magnitude or three times the observed MS scatter (e.g., \citealt{schreiber2015}), these galaxies must have experienced a particular event in their history which suppressed star formation (either permanently or temporarily). At any given epoch, this is equivalent to selecting galaxies with a low specific $\sfr$ ($\ssfr = \sfr/\mstar$).

Regardless of how they are defined, the evolution of the number density of QGs has been a long standing debate, and has proven an important tool to constrain galaxy evolution models (see \citealt{daddi2000,glazebrook2004,cimatti2004,glazebrook2017}, discussions therein, and below). After two decades of observations, solid evidence now show that QGs already existed in significant numbers in the young Universe at all epochs, now up to $z\sim4$ \citep[e.g.,][]{franx2003,glazebrook2004,cimatti2004,kriek2006,kriek2009,gobat2012,hill2016,glazebrook2017}, and that their number density has been rising continuously until the present day \citep[e.g.,][]{faber2007,ilbert2010,brammer2011,muzzin2013,ilbert2013,stefanon2013,tomczak2014,straatman2014}. Spectroscopic observations confirmed their low current $\sfr$s from faint or absent emission lines, their old effective ages (mass- or light-weighted) of more than half a Gyr from absorption lines, or their large masses from kinematics \citep[e.g.,][]{kriek2006,kriek2009,vandesande2013,hill2016,belli2017-a,belli2017,glazebrook2017}. High-resolution imaging from the {\it Hubble Space Telescope} (HST) simultaneously showed that distant QGs also display ``\cite{devaucouleurs1948}''-type density profiles, and effective radii getting increasingly larger with time possibly as a result of dry merging \citep[e.g.,][]{vandokkum2008,newman2012,muzzin2012,vanderwel2014-a}.

The existence of these galaxies in the young Universe poses a number of interesting and still unanswered questions. Chief among them is probably the fact that, according to our current understanding of cosmology, galaxies are not closed boxes but are continuously receiving additional gas from the intergalactic medium through infall \citep[e.g.,][]{press1974,audouze1976,rees1977,white1978,tacconi2010}. While the specific infall rate should go down with time as the density contrast in the Universe sharpens and the merger rate decreases \citep[e.g.,][]{lacey1993}, gas flows still remain large enough to sustain substantial star formation in massive galaxies, where feedback from supernovae is inefficient \citep[e.g.,][]{benson2003}, and also in clusters \citep{fabian1994}. A mechanism must therefore be invoked in massive galaxies, either to remove this gas from the galaxies, or to prevent it from cooling down to temperatures suitable for star formation. To produce observationally-identifiable QGs, this mechanism must act over at least the lifetime of OB stars, a few tens of Myr, and should be allowed to persist over longer periods to explain their observed ages of up to several Gyrs (e.g., \citealt{kauffmann2003}). This mechanism has been dubbed ``quenching'' \citep[e.g.,][]{bower2006,faber2007}.

Nowadays, the most favored actor for quenching is the feedback that slow and fast-growing supermassive black holes can apply on their host galaxies \citep[e.g.,][]{silk1998,bower2006,croton2006,hopkins2008,cattaneo2009}. During the fastest accretion events (e.g., during a galaxy merger), the energetics of these active galactic nuclei (AGNs) is such that they are capable of driving powerful winds and remove gas from the galaxy, resulting in so-called quasar-mode feedback. However this mechanism alone cannot prevent star formation over long periods of time. Indeed, the expelled gas eventually re-enters the galaxy. This gas must first cool down (hence form stars) before reaching the galaxy's center, fueling black hole growth, and triggering a new quasar event. There is therefore a need to introduce a heating source to prevent the gas infalling on quiescent galaxies from cooling (this need was first identified in the core of galaxy clusters, e.g., \citealt{blanton2001}). This long-lasting, less violent mechanism could then maintain the quiescence established by a quasar episode.

Lower levels of accretion onto central black holes can fulfill this role, by injecting energy into the halo of their host galaxy with jets (see \citealt{croton2006}). However this is not the sole possible explanation. In particular, ``gravitational heating'' of infalling gas in massive dark matter halos can have the same net effect \citep{birnboim2003,dekel2008}, while stabilization of extended gas disks by high stellar density in bulges can also prevent star formation on long timescales \citep{martig2009}.

While all of these phenomena have been shown to play some role in quenching galaxies, it remains unknown which (if any) is the dominant process. For example, recent simulations show that the QG population up to $z\sim2$ can be reproduced without the violent feedback of AGNs and instead simply shutting off cold gas infall, leaving existing gas to be consumed by star formation \citep{gabor2012,dave2017}. Furthermore, the observation of significant gas reservoirs in higher redshifts QGs, as well as SFGs transitioning to quiescence, suggests that quenching is not simply caused by a full removal of the gas, but is accompanied (and, perhaps, driven) by a reduced star-formation efficiency \citep[e.g.,][]{davis2014,alatalo2014,alatalo2015,french2015,schreiber2016,suess2017,lin2017,gobat2018}. A complete census of QGs across cosmic time and a better understanding of their star formation histories are required to differentiate these different mechanisms.

Because of their low $\ssfr$ and the lack of young OB stars, QGs necessarily have red optical colors. For this reason they are usually identified from said colors, as seen in broadband photometry either directly with observed bands \citep[e.g.,][]{franx2003,daddi2004-a,labbe2005} or by computing rest-frame colors when the redshift is known \citep[e.g.,][]{faber2007,williams2009,ilbert2010}. However, dusty SFGs can contaminate such color-selected samples: while quiescent galaxies are red, red galaxies are not necessarily quiescent. The rate of contamination probably depends on the adopted method and the quality of the data. Selection methods based on a single color (such as color-magnitude diagrams) were very successful in the local Universe, but suffer from high contamination at higher redshifts owing to the increasing prevalence of dusty red galaxies \citep[e.g.,][]{labbe2005,papovich2006}. Two-color criteria were later introduced to break the degeneracy between dust and age to first order, and allow the construction of purer samples (\citealt{williams2009,ilbert2010}). Compared to full spectral modeling coupled to a more direct $\ssfr$ selection, these color criteria are less model-dependent, particularly so in deep fields where the wavelength coverage is rich and interpolation errors are negligible. Because they are so simple to compute, observational effects are also simpler to understand. But as a trade of, the comparison with theoretical models is harder than with a more direct $\ssfr$ selection, since it requires models to predict synthetic photometry.

Recently, a number of QGs were identified at $z>3$ with such color selection technique \citep{straatman2014,mawatari2016}. Their observed number density significantly exceeds that predicted by state-of-the-art cosmological simulations, with and without AGN feedback \citep[e.g.,][]{wellons2015,sparre2015,dave2016}, and requires a formation channel at $z>5$ with $\sfr$s larger than observed in the mostly dust-free Lyman-break galaxies (LBGs; e.g., \citealt{smit2012,smit2016}). However, at the time the accuracy of color selections of QGs had not been tested beyond $z\sim2$, and spectroscopic confirmation of their redshifts and properties was needed to back up these unexpected results.

For this reason, we have designed several observing campaigns to obtain near-infrared spectra of these color-selected $z>3$ massive QGs with Keck--MOSFIRE. The first results from this data set were described in \cite{glazebrook2017} (hereafter \citetalias{glazebrook2017}), where we reported the spectroscopic confirmation of the most distant QG at $z\sim3.7$, the first at $z>3$, using Balmer absorption lines. While flags were raised owing to the detection of sub-millimeter emission toward this galaxy by ALMA \citep{simpson2017}, we later demonstrated this emission originates from a neighboring dusty SFG, and provided a deep upper limit on obscured star formation in the QG \citep{schreiber2018}. The confirmed redshift and quiescence of this galaxy (ZF-COS-20115, nicknamed ``Jekyll'') provided the first definite proof that QGs do exist at $z>3$, and the fact that these were found in cosmological surveys of small area (a fraction of a square degree) implies they are not particularly rare.

In this paper, we describe the observations and results for the entire sample of galaxies observed with MOSFIRE. Using this sample, we derived statistics on the completeness and purity of the \uvj color selection at $z>3$, and used this information to derive updated number densities and star formation histories for QGs at these early epochs, to compare them against models.

In \rsec{SEC:sample}, we describe our observations and sample, including in particular the sample selection, the spectral energy distribution (SED) modeling, and the reduction of the spectra. In \rsec{SEC:results_spec} we describe our methodology for the analysis of the spectra, and make an inventory of the observed spectral features, the line properties, and the measured redshifts. In particular, \rsec{SEC:new_colors} discusses the revised \uvj colors. In \rsec{SEC:results_sfh} we discuss the quiescence and inferred star-formation histories for the galaxies with MOSFIRE spectra. In \rsec{SEC:rhogal}, we build on the results of the previous sections to update the number density of quiescent galaxies, using the full ZFOURGE catalogs, and discuss the link between the \uvj selection and the specific $\sfr$. Lastly, \rsec{SEC:discussion} compares our observed number densities and star formation histories to state-of-the-art galaxy evolution models, while \rsec{SEC:conclusion} summarizes our conclusions and lists possibilities for future works.

In the following, we assumed a $\Lambda$CDM cosmology with $H_0 = 70\ {\rm km}\,{\rm s}^{-1} {\rm Mpc}^{-1}$, $\Omega_{\rm M} = 0.3$, $\Omega_\Lambda = 0.7$ and a \cite{chabrier2003} initial mass function (IMF) to derive physical parameters from the photometry and spectra. All magnitudes are quoted in the AB system, such that $m_{\rm AB} = 23.9 - 2.5\log_{10}(S_{\!\nu}\ [\uJy])$.

\section{Sample selection and observations \label{SEC:sample}}

This section describes the sample of galaxies we analyzed in this paper, the new MOSFIRE observations, the associated data reduction, and the analysis of the spectra.

\subsection{Parent catalogs}

The sample studied in this paper consists of $3 < z < 4$, massive ($\mstar\ge2\times10^{10}\,\msun$) galaxies identified using photometric redshifts. The \uvj color-color diagram was then used to separate star-forming and quiescent galaxies \citep{williams2009}. The galaxies were selected either from the ZFOURGE or 3DHST catalogs \citep{skelton2014,straatman2016} in the CANDELS fields EGS/AEGIS, GOODS--South, COSMOS, and UDS \citep{grogin2011,koekemoer2011}. All fields include a wide variety of broadband imaging ranging from the $U$ band up to the \spitzer $8\,\um$ channel. This includes in particular ($5\sigma$ limiting magnitudes quoted for EGS, GOODS-S, COSMOS, and UDS, respectively): deep \hubble imaging in the F606W ($R<26.8$, 27.4, 26.7, 26.8); F814W ($I<26.4$, 27.2, 26.5, 26.8); F125W ($J<26.3$, 26.1, 26.1, 25.8); F160W ($H<26.1$, 26.4, 25.8, 25.9); deep \Ks or $K$-band imaging ($K<24$, 24.8, 25, 24.9); and deep \spitzer $3.6$ and $4.5\,\um$ imaging ($[3.6]<25.2$, 24.8, 25.1, 24.6). The photometry in these catalogs was assembled with the same tools and approaches, namely aperture photometry on residual images cleaned of neighboring sources \citep[see][]{skelton2014,straatman2016}.

The ZFOURGE catalogs supersede the 3DHST catalogs by bringing in additional medium bands from $\lambda=1.05$ to $1.70\,\um$ and deeper imaging in the \Ks band (obtained with the Magellan FourStar camera). The additional near-infrared filters allow a finer sampling of the Balmer break at $z\sim2$--$3$, and more accurate photometric redshifts. However, they only cover a $11\arcmin\times11\arcmin$ region within each of the southern CANDELS fields (GOODS--South, UDS, and COSMOS). We thus used the higher quality data from ZFOURGE whenever possible, and resorted to the 3DHST catalogs outside of the ZFOURGE area. In both cases, we only used galaxies with a flag \texttt{use=1}. In ZFOURGE, we used an older version of the \texttt{use} flags than that provided in the DR1. Indeed, the latter were defined to be most conservative, in that they flag all galaxies which are not covered in all FourStar bands, those missing \hst imaging, or those too close to star spikes in optical ground-based imaging \citep{straatman2016}. This would effectively reduce the covered sky area by excluding galaxies which, albeit missing a few photometric bands, are otherwise well characterized. Instead, we adopted the earlier \texttt{use} flags from \cite{straatman2014}, which are more inclusive.

After the sample was assembled, a few source-specific adjustments were applied to the catalog fluxes. For ZF-COS-17779, we discarded the CFHT photometry which had negative fluxes with high significance (although inspecting the images did not uncover any particular issue). For 3D-EGS-26047 we removed the WirCAM $J$ band which was incompatible with the flux in the surrounding passbands (including the Newfirm $J$ medium bands), and for which the image showed some artifacts close to the source. For 3D-EGS-40032, we discarded the Newfirm photometry because the galaxy was at the edge of the FOV; the noise in the image at this location is higher but the error bars reported in the catalog were severely underestimated, visual inspection of the image revealed no detection. For 3D-EGS-31322, we removed the \spitzer 5.8 and 8$\,\um$ fluxes which were abnormally low; the galaxy is located in a crowded region, and the photometry in these bands may have been poorly de-blended. These modifications are minor, and do not impact our results significantly. Lastly, for ZF-COS-20115 (the \citetalias{glazebrook2017} galaxy) we used the photometry derived in \cite{schreiber2018}, where the contamination from a dusty neighbor ({\it Hyde}) was removed. This reduced the stellar mass of ZF-COS-20115 by $30\%$ and had no impact on its inferred star formation history \citep[see][]{schreiber2018}.

In this paper, our main focus is placed on quiescent galaxies observed with MOSFIRE (this sample is described later in \rsec{SEC:sample_final}). However, to place these galaxies in a wider context, we also considered all massive galaxies in the parent sample at $3<z<4$. For this purpose, we only used the ZFOURGE catalogs (in GOODS--South, COSMOS, and UDS) since they have data of similarly high quality, and are all \Ks-selected (while the 3DHST catalogs were built from a detection image in F125W+F140W+F160W). To further ensure reliable photometry, we only considered galaxies with $K_{\rm s}<24.5$; the impact of this magnitude cut on the completeness is addressed in the next section. We visually inspected the SEDs and images of all galaxies with $\mstar > 10^{10}\,\msun$ to reject those with problematic photometry (3\% of the inspected galaxies). In the end, the covered area was $139$, $150$, and $153\,{\rm arcmin}^2$ in GOODS--South, COSMOS, and UDS, respectively.

\subsection{Initial photometric redshifts and galaxy properties \label{SEC:zphot}}

\begin{table}
\begin{center}
\caption{List of model parameters in our SED modeling. \label{TAB:params}}
\begin{tabular}{lllll}
\hline\hline \\[-0.3cm]
Parameter (unit) & Low & Up & Step & Values \\
\hline \\[-0.3cm]
$t_{\rm burst}$ ($\Gyr$)   & 0.01      & $t_H(z)$ & 0.05$^a$ & 45--50 \\
$\tau_{\rm rise}$ ($\Gyr$) & 0.01      & 3        & 0.1$^a$  & 26 \\
$\tau_{\rm decl}$ ($\Gyr$) & 0.01      & 3        & 0.1$^a$  & 26 \\
$R_{\sfr}$                 & $10^{-2}$ & $10^5$   & 0.2$^a$  & 36 \\
$t_{\rm free}$ ($\Myr$)    & 10        & 300      & 0.5$^a$  & 4 \\
$A_{\rm V}$ (mag)          & 0         & 6        & 0.02     & 61 \\
\hline \\[-0.3cm]
$Z$                        & \multicolumn{4}{c}{$Z_\sun$} \\
$z$                        & \multicolumn{4}{c}{$\zphot$ or $\zspec$} \\
IMF                        & \multicolumn{4}{c}{\cite{chabrier2003}} \\
Attenuation curve          & \multicolumn{4}{c}{\cite{calzetti2000}} \\
Stellar population         & \multicolumn{4}{c}{\cite{bruzual2003}} \\
\hline
\end{tabular}
\end{center}
{\footnotesize $^a$ Logarithmic step, in dex.}
\end{table}

\begin{figure}
\begin{center}
\includegraphics[width=0.5\textwidth]{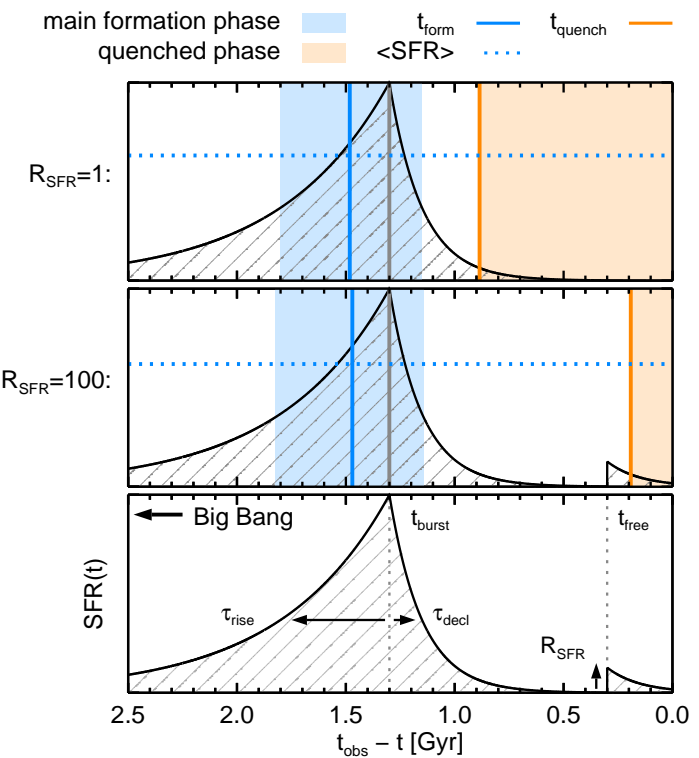}
\end{center}
\caption{Illustration of the adopted star formation history parametrization (bottom) and the marginalized parameters (middle and top). We show the time of peak $\sfr$ (solid gray line, here coinciding with $t_{\rm burst}$), the star-forming phase surrounding it (shaded in pale blue), and the mean $\sfr$ during this phase (horizontal blue dotted line). We also display the time of quenching $t_{\rm quench}$ (orange solid line) and the following quenched phase (shaded in pale orange). Finally, the time at which the galaxy had formed $50\%$ its stars ($t_{\rm form}$) is shown with a blue solid line. \label{FIG:sfh}}
\end{figure}

The photometric redshifts ($\zphot$), rest-frame colors ($U-V$ and $V-J$), and stellar masses ($\mstar$) provided in the ZFOURGE and 3DHST catalogs were computed with the same softwares, namely EAzY \citep{brammer2008} and FAST \citep{kriek2009}, albeit with slightly different input parameters. These values were used to build the MOSFIRE masks in the different observing programs. However, to ensure the most homogeneous data set for our analysis, we recomputed redshifts, colors, and masses for all galaxies once the sample was compiled, using a uniform setup for all fields and taking advantage of all the available photometry. This setup is described below.

Photometric redshifts and rest-frame colors were obtained with the latest version of EAzY\footnote{Commit \texttt{\#5590c4a} (19/12/2017) on \url{https://github.com/gbrammer/eazy-photoz}.} and the galaxy template set ``\texttt{eazy\_v1.3}'', which includes in particular a ``old and dusty'' and a ``high-equivalent-width emission line'' template. These additional templates were also used in the original ZFOURGE catalogs, but not in 3DHST. We also did not enable the redshift prior based on the $K$-band magnitude since this prior is based on models which do not reproduce the high redshift mass functions (see discussion in \rsec{SEC:discussion}). The resulting scatter in photometric redshifts was $5\%$ when comparing our new redshifts to that published by ZFOURGE for the entire catalog at $z>3$, and $7\%$ for the quiescent galaxies (described later in \rsec{SEC:sample_final}).

Stellar masses and $\sfr$s were re-computed using FAST++\footnote{\url{https://github.com/cschreib/fastpp}} v1.2 with the same setup as in \cite{schreiber2018}, but with more refined star-formation histories. Briefly, we assumed $z=\zphot$, the \cite{bruzual2003} stellar population model, the \cite{chabrier2003} initial mass function (IMF), and the dust screen model of \cite{calzetti2000} with $A_{\rm V}$ up to $6$ mag. The only notable difference with the published ZFOURGE and 3DHST catalogs is that we assumed a more elaborate functional form for the star formation history (SFH), which consisted of two main phases: an exponentially rising phase followed by an exponentially declining phase, both with variable $e$-folding times $\tau_{\rm rise}$ and $\tau_{\rm decl}$, respectively:
\begin{align}
\sfr_{\rm base}(t) \propto \left\{\begin{array}{ll}
e^{(t_{\rm burst} - t)/\tau_{\rm rise}} & \text{for $t > t_{\rm burst}$,} \\
e^{(t - t_{\rm burst})/\tau_{\rm decl}} & \text{for $t \le t_{\rm burst}$,} \\
\end{array}\right.\label{EQ:sfh_noburst}
\end{align}
where $t$ is the ``lookback'' time ($t=0$ is the point in time when the galaxy is observed, $t>0$ is in the galaxy's past). This was performed assuming $z=\zphot$ initially, and later on with $z=\zspec$ (\rsec{SEC:fast_fit}). Varying the lookback time $t_{\rm burst}$ that separates these two epochs, this allowed us to describe a large variety of SFHs, including rapidly or slowly rising SFHs, constant SFHs, and rapidly or slowly quenched SFHs (see \citealt{schreiber2018} for a more detailed description of this model). Allowing rising SFHs in particular can prove crucial to properly characterize massive SFGs at high redshift \citep{papovich2011}. We varied $t_{\rm burst}$ from $10\,\Myr$ to the age of the Universe (at most $2\,\Gyr$ at $z>3$), and $\tau_{\rm rise}$ and $\tau_{\rm decl}$ from $10\,\Myr$ to $3\,\Gyr$, all with logarithmic steps ($0.05\,\dex$ for $t_{\rm burst}$, $0.1\,\dex$ for $\tau_{\rm rise}$ and $\tau_{\rm decl}$).

In addition, following \cite{ciesla2016,ciesla2017} and \citetalias{glazebrook2017}, we decoupled the current $\sfr$ from the past history of the galaxy by introducing a free multiplicative factor to the instantaneous $\sfr$ within a short period, of length $t_{\rm free}$, preceding observation:
\begin{align}
\sfr(t) = \sfr_{\rm base}(t)\times\left\{\begin{array}{ll}
1 & \text{for $t > t_{\rm free}$,} \\
R_{\rm SFR} & \text{for $t \le t_{\rm free}$.} \\
\end{array}\right.\label{EQ:sfh}
\end{align}
We considered values of $t_{\rm free}$ ranging from $10$ to $300\,\Myr$, and values of $R_{\rm SFR}$ ranging from $10^{-2}$ to $10^5$ (i.e., either abrupt quenching or bursting), with logarithmic steps of $0.5$ and $0.2\,\dex$, respectively. We emphasize that this additional parameter is not directly linked to quenching, as a galaxy may still have a very low $\ssfr$ from \req{EQ:sfh_noburst} alone (see \rfig{FIG:sfh}). In fact, as discussed later in \rsec{SEC:two_sfh}, this additional freedom had little impact on the quiescent galaxies beside marginally increasing the uncertainty on the SFH, however we find it is necessary to properly reproduce the bulk properties of the star-forming galaxies. In particular, without this extra freedom the mean $\ssfr$ of main-sequence galaxies was too low by a factor of about three compared to stacked \herschel and ALMA measurements (this is also an issue affecting the $\sfr$s provided in the original ZFOURGE and 3DHST catalogs).

This model is illustrated in \rfig{FIG:sfh}, and the parameters with their respective bounds are listed in \rtab{TAB:params}. Over $200$ million models were considered for each galaxy, and the fit could be performed on a regular desktop machine in less than a day thanks to the numerous optimizations in FAST++. The adopted parametrization described above may seem overly complex, and indeed most of the free parameters in \reqs{EQ:sfh_noburst} and \ref{EQ:sfh} have little chance to be constrained accurately. This was not our goal however, since we eventually marginalized over all these parameters to compute more meaningful quantities, such as the current $\sfr$ and stellar mass, and non-parametric quantities describing the SFH (see \rfig{FIG:sfh} and \rsec{SEC:fast_fit}). The point of introducing such complexity is therefore to allow significant freedom on the SFH, to avoid forcing too strong links between the current and past $\sfr$, as well as to obtain accurate error bars on the aforementioned quantities. A similar approach was adopted in \citetalias{glazebrook2017}.

We then compared our best-fit values to that initially given in the ZFOURGE and 3DHST catalogs. Considering all galaxies at $3<\zphot<4$ and $\mstar > 10^{10}\,\msun$, we find a scatter in stellar masses of $0.07\,\dex$ with a median increase of $+0.04\,\dex$ (our new masses are slightly larger), while the scatter in $\sfr$ is $0.34\,\dex$ and a median increase of $+0.26\,\dex$ (our $\sfr$s are substantially larger).

To estimate the completeness in mass of our sample resulting from our $K_{\rm s}<24.5$ magnitude cut, we binned galaxies in $\ssfr$ and computed in each bin the 80th percentile of the mass-to-light ratio in $K$, $\mean{\mstar/L_K}$ (where $L_K$ is the luminosity in the observed \Ks band and $\mstar$ is the best-fit stellar mass obtained with FAST++). We note that this method accounts for changes in $M/L$ caused both by variations in stellar populations as well as variation in dust obscuration. Since galaxies with low $\ssfr$ tend to be less obscured at fixed mass (\citealt{wuyts2011-a}), these two effects work in opposite directions and can lead to a weaker evolution of $M/L$ with $\ssfr$. In practice, we find $\mean{\mstar/L_K} = 1.6\,\msun/\lsun$ for $\ssfr = 10^{-3}\,\Gyr^{-1}$, and $0.24\,\msun/\lsun$ for $\ssfr = 10\,\Gyr^{-1}$. Our adopted magnitude cut of $K_{\rm s}<24.5$ implies $\mstar > 2.3\times10^{10}\,\mean{\mstar/L_K}$ at $z=3.5$, hence a $80\%$ completeness down to $3.7\times10^{10}\,\msun$ for $\ssfr < 10^{-3}\,\Gyr^{-1}$ (this is consistent with the value obtained in \citealt{straatman2014}), and a factor seven lower at $\ssfr = 10\,\Gyr^{-1}$.

\subsection{MOSFIRE masks and runs \label{SEC:masks}}

\begin{table*}
\begin{center}
\caption{MOSFIRE masks used in this paper.\label{TAB:masks}}
\begin{tabular}{llllccccc}
\hline\hline \\[-0.3cm]
Mask    & PI         &  \multicolumn{2}{l}{Observing date}   & \multicolumn{2}{c}{Integration time} &
\multicolumn{2}{c}{Average seeing} & Quiescent \\
        &            & $H$                 & $K$                    & $H$   & $K$   & $H$         & $K$         & candidates \\
\hline \\[-0.3cm]
COS-W182 & Glazebrook & 2016-Feb-26, 27 & 2016-Jan-8, 2016-Feb-27 & 3.9h & 7.2h & 0.75" & 0.61" & 5 \\
COS-U069 & Illingworth & 2014-Dec-16 & 2014-Dec-16 & 0.3h & 3.6h & 0.80" & 0.55" & 2 \\
COS-Z245 & Kewley & -- & 2017-Feb-14 & -- & 1.6h & -- & 0.61" & 2 \\
COS-Y259-A & Oesch & -- & 2014-Dec-13 & -- & 3.3h & -- & 0.71" & 1 \\
COS-Y259-B & Oesch & -- & 2014-Dec-14 & -- & 2.0h & -- & 0.57" & 1 \\
EGS-W057 & Glazebrook & 2017-Feb-13, 14 & 2016-Feb-26, 27 & 0.8h & 4.8h & 0.63" & 0.65" & 6 \\
UDS-W182 & Glazebrook & 2016-Jan-8 & 2016-Jan-8 & 0.3h & 2.4h & 0.69" & 0.65" & 4 \\
UDS-U069 & Illingworth & -- & 2014-Dec-16 & -- & 4.7h & -- & 0.66" & 1 \\
UDS-Y259-A & Oesch & -- & 2014-Dec-13 & -- & 4.9h & -- & 0.63" & 5 \\
UDS-Y259-B & Oesch & -- & 2014-Dec-14 & -- & 4.0h & -- & 0.75" & 4 \\
\hline
\end{tabular}
\end{center}
\end{table*}

MOSFIRE \citep{mclean2012} is a multi-object infrared spectrograph installed on the Keck I telescope, on top of Mauna Kea in Hawaii. Its field of view of $6\arcmin\times3\arcmin$ can be used to simultaneously observe up to 46 slits per mask, with a resolving power of $R\sim3500$ in a single band ranging from $Y$ ($0.97\,\um$) to $K$ ($2.41\,\um$). The data presented here make use only of the $H$ and $K$ bands.

The quiescent galaxies studied in this paper were observed by four separate MOSFIRE programs comprising 10 different masks, listed in \rtab{TAB:masks}. All masks contained a bright ``slit star'', detected in each exposure, which was used a posteriori to measure the variations of seeing, alignment, and effective transmission with time (see \rapp{APP:reduction}). Slits were configured with the same width of $0.7\arcsec$ (except for the mask COS-Y259-A which had $0.9\arcsec$ slits), and masks were observed with the standard ``ABBA'' pattern, nodding along the slit by $\pm1.5\arcsec$ around the target position. Individual exposures lasted $120$ and $180$s in the $H$ and $K$ bands, respectively.

The first program was primarily targeting $z\sim3.5$ quiescent galaxies (PI: Glazebrook), and observed one mask in EGS, one mask in COSMOS, and one mask in UDS (masks COS-W182, UDS-W182, and EGS-W057). Each mask was observed in the $H$ and $K$ filters, with on-source integration times ranging from $0.3$ to $3.9$ hours in $H$, and $2.4$ to $7.2$ hours in $K$. The masks were filled in priority with quiescent galaxy candidates identified in \cite{straatman2014} (or from the 3DHST catalogs for EGS), and our MOSFIRE observations for the brightest of these galaxies were already discussed in \citetalias{glazebrook2017}. The remaining slits were filled with massive $z\sim4$ star-forming galaxies, and $z\sim2$ galaxies; these fillers are not discussed in the present paper, and were only used for alignment correction and data quality tests. The SEDs of all galaxies were visually inspected, and this determined their relative priorities in the mask design.

The second and third programs (PIs: Oesch, Illingworth) were more broadly targeting massive galaxies at $2<z<3.6$ identified in the 3DHST catalogs, and quiescent galaxies were not prioritized over star-forming ones (see \citealt{vandokkum2015}). These programs consisted of multiple masks in EGS, COSMOS and UDS, however all the quiescent candidates in EGS were at $z<3$. We thus only used a total of three masks in COSMOS, and three masks in UDS (masks COS-Y259-A, COS-Y259-B, UDS-Y259-A, UDS-Y259-B, COS-U069, and UDS-U069). Only one mask was observed in the $H$ band for $0.3$h, and all masks were observed in $K$ with integration times ranging from $2.0$ to $4.9$h.

The fourth and last program is the MOSEL emission line survey (PI: Kewley). This program observed several masks, in which massive $z\sim4$ galaxies from ZFOURGE were only observed as fillers. Only two quiescent galaxy candidates were actually observed in one mask of the COSMOS field (mask COS-Z245), where $1.6$h was spent observing in the $K$ band. One of them was the galaxy described in \citetalias{glazebrook2017}, for which the red end of the $K$ was observed to cover the absent $\oiii$ emission line.

\subsection{Observed sample \label{SEC:sample_final}}

\begin{figure*}
\begin{center}
\includegraphics[width=0.9\textwidth]{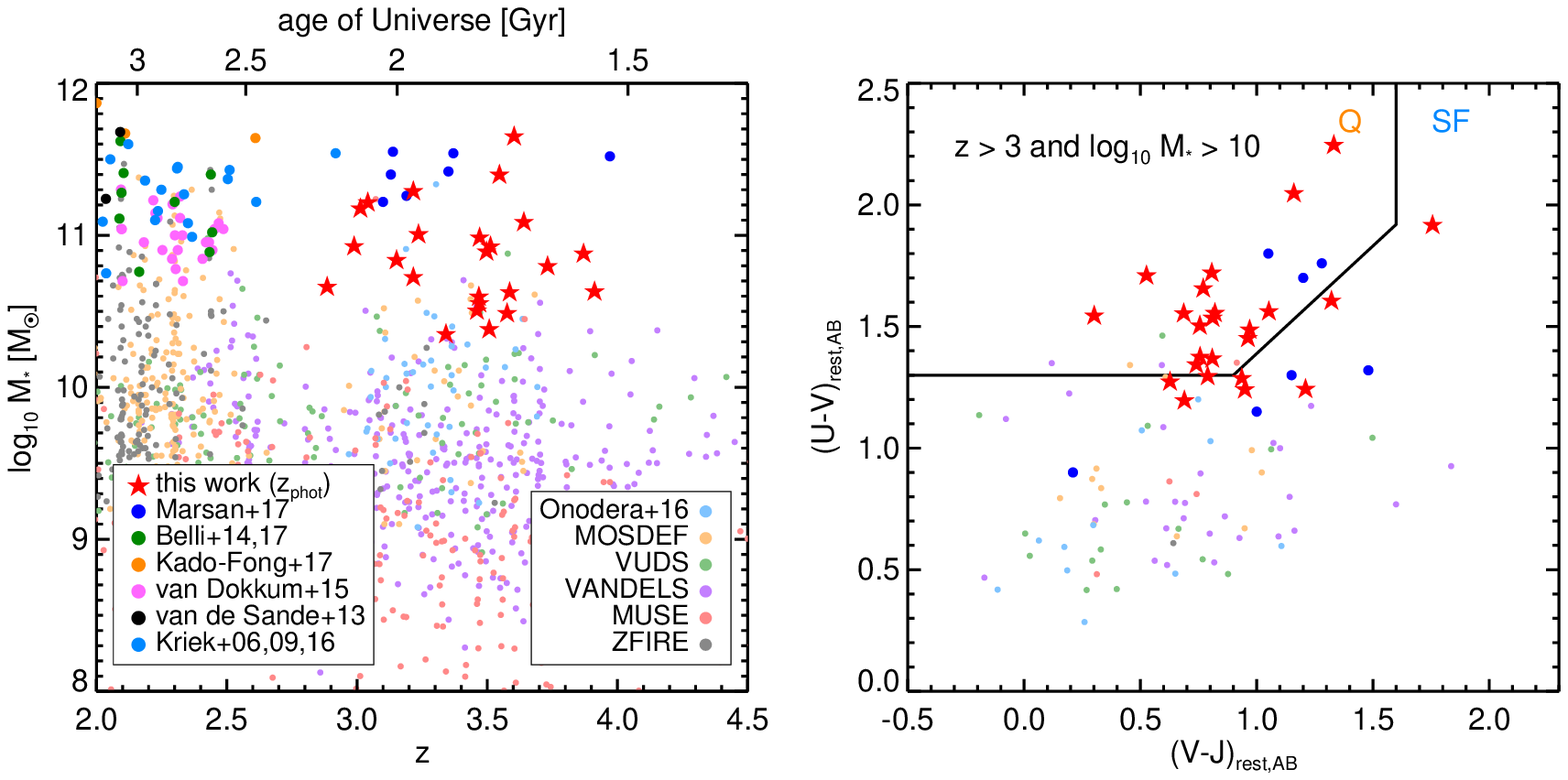}
\end{center}
\caption{{\bf Left:} Stellar mass as a function of redshift for galaxies with public spectroscopic redshifts from the literature (circles) and galaxies from our sample with photometric redshifts (red stars, using photometric redshifts). We show the sample of massive $z>3$ galaxies from \cite{marsan2017} in dark blue, the sample of \cite{onodera2016} in light blue, the quiescent $z\sim2$ galaxies of \cite{belli2014,belli2017} in dark green and \cite{kado-fong2017} in orange, the compact $z\sim2$ star-forming galaxies of \cite{vandokkum2015} in pink, the quiescent galaxies observed in \cite{kriek2006,kriek2009,kriek2016} in medium blue, the quiescent galaxies from \cite{vandesande2013} in black, the galaxies observed by MOSDEF \citep{kriek2015} in orange, the galaxies observed by VUDS \citep{tasca2017} in green, the galaxies observed by VANDELS \citep{mclure2017} in purple, the galaxies in the MUSE deep fields \citep{inami2017} in light pink, and the targets of the ZFIRE program in gray \citep{nanayakkara2016}. {\bf Right:} \uvj color-color diagram for a subset of galaxies shown on the left, limited to $z>3$ and $\mstar > 10^{10}\,\msun$. The $(U-V)$ and $(V-J)$ colors were computed in the rest frame, in the AB system. The black line delineates the standard dividing line between quiescent (Q) and star-forming (SF) galaxies, as defined in \cite{williams2009}.\label{FIG:mz}}
\end{figure*}

\begin{figure*}
\begin{center}
\includegraphics[width=\textwidth]{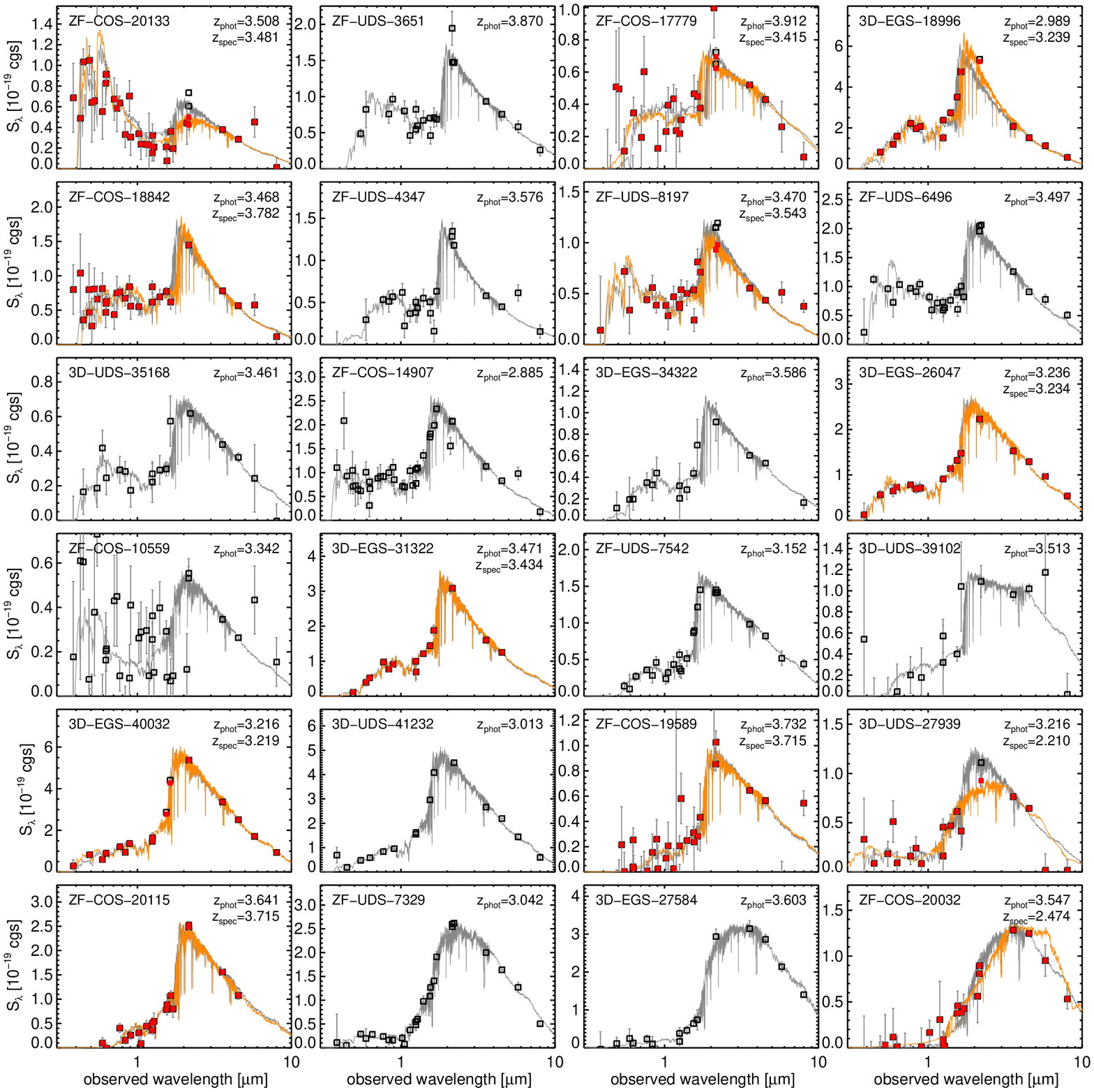}
\end{center}
\caption{Spectral energy distributions of the galaxies in our target sample, sorted by increasing observer-frame $z-K$ color (rest-frame ${\rm NUV} - g$ at $z=3.5$). The observed photometry is shown with open black squares and gray error bars, and the best-fitting stellar continuum template from FAST++ obtained assuming $z=\zphot$ is shown in gray in the background. For galaxies with a measured spectroscopic redshift (see \rsec{SEC:redshifts}), we display the best-fitting template at $z=\zspec$ in orange, and the photometry corrected for emission line contamination with red squares. \label{FIG:allseds}}
\end{figure*}

From the MOSFIRE masks described in the previous section, we extracted all the galaxies with $\zphot>2.8$, $\mstar\ge2\times10^{10}\,\msun$ and \uvj colors satisfying the \cite{williams2009} criterion with a tolerance threshold of $0.2\,{\rm mag}$. The resulting $24$ quiescent galaxy candidates are listed in \rtab{TAB:galaxies}, and their properties determined from the photometry alone (\rsec{SEC:zphot}) are listed in \rtab{TAB:galaxies_zphot}. The photometric redshifts ranged from $\zphot=2.89$ up to $3.91$, and stellar masses ranged from $\mstar=2.3\times10^{10}$ to $4.5\times10^{11}\,\msun$, as illustrated in \rfig{FIG:mz}. The broadband SEDs and best fit models using $\zphot$ are shown in \rfig{FIG:allseds}.

Some of our targets were observed in multiple MOSFIRE masks, and have accumulated more exposure time than the rest of the sample. In particular, ZF-COS-20115 (already described in \citetalias{glazebrook2017}) was observed for a total of $14.4$h in the $K$ band and $4.2$h in $H$. Other galaxies have exposure times ranging from $1.6$h to $7.3$h in the $K$ band, and zero to $3.9$h in $H$. The resulting line sensitivities are discussed in \rsec{SEC:sensitivity}.

In \rfig{FIG:mz} we compare this sample to recent spectroscopic campaigns targeting high-redshift galaxies. With the exception of the sample studied in \cite{marsan2017}, massive galaxies at $z>3$ have so far received very limited spectroscopic coverage, and the situation is even worse for quiescent galaxies. Priority is often given to lower mass, bluer galaxies, for which redshifts can be more easily obtained with emission lines. Indeed, we checked that, despite being selected in the well studied CANDELS fields, none of our targets were observed by the largest spectroscopic programs (MOSDEF, VUDS, and VANDELS). The only exception is ZF-COS-20115 which was observed by MOSDEF for $1.6$h in $K$; we did not attempt to combine these data with our own given that this galaxy was already observed for $14$ hours and such a small increment would not bring significant improvement.

\begin{figure}
\begin{center}
\includegraphics[width=0.5\textwidth]{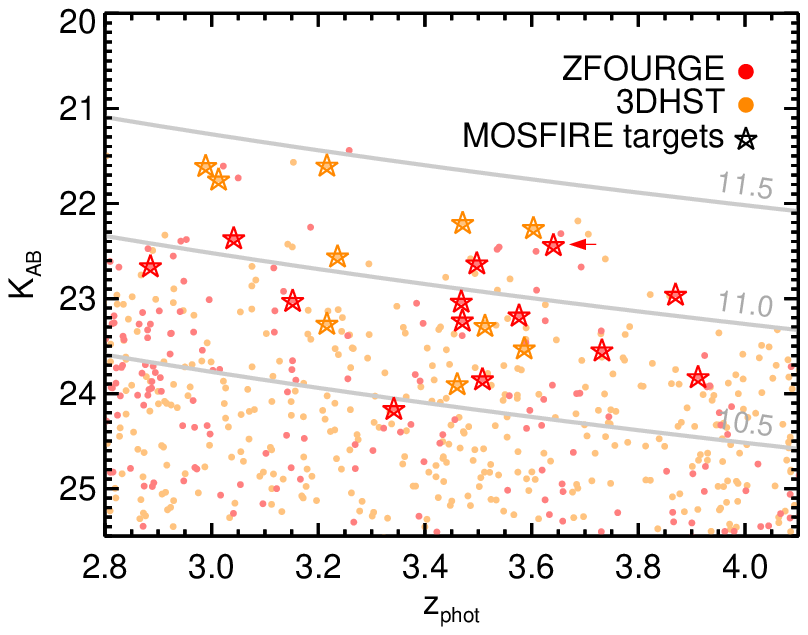}
\end{center}
\caption{$K$-band magnitude as a function of photometric redshift for \uvj quiescent galaxies (with a $0.2\,{\rm mag}$ threshold on the \uvj diagram). The quiescent galaxies from ZFOURGE are shown in red, while those from 3DHST in UDS and EGS are shown in orange. The stars indicate the galaxies for which we collected MOSFIRE spectra. The red arrow indicates the position of the $z=3.7$ galaxy first discussed in \citetalias{glazebrook2017}. The gray solid lines show the $K$-band magnitude corresponding to different stellar masses, $3\times10^{10}$, $10^{11}$, and $3\times10^{11}\,\msun$ (assuming $\mstar/L_K=\msun/\lsun$). \label{FIG:selection}}
\end{figure}

Combining data from these different programs, the collected MOSFIRE data have a non-trivial selection function. In some programs, galaxies were prioritized based on how clean their SEDs looked, which can bias our sample toward those quiescent candidates with the best photometry, or those with a more pronounced Balmer break. In addition, samples drawn from the 3DHST catalogs also tend to have lower redshifts and brighter magnitudes than that drawn from the ZFOURGE catalogs, as could be expected based on the different selection bands and depths in these two catalogs. Yet, as shown in \rfig{FIG:selection}, the combined sample does homogeneously cover the magnitude-redshift or mass-redshift space for quiescent galaxies, within $3<z<4$ and $\mstar>4\times10^{10}\,\msun$ (or $K<23.5$). We thus considered this spectroscopic sample to be fairly representative of the overall \uvj-quiescent population at these redshifts.

\subsection{Reduction of the spectra}

The reduction of the raw frames into 2D spectra was performed using the MOSFIRE pipeline as in \cite{nanayakkara2016}. However, since we were mostly interested in faint continuum emission, we performed additional steps in the reduction to improve the signal-to-noise and the correction for telluric absorption. The full procedure is described in \rapp{APP:reduction}, and can be summarized as follows.

All masks were observed with a series of standard ABBA exposures, nodding along the slit. For each target, rather than stacking all these exposures into a combined 2D spectrum, we reduced all the individual ``${\rm A}-{\rm B}$'' exposures separately and extracted a 1D spectrum for each pair of exposures. These spectra were optimally extracted with a Gaussian profile of width determined by the time-dependent seeing (hence, assuming the galaxies are unresolved), and were individually corrected for telluric correction and effective transmission using the slit star. Using the slit star rather than a telluric standard observed during the same night, we could perform the telluric and transmission correction for each exposure separately, rather than on the final data. This correction included slit loss correction, calibrated for point sources (see next section for the correction to total flux). The individual spectra were then optimally combined, weighted by inverse variance, to form the final spectrum. This approach allows to automatically down-weight exposures with poorer seeing. Flux uncertainties in each spectral element were determined by bootstrapping the exposures, and a binning of three spectral elements was adopted to avoid spectrally-correlated noise. This resulted in an average dispersion of $\lambda/\Delta\lambda\sim3000$, which is close to the nominal resolution of MOSFIRE with $0.7\arcsec$ slits. Further binning or smoothing were used for diagnostic and display purposes, but all the science analysis was performed on these $\lambda/\Delta\lambda\sim3000$ spectra. For this and in all that follows, binning was performed with inverse variance weighting, in which regions of strong OH line residuals were given zero weight.

\subsection{Rescaling to total flux}

Our procedure for the transmission correction includes the flux calibration, as well as slit loss correction. However, because the star used for the flux calibration is a point source, the slit loss corrections are only valid if our science targets are also point-like (angular size $\ll 0.6\arcsec$, the typical seeing, see \rtab{TAB:masks}). If not, additional flux is lost outside of the slit and has to be accounted for.

We estimated this additional flux loss by analyzing the $H$ and $K$ broadband images of our targets, convolved with a Gaussian kernel if necessary to match our average seeing (see \citealt{nanayakkara2016}). We simulated the effect of the slit by measuring the broadband flux $S_{\rm slit}$ in a rectangular aperture centered on each target and with the same position angle as in the MOSFIRE mask, and by measuring the ``total'' flux in a $2\arcsec$ diameter aperture, $S_{\rm tot}$. Since our transmission correction already accounted for slit loss for a point source, we also measured the fraction of the flux in the slit for a Gaussian profile of width equal to the seeing, $f_{\rm PSF, slit}$. We then computed the expected slit loss correction for extended emission as $S_{\rm tot} \times f_{\rm PSF, slit}/S_{\rm slit}$. The obtained values ranged from $1.0$ (no correction) to $1.8$ with a median of $1.2$, and were multiplied to the 1D spectra.

We then compared the broadband fluxes from the ZFOURGE or 3DHST catalogs against synthetic fluxes generated from our spectra, integrating flux within the filter response curve of the corresponding broadband. Selecting targets which have a synthetic broadband flux detected at $>$$10\sigma$, we find that our corrections missed no more than $30\%$ of the total flux, with an average of $10\%$. For fainter targets, this number reached at most $150\%$, and the highest values are found for the three faintest targets of the EGS-W057 mask (3D-EGS-26047, 3D-EGS-27584 and 3D-EGS-34322). While one of these three is intrinsically faint and thus has an uncertain total flux, the other two were expected to be detected with a synthetic broadband $S/N$ of $19$ and $25$, but we find only $9$ and $7$, respectively. This may suggest a misalignment of the slits for these particular targets. To account for this and other residual flux loss, we finally rescaled all our spectra to match the ZFOURGE or 3DHST photometry. We only performed this correction if the continuum was detected at more than $5\sigma$ in the spectrum, to avoid introducing additional noise.

We noted that one galaxy's average flux in the $H$ band was negative (3D-EGS-27584), and we also observed a strong negative trace in its stacked 2D spectrum. Because this galaxy is close ($1.5\arcsec$) to a bright $z\sim1$ galaxy, we suspect that some of the bright galaxy's flux contaminated the $H$ band. Regardless of the cause, this $H$-band spectrum was unusable. However, and perhaps owing to the neighboring galaxy being fainter in $K$, the $K$-band spectrum appeared unaffected and the target galaxy's continuum was well detected; we thus kept it in our sample and simply discarded the $H$-band spectrum.

\subsection{Achieved sensitivities \label{SEC:sensitivity}}

The achieved spectral sensitivities and $S/N$ in coarse $70\,\AA$ bins ($\sim$$1000\,\kms$) are listed for all our targets in \rtab{TAB:galaxies_limits}. We describe in more detail the derivation of these uncertainties and their link to spectral binning in \rapp{SEC:uncertainty}. Because our sample is built from masks with different exposure times, the average sensitivity can vary from one galaxy to the next. In practice, the median sensitivity ($1\sigma$, [min, median, max]) ranges between $[0.4,0.7,0.9]\times10^{-19}\,{\rm erg/s/cm^2/\AA}$ in $H$ band, and $[0.2,0.5,0.9]\times10^{-19}\,{\rm erg/s/cm^2/\AA}$ in $K$, which resulted in continuum $S/N$ of $[0.4,1.3,7.1]$ and $[0.7,3.6,12]$, respectively (these ranges reflect variation within our sample, and not variations of sensitivity within a given spectrum).

In terms of line luminosity at $z=3.5$, assuming a width of $\sigma=300\,\kms$, these correspond to $3\sigma$ detection limits of $[0.5,0.9,1.1]\times10^{42}\,{\rm erg/s}$ in $H$ band, and $[0.3,0.6,1.1]\times10^{42}\,{\rm erg/s/cm^2/\AA}$ in $K$. For $\hbeta$ in $K$ and assuming no dust obscuration, this translates into $3\sigma$ limits on the $\sfr$ of $[4,9,16]\,\msun/\yr$ (see \rsec{SEC:sfr_comp} for the conversion to $\sfr$). For the massive galaxies in our sample, this is a factor $[0.04,0.11,0.20]$ times the main sequence $\sfr$. With $A_V=2\,{\rm mag}$, this is increased to a factor $[0.39,0.98,1.8]$. Therefore, on average, our spectra are deep enough to detect low levels of unobscured star-formation, or obscured star-formation in main-sequence galaxies.

Finally, given the observed $K$-band magnitudes of our targets and considering the median uncertainties listed above, these spectra allow us to detect lines contributing at least [0.3,1.1,3.8]\% of the observed broadband flux (resp.~[min,median,max] of our sample). This suggests we should be able to determine, in all our targets, if emission lines contribute significantly to their observed Balmer breaks. However this is assuming a constant uncertainty over the entire $K$ band, which is optimistic. Indeed, a fraction of the wavelength range covered by the MOSFIRE spectra is rendered un-exploitable because of bright sky line residuals.

To quantify this effect for each galaxy, we set up a line detection experiment in which we simulated the detection of a single line, of which we varied the full width from $\Delta\lambda=100$ to $1000\,\kms$, and the central wavelength $\lambda_0$ within the boundaries of the $K$ filter passband. In each case, we computed the line flux required for the line to contribute $f=10\%$ to the observed broadband flux, accounting for the broadband filter transmission at the line's central wavelength. For simplicity, here we assume that the line has a tophat velocity profile and that the filter response does not vary over the wavelength extents of the line. By definition,
\begin{equation}
S_{\rm BB} = \frac{\int \dd \lambda\,R(\lambda)\,S_\lambda(\lambda)}{\int \dd \lambda\,R(\lambda)}\,
\end{equation}
where $S_{\rm BB}$ is the observed broadband flux density (e.g., in ${\rm erg/s/cm^2/\AA}$), $R(\lambda)$ is the broadband filter response, and $S_\lambda(\lambda)$ is the spectral energy distribution of the galaxy. Decomposing $S_\lambda$ into a line and a continuum components, and with the above assumptions, we can extract the line peak flux density
\begin{equation}
S_{\rm line}(f, \lambda_0, \Delta\lambda) = f\,S_{\rm BB}\,\frac{\int \dd \lambda\,R(\lambda)}{\Delta \lambda\,R(\lambda_0)}\,.
\end{equation}

For each galaxy, we then compared this line flux against the observed error spectrum, and computed the fraction of the $K$ passband where such a line could be detected at more than $5\sigma$ significance. At fixed integrated flux, narrower lines should have a higher peak flux and thus be easier to detect, but they can also totally overlap with a sky line and become practically undetectable, contrary to broader lines. As we show below, in practice these two effects compensate such that the line detection probability does not depend much on the line width.

We find that narrow lines ($100\,\kms$) can be detected over [73,82,92]\% of the $K$ passband, while broad lines ($500\,\kms$) can be detected over [77,86,96]\% (resp.~[min,median,max] of our sample). Therefore the probability of missing a bright emission line, which we adopted as the average probability for the narrow and broad lines, is typically $15\%$ per galaxy. The highest value is $27\%$ (3D-UDS-35168) and is in fact more caused by lack of overall sensitivity toward the red end of the $K$ band rather than by sky lines. We used these numbers later on, when estimating detection rates, by attributing a probability of missed emission line to each galaxy.

\subsection{Archival ALMA observations \label{SEC:alma}}

We cross matched our sample of quiescent galaxies with the ALMA archive and find that nine galaxies were observed, all in Band 7 except ZF-COS-20115 which was also observed in Band 8. The majority (ZF-COS-10559, ZF-COS-20032, ZF-COS-20115, ZF-UDS-3651, ZF-UDS-4347, ZF-UDS-6496, and 3D-UDS-39102) were observed as part of the ALMA program 2013.1.01292.S (PI: Leiton), which we introduced in \cite{schreiber2017}. ZF-COS-20115 was also observed in Band 8 in 2015.A.00026.S (PI: Schreiber; \citealt{schreiber2018}), ZF-UDS-6496 was also observed in 2015.1.01528.S (PI: Smail), while 3D-UDS-27939 and 3D-UDS-41232 were observed in 2015.1.01074.S (PI: Inami).

We measured the peak fluxes of all galaxies on the primary-beam-corrected ALMA images, and determined the associated uncertainties from the pixel RMS within a $5\arcsec$ diameter annulus around the source. Parts of the programs 2015.1.01528.S and 2015.1.01074.S were observed at high resolution (FWHM of $0.2\arcsec$) which may resolve the galaxies, therefore we re-reduced the images from these two programs with a tapering to $0.4\arcsec$ and $0.7\arcsec$ resolution, respectively, before measuring the fluxes (these were the highest values we could pick while still providing a reasonable sensitivity of about $0.3\,\mJy$ RMS). For ZF-COS-20115 we used the flux reported in \cite{schreiber2018}, after de-blending it from its dusty neighbor, resulting in a non-detection. In total, two quiescent galaxiy candidates were thus detected, ZF-COS-20032 and 3D-UDS-27939, with no significant spatial offset ($<0.2\arcsec$). As we show below, these are dusty redshift interlopers for which we detected $\halpha$ emission; we kept them in our analysis regardless, since they provide important statistics on the rate of interlopers. Since both galaxies are spatially extended, we used their integrated flux as measured from $(u,v)$ plane fitting using \texttt{uvmodelfit} (as in \citealt{schreiber2017}). Excluding ZF-COS-20032, 3D-UDS-27939, and ZF-COS-20115, the stacked ALMA flux of the remaining galaxies is $0.07\pm0.11\,\mJy$ (using inverse variance weighting), indicating no detection. The collected fluxes are listed in \rtab{TAB:galaxies}.

\section{Redshifts and line properties \label{SEC:results_spec}}

Here we describe the newly obtained spectroscopic redshifts, how they were measured, and how they compare to photometric redshifts. We also discuss the properties of the identified emission and absorption lines, and what information they provide on the associated galaxies.

\subsection{Redshift identification method and line measurements \label{SEC:redshifts_method}}

The spectra were analyzed with \texttt{slinefit}\footnote{\url{https://github.com/cschreib/slinefit}} to measure the spectroscopic redshifts. Using this tool, we modeled the observed spectrum of each galaxy as a combination of a stellar continuum model and a set of emission lines. The continuum model was chosen to be the best-fit FAST++ template obtained at $z=\zphot$ (see \rsec{SEC:zphot}). The emission lines were assumed to have a single-component Gaussian velocity profiles, and to share the same velocity dispersion. The line doublets of $\oiii$ and $\nii$ were fit with fixed flux ratios of $0.3$, $\oii$ with a flux ratio of one, and $\sii$ with a flux ratio of $0.75$, otherwise the line ratios were left free to vary. Emission lines with a negative best-fit flux were assumed to have zero flux, and the fit was repeated without these lines; we therefore assumed that the only allowed absorption lines had to come from the stellar continuum model from FAST++. This continuum model was convolved with a Gaussian velocity profile to account for the stellar velocity dispersion $\sigma_\ast$. Based on the empirical relation with the stellar mass observed at $z\sim2$ in \cite{belli2017}, we assumed $\log_{10}(\sigma_\ast/(\kms)) = 2.4 + 0.33 \times \log_{10}(\mstar/10^{11}\,\msun)$.

The photometry was not used in the fit. Since we took particular care in the flux and telluric calibration of our spectra, we did not fit any additional color term to describe the continuum flux, a method sometimes introduced to address shortcomings in the continuum shape of observed spectra \citep[e.g.,][]{cappellari2004}. Even without such corrections, the reduced $\chi^2$ of our fits are already close to unity (\rtab{TAB:galaxies_zspecs}), indicating that the quality of the fits are excellent and further corrections are not required. Furthermore, as discussed below, all the spectroscopic redshifts we measured are anchored on emission or absorption features anyway, which are not affected by such problems.

\begin{figure}
\begin{center}
\includegraphics[width=0.5\textwidth]{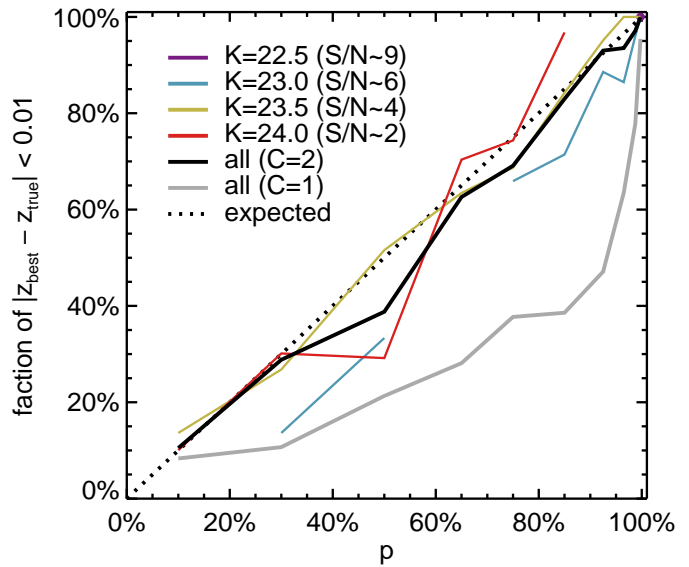}
\end{center}
\caption{Calibration of the criterion for redshift reliability, $p$, using simulated spectra. The $p$ value quantifies the probability that the measured redshift lies within $\Delta z = 0.01$ of the true redshift. The $x$-axis shows the $p$ value estimated from the $P(z)$ of the simulated spectra, and the $y$-axis is the actual fraction of the simulated redshift measurements that lie within $\Delta z= 0.01$ of the true redshift. The line of perfect agreement is shown wish a dashed black line. The relation obtained with $C=2$ (see text) is shown with colored lines for simulated spectra of different $K$-band magnitude (the $S/N$ given in parentheses corresponds to $70\,\AA$ bins), and for all magnitudes combined in black. All simulated galaxies with $K=22$ had an estimated $p\sim100\%$ and are therefore shown as a single data point in the top-right corner. The relation for all magnitudes and $C=1$ is shown in gray for comparison. \label{FIG:simu_odds}}
\end{figure}

For each source, we systematically explored a fixed grid of redshifts covering $2<z<5$ in steps of $\Delta z = 0.0003$, fitting a linear combination of the continuum model and the emission lines and computing the $\chi^2$. The redshift probability distribution was then determined from \citep[e.g.,][]{benitez2000}
\begin{equation}
P(z) \propto \exp\left[-\frac{\chi^2(z) - \chi^2_{\rm min}}{2\,C}\right]\,. \label{EQ:pz}
\end{equation}
The constant $C$ is an empirical rescaling factor described below. From this $P(z)$, we then estimated the probability $p$ that the true redshift lies within $\pm0.01$ of the best-fit redshift, namely:
\begin{equation}
p = \int_{-0.01}^{+0.01} \dd u\,P(z_{\rm peak} + u)\,.
\end{equation}
We considered as ``robust'', ``uncertain'' and ``rejected'' spectroscopic identifications those for with we computed $p>90\%$, $50\% < p < 90\%$ and $p < 50\%$, respectively. The reliability of this classification is assessed in the next section.

Since not all our targets were expected to have detectable emission lines, we ran \texttt{slinefit} twice: with and without including emission lines. Doing so solved cases where the redshift got hooked on spurious positive flux fluctuations while the continuum was otherwise well detected (e.g., for 3D-EGS-31322). Comparing the outcome of this run to the run with emission lines, we kept the redshift determination with the highest $p$ value.

To speed up computations and avoid unphysical fits, we first performed the fit only including the brightest emission lines, namely $\sii$, $\halpha$, $\nii$, $\oiii$, $\hbeta$ and $\oii$, and only allowing two velocity dispersion values: $\sigma=60\,\kms$, which is essentially unresolved, and $300\,\kms$, the expected dispersion for galaxies of these masses. Once the redshift was determined, we ran \texttt{slinefit} again fixing $z=\zspec$, leaving the velocity dispersion free to vary from $\sigma=60$ to $1000\,\kms$, and adding fainter lines to the fit, namely $\neiii$, $\neiv$, $\nev$, $\mgii$, $\heii$, $\oone$, $\hei$, and $\oiiifour$. From this run we computed the velocity dispersions, total fluxes and rest-frame equivalent widths of all lines. Uncertainties on all these parameters were determined from Monte Carlo simulations where the input spectrum was randomly perturbed within the uncertainties. We note that since we fit the lines jointly with a stellar continuum model, our line fluxes were automatically corrected for stellar absorption.

To make sure that our redshifts and line properties were not biased because the continuum models were obtained at $z=\zphot$ rather than $z=\zspec$, in a second step we re-launched the entire procedure described above, this time using the best-fit stellar continuum model obtained at $z=\zspec$ (see \rsec{SEC:fast_fit}). The best-fit redshifts did not change significantly, except for one galaxy (ZF-UDS-6496, $\zspec=3.207$ became $2.033$) for which the redshift was anyway rejected ($p < 50\%$). No galaxy changed its classification category (e.g., from robust to uncertain) in the process, while line fluxes and equivalent widths changed by at most $2\%$. The differences were thus insignificant, but for the sake of consistency we used the results of this second run in all that follows.

\subsection{Accuracy of the derived redshifts}

In ideal conditions, namely if our search method was perfect and the noise in each spectral element of the spectrum was uncorrelated, Gaussian, and with an RMS equal to the corresponding value in the uncertainty spectrum, then the constant $C$ in \req{EQ:pz} should be set to one. However any of these conditions may be untrue, in which case we could attempt to compensate by setting $C>1$ (which would effectively broaden the probability distribution). The reduced $\chi^2$ we obtain are very close to one (see \rtab{TAB:galaxies_zspecs}), which should be a sign that our uncertainty spectra are in good agreement with the observed noise. However the reduced $\chi^2$ is always dominated by the noise of the highest frequency (in the Fourier sense, i.e., one spectral element), while the continuum spectral features useful for the redshift determination actually span multiple spectral elements. Therefore this constant $C$ has a different sensitivity to the noise properties compared to the reduced $\chi^2$.

We thus calibrated $C$ by simulating redshift measurements of artificial galaxies of various $K$-band magnitude added to pure sky spectra. We find that setting $C=2$ is required to obtain accurate $p$ values, as illustrated in \rfig{FIG:simu_odds}.

In an attempt to investigate the source of this correction, we also performed an identical test on mock spectra produced with ideal Gaussian noise. Despite the ideal noise, we find that a correction is still required, with $C=C_{\rm ideal}=1.25$. This suggests that part of the needed correction is intrinsic to our redshift measurement method, and not related to the quality of the data. If we decompose $C=C_{\rm ideal}\times C_{\rm noise}$, we find $C_{\rm noise}=1.6$, which would be equivalent to stating that our uncertainty spectrum is underestimating the noise (on the relevant scales) by $\sqrt{C_{\rm noise}}=26\%$. This value is close to our estimate of the residual correlated noise in \rapp{APP:reduction}.

Finally, we compared this automatic identification method to visual identification: all the redshifts that were visually identified (looking mostly for the $\oiii$ doublet and Balmer absorption lines) were recovered with $p>90\%$, except 3D-EGS-31322 for which $p=84\%$. In addition, the automatic identification allowed us to obtain additional redshifts for galaxies with no detectable emission lines and with weak continuum emission, albeit with a reduced (but quantified) reliability.

\subsection{Measured redshifts \label{SEC:redshifts}}

\begin{figure*}
\begin{center}
\includegraphics[width=\textwidth]{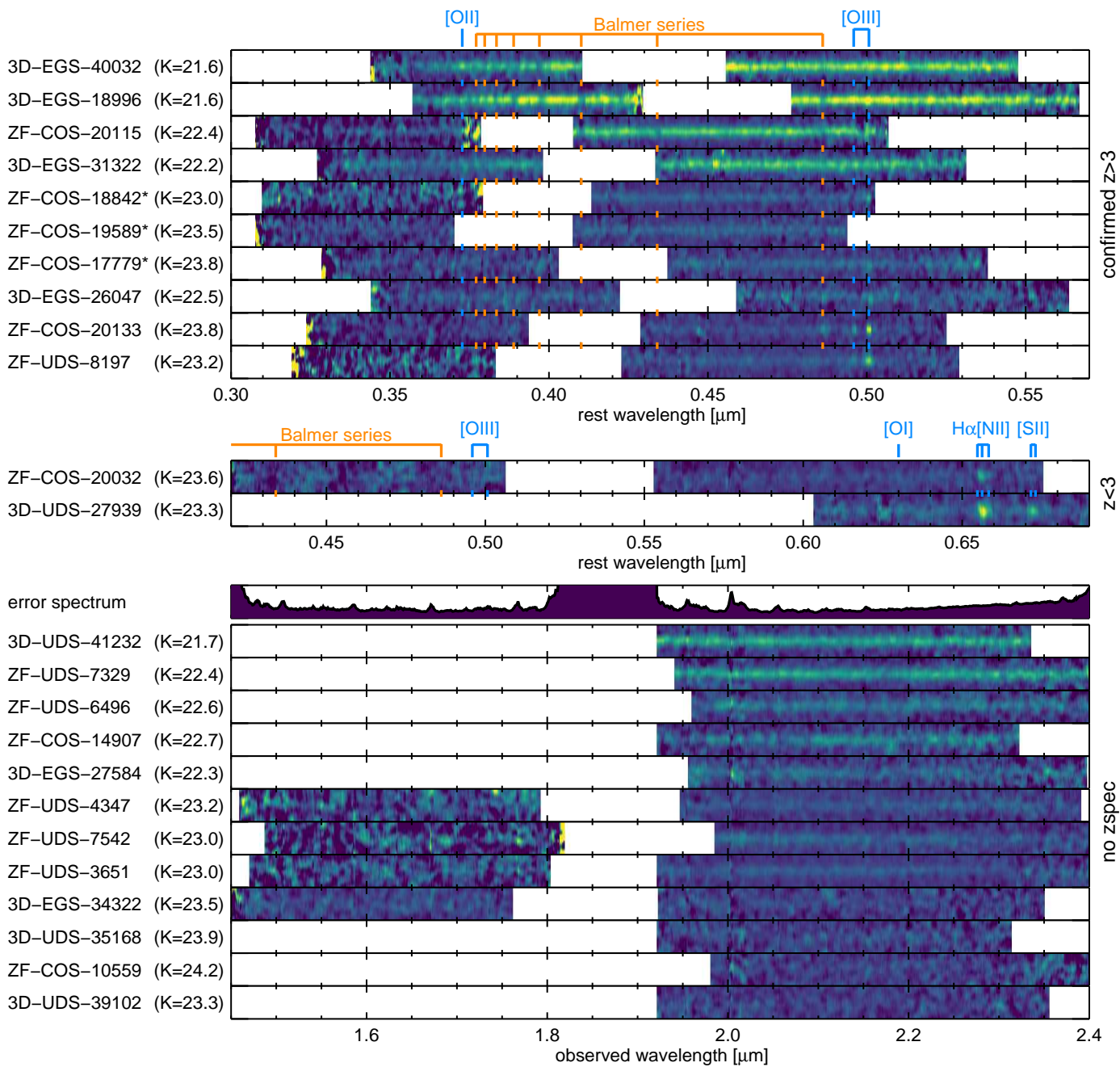}
\end{center}
\caption{2D spectra of our sample, flux-calibrated and corrected for telluric absorption. For display purposes only, these spectra were smoothed with a $70\,\AA$ boxcar filter in wavelength and $0.7\arcsec$ FWHM Gaussian along the slit. The galaxies are sorted by decreasing $K$-band continuum $S/N$. {\bf Top:} Galaxies with a spectroscopic redshift $\zspec>3$, aligned on the same rest frame wavelength grid. The most prominent emission and absorption lines are labeled in blue and orange ticks, respectively. Galaxies with an uncertain redshift (see text) are marked with an asterisk. {\bf Middle:} Same as top but for $\zspec<3$. {\bf Bottom:} Galaxies without spectroscopic redshift, aligned on the same observed wavelength grid. The average error spectrum is shown at the the top, to illustrate the regions with the strongest atmospheric features (telluric absorption and OH lines).\label{FIG:spec2d}}
\end{figure*}

A condensed overview of the outcome of the automatic redshift search is provided in \rfig{FIG:spec2d}. The results are listed in full detail in \rtab{TAB:galaxies_zspecs}, and illustrated for each galaxy in \rfigs{FIG:allspecs} and \ref{FIG:allspecs2}. In summary, we obtain a spectroscopic identification for $50\%$ of our sample, with eight robust redshifts and four uncertain redshifts, and find a $\zphot$ catastrophic failure rate of $8\%$, where the contaminants are $z\sim2.5$ dusty galaxies. We quantify the accuracy of the photometric redshifts to a median $|z-\zphot|$ of $1.2\%$, which implies that even the galaxies without $\zspec$ should be reliable. We describe these results in more detail in the following sub-sections.

\subsubsection{Robust redshifts}

\begin{figure*}
\begin{center}
\includegraphics[width=\textwidth]{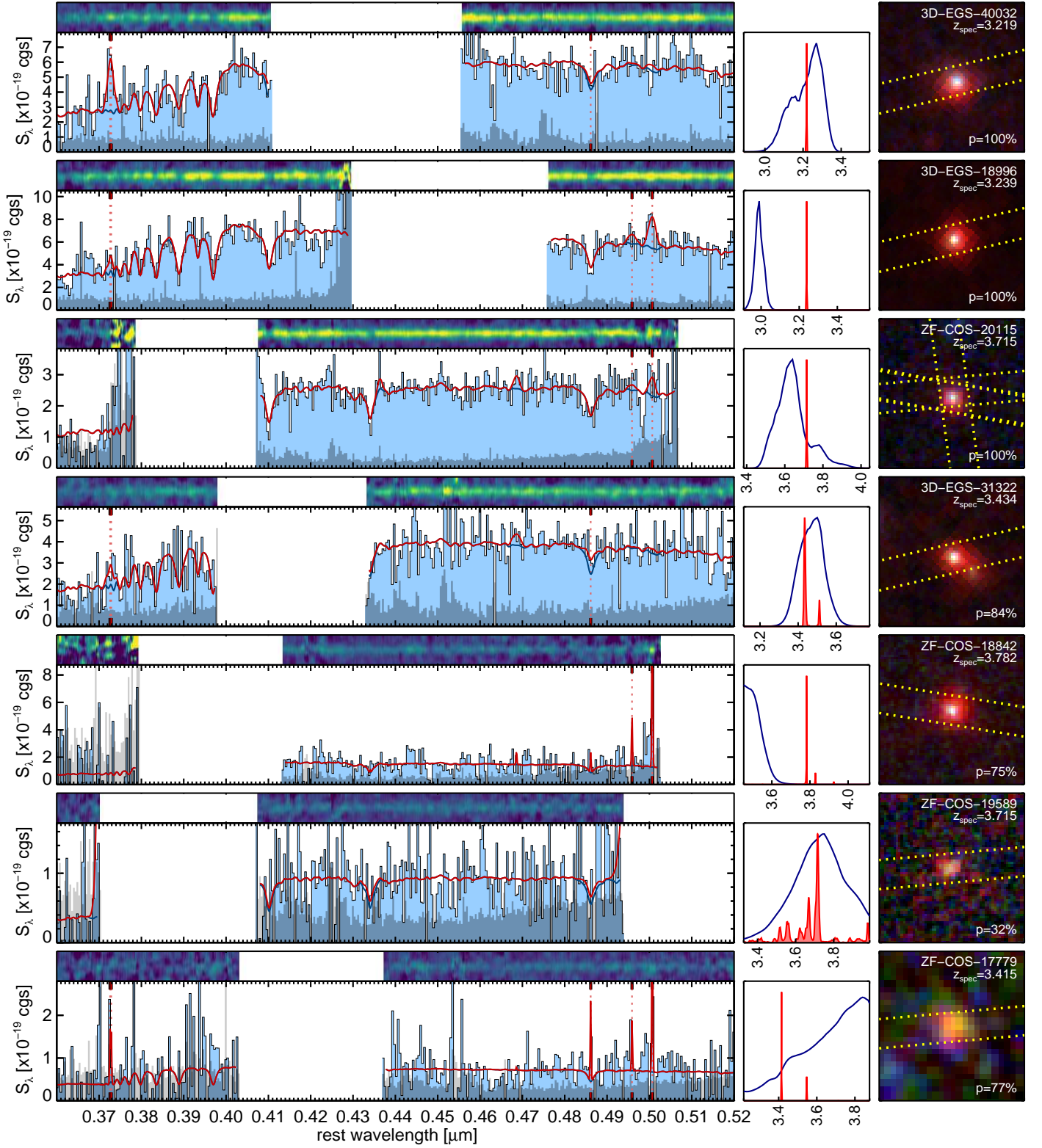}
\end{center}
\caption{From left to right: MOSFIRE spectrum, redshift probability distribution, and false-color image of the galaxies with a measured $\zspec$. The galaxies are sorted as in \rfig{FIG:spec2d}. The spectrum on the left is displayed as a function of rest-frame wavelength. The observed spectrum is shown with a black solid line and blue shading, and the best-fit model obtained at the end of the redshift fitting procedure is shown in red. The uncertainty is shown as a dark shaded area at the bottom of each plot, and the 2D spectrum is displayed at the top, with smoothing to enhance the display. For the redshift probability distribution, the $p(z)$ from the spectra are shown in red, while the $p(z)$ from the photometry (EAzY) are shown in dark blue. Finally, the false-color images are composed of the WFC3-F125W (blue), WFC3-F160W (green) and \Ks bands (red, either from ZFOURGE, HawK-I, or Ultra-VISTA), with linear scaling. Each image is $3.6\arcsec\times3.6\arcsec$ across. We also show the extents of the MOSFIRE slits as a dotted yellow rectangles. \label{FIG:allspecs}}
\end{figure*}

In total, we obtain eight robust spectroscopic identifications, with $\zspec$ ranging from $2.210$ to $3.715$. The highest measured redshift, $\zspec=3.715$, is that of ZF-COS-20115, which was first reported in \citetalias{glazebrook2017}, and is based on the detection of $\hbeta$, $\hgamma$, and $\hdelta$ absorption. We note that this value is slightly lower than the redshift obtained in \citetalias{glazebrook2017} ($\zspec=3.717$); this results from the accumulation of more data, and a slightly different measurement method. The change, contained within the error bars, has no implication on the nature of the neighboring dusty source \citep{schreiber2018}. Balmer absorption lines are found in two other galaxies, 3D-EGS-18996 at $\zspec=3.239$ and 3D-EGS-40032 at $\zspec=3.219$ (see \rfig{FIG:allspecs}, top). These two galaxies being at slightly lower redshifts, the rest of the Balmer series appears at the red end of the $H$ band. Although at this stage the continuum model was not yet fine-tuned to reproduce the strength of the Balmer absorption lines (this is done later in \rsec{SEC:fast_fit}), the quality of the fit is already excellent. This illustrates the good agreement between the photometric and spectroscopic age-dating, which was already pointed out in \citetalias{glazebrook2017} and \cite{schreiber2018} when studying the case of ZF-COS-20115.

\begin{figure*}
\begin{center}
\includegraphics[width=\textwidth]{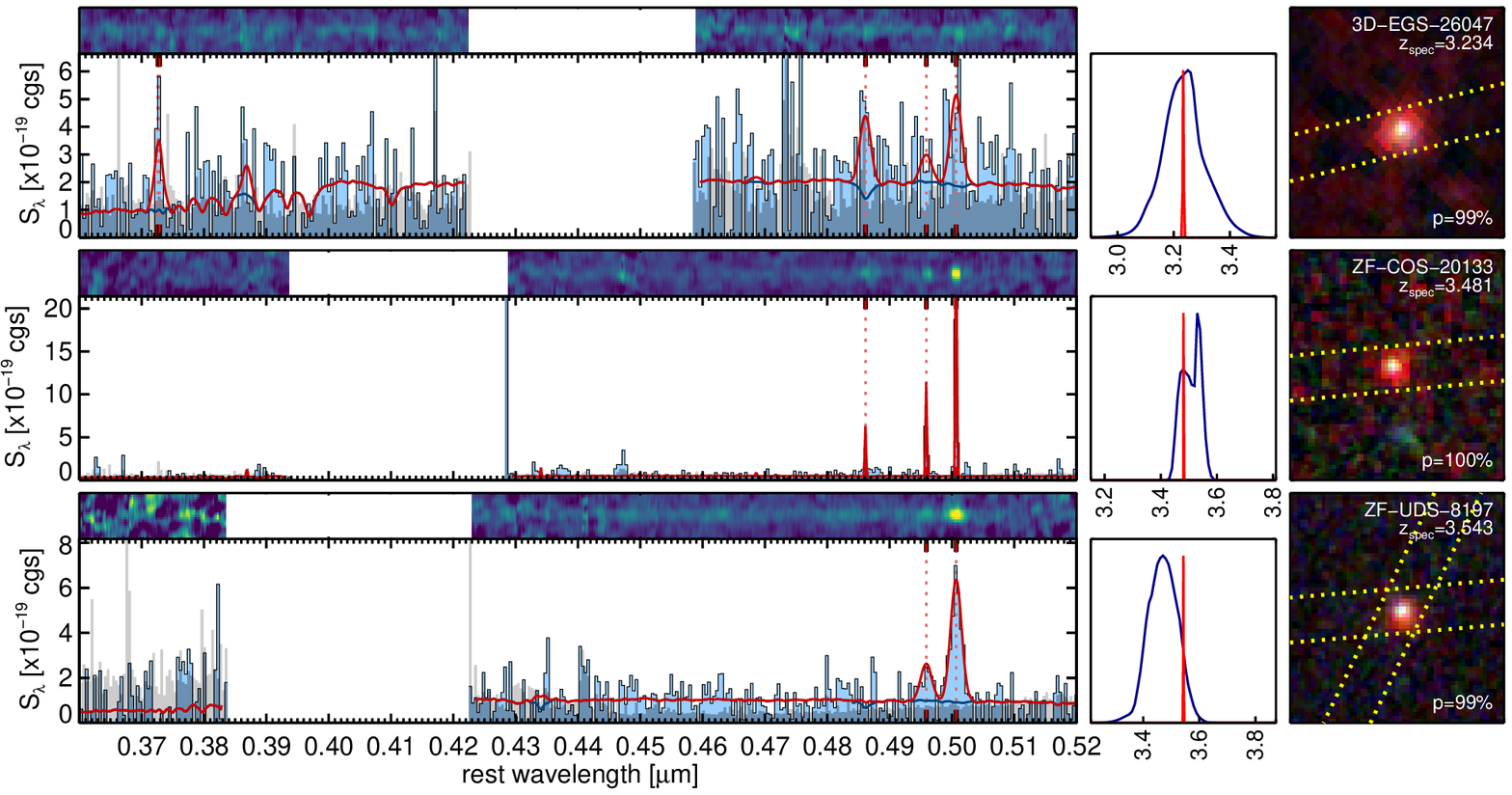}
\includegraphics[width=\textwidth]{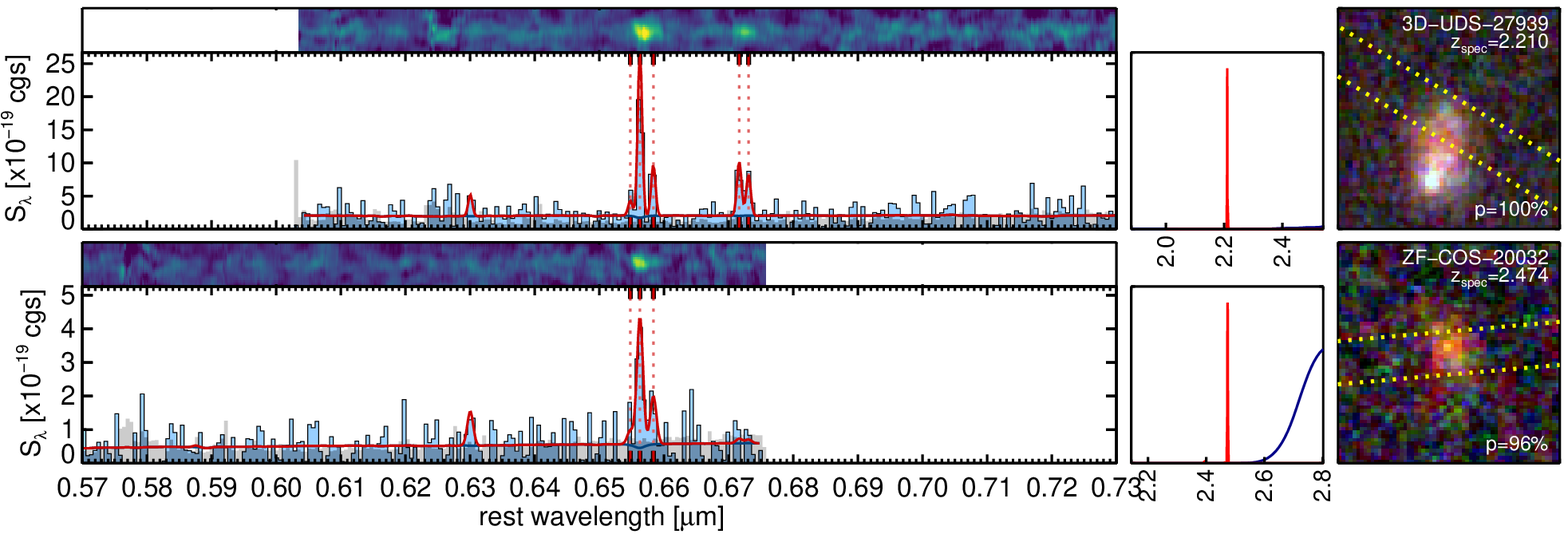}
\end{center}
\caption{\rfig{FIG:allspecs} continued. \label{FIG:allspecs2}}
\end{figure*}

Beside these three galaxies, the rest of the redshifts were determined using emission lines. Two galaxies turn out to be redshift interlopers, ZF-COS-20032 at $\zspec=2.474$ and 3D-UDS-27939 at $\zspec=2.210$, for which we detected $\halpha$ and $\nii$. These two galaxies are shown in \rfig{FIG:allspecs2} (bottom). ZF-COS-20032 is significantly extended in the F160W image and is detected by ALMA at $890\,\um$, which indicates it might be an obscured disk. 3D-UDS-27939 is also extended, and blended with another galaxy. In the 3DHST catalog, this blended system was split in two galaxies, one of which was our target with $\zphot=3.22$, while the other was attributed a lower $\zphot=2.24\pm0.02$. This value is in fact consistent with our measured $\zspec$ for the quiescent candidate, which suggests the two objects are either a major merger, or two parts of the same galaxy with a strong attenuation gradient. Regardless, as can be seen in \rfigs{FIG:allspecs} and \ref{FIG:allspecs2}, the morphologies of both ZF-COS-20032 and 3D-UDS-27939 stands apart from that of the rest of the sample, where galaxies are typically more compact; this could be a natural consequence of the different mass-size relation for star-forming and quiescent galaxies \citep[e.g.,][]{vanderwel2014,straatman2015}.

The redshifts for the remaining three galaxies (ZF-COS-20133, 3D-EGS-26047, and ZF-UDS-8197) were obtained using a combination of the $\oiii$ doublet, $\hbeta$, and $\oii$. ZF-COS-20133 and ZF-UDS-8197 are both found to have particularly bright $\oiii$ emission and little to no $\hbeta$ and $\oii$, as shown in \rfig{FIG:allspecs2}. Their line widths, however, are very different: the former has unresolved line profiles ($\sigma_{v} \leq 60\,\kms$) in both $\oiii$ and $\hbeta$, while the latter has extremely broad $\oiii$ ($\sigma_{v} = 530\pm54\,\kms$). A more detailed description of the emission line properties of these galaxies is provided later in \rsec{SEC:lines}. Lastly, 3D-EGS-26047 has faint $\oiii$ and $\hbeta$ lines of comparable fluxes, as a well as $\oii$. As shown in \rfig{FIG:allspecs2}, the lines of this galaxy are only marginally detected, and it is only by combining them in the redshift search that we could obtain a measure of the redshift (which is in excellent agreement with the $\zphot$).

\begin{figure}
\begin{center}
\includegraphics[width=0.5\textwidth]{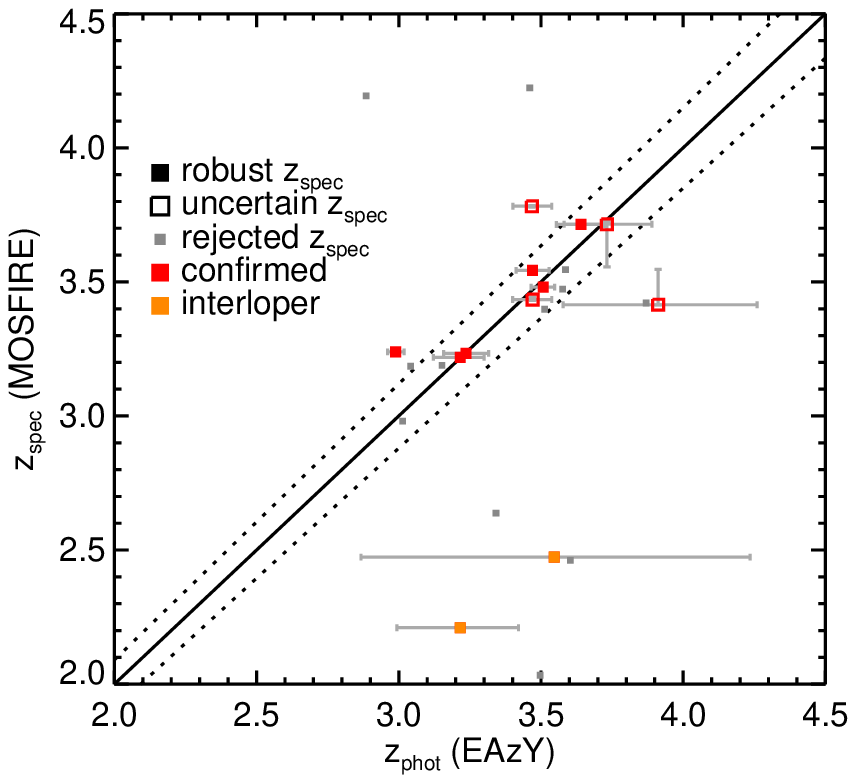}
\end{center}
\caption{Comparison between the photometric redshifts obtained with EAzY from the broadband photometry alone ($\zphot$) and the spectroscopic redshifts determined from the MOSFIRE spectra ($\zspec$). Robust redshifts ($p>90\%$) are shown with large filled squares, uncertain redshifts ($50\%<p<90\%$) are shown with open squares, and galaxies with rejected $\zspec$ are shown with small gray squares. Galaxies confirmed to be at $z\ge3$ are displayed in red, and interlopers are displayed in orange. The error bars show the $68\%$ confidence intervals. The solid line shows the one-to-one match, while the dotted lines above and below indicate the typical $\zphot$ uncertainty of $3\%$, as estimated for ZFOURGE at $3<z<4$ in \cite{straatman2016}. These values are listed in \rtab{TAB:galaxies_zspecs}.  \label{FIG:zphotzspec}}
\end{figure}

The comparison of our spectroscopic redshifts against the photometric redshifts from EAzY is presented in \rfig{FIG:zphotzspec}. Excluding the two outliers, the agreement between the $\zspec$ and $\zphot$ is excellent: the largest $|\zspec - \zphot|/(1+z)$ is $6.3\%$ for 3D-18996, and the median is $1.1\%$.

\subsubsection{Uncertain redshifts \label{SEC:zspec_uncertain}}

A further four galaxies were attributed an uncertain redshift: ZF-COS-17779, ZF-COS-18842, ZF-COS-19589, and 3D-EGS-31322. We included in this list the galaxy ZF-COS-19589, whose redshift of $\zspec=3.715$ (identified using Balmer absorption features, see \rfig{FIG:allspecs}) should have been rejected on the basis of its $p=32\%$. Indeed, this redshift lies within $\Delta z < 0.01$ of ZF-COS-20115, which has a robust $\zspec$ and is located only $23\arcsec$ away. Based on the possibility of these two galaxies being physically associated, we gave extra credit to this $\zspec$ and promoted it to the uncertain category. Otherwise, the constraints on the redshift form the spectrum alone are relatively poor, but the absence of a break in the $K$ band rules out redshifts $z>3.8$.

The galaxy 3D-EGS-31322 (shown in \rfig{FIG:allspecs}) has a well-detected continuum emission and a significant break at the red end of the $H$ band. This break is sufficient to confirm that the redshift is indeed $z\sim3.5$, but a more precise redshift requires line identifications. Balmer absorption features may be identified, in particular $\hepsilon$ at the red edge of the $H$ band and $\hgamma$ at the blue edge of the $K$ band, along with weak $\oii$ emission. However the $S/N$ is low enough that these identifications are ambiguous. In all cases however, $\hbeta$ absorption and $\oiii$ emission are weak or non-existent.

The last two galaxies, ZF-COS-17779 and ZF-COS-18842 (shown in \rfig{FIG:allspecs}), are essentially identified using a single narrow emission line. While this emission is securely detected in both cases ($7.7$ and $5.6\sigma$, respectively), the identification of the corresponding emission line is partly degenerate. For both galaxies, the automated redshift search attributed this emission to the brightest line of the $\oiii$ doublet, $\oiii_{5007}$. For ZF-COS-17779, this solution is also backed up by a plausible detection of $\hbeta$ ($4.1\sigma$) and tentative $\oii$ ($1.7\sigma$). For ZF-COS-18842 on the other hand, $\hbeta$ is detected at only $1.0\sigma$, and since the line is located almost at the edge of the $K$ band, the detected emission could also be attributed either to the fainter line of the $\oiii$ doublet, $\oiii_{4959}$, or to $\hbeta$. The only reason why these alternative solutions are disfavored is because they provided a poorer fit to the continuum emission, in particular regarding the presence of absorption features. Indeed, at these higher redshifts, the $\hdelta$ line enters the $K$ band but does not correspond to any absorption feature in the observed spectrum, and thus would have created a tension of $2.0$ and $2.9\sigma$ (if the detected emission line is $\oiii_{4959}$ or $\hbeta$, respectively). Likewise, the $\hgamma$ line is covered for all three solutions, and although there is no clear evidence that this absorption line is actually detected, the $\oiii_{5007}$ solution provides the smallest tension ($1.1\sigma$, versus $2.4\sigma$ for the other two solutions). This evidence is however marginal, since an alternative possibility is that we overestimated the strength of the absorption lines in the continuum template, that is, if the galaxy is younger (or older) than its broadband photometry initially suggested.

Lastly, we manually rejected from the uncertain category the $\zspec=4.194$ for the galaxy ZF-COS-14907 which had $p=63\%$; its surprisingly high value (highest $\zspec$ of all the sample), poor fit (reduced $\chi^2=1.2$, highest of all the sample), and blatant inconsistency with the photometric redshift ($\zphot = 2.89^{+0.06}_{-0.06}$, $\sim$20$\sigma$ difference, again the highest of the sample) suggested an issue with the spectrum.

In the end, for the four galaxies in the uncertain category, the highest $|\zspec - \zphot|/(1+z)$ is $10\%$ for ZF-COS-17779, with a median of $3.9\%$. This is higher than for the galaxies of the robust sample, and could be expected since the uncertain galaxies are on average $0.7\,{\rm mag}$ fainter in $K$. Considering the combined robust and uncertain sample, the median $|\zspec - \zphot|/(1+z)$ is $1.2\%$; we can therefore conclude that, save for the few galaxies with strong emission lines, the $\zphot$ were highly accurate, confirming the results of \cite{straatman2016} obtained with galaxy pairs.

\subsubsection{Unconfirmed redshifts}

We could not determine spectroscopic redshifts for the remaining $12$ galaxies. As can be seen on \rfig{FIG:spec2d}, these are not particularly fainter, and a Kolmogorov-Smirnov test gives $p=99\%$ of the two samples having the same $K$-band magnitude distribution. Likewise, their photometric redshift distribution is consistent with being the same as that of the spectroscopically confirmed galaxies (KS test: $p=99\%$). However, the five brightest of these $12$ galaxies have no $H$-band coverage from MOSFIRE. As demonstrated with 3D-EGS-40032 and 3D-EGS-18996, the $H$ band can prove particularly useful in determining redshifts when the high-order lines from the Balmer series are observed, or simply to confirm the absence of continuum emission (e.g., ZF-COS-20115). For the bright but unconfirmed galaxies, a possible explanation for their lack of identification would be that they have weaker Balmer absorption owing to them having older or younger stellar populations. But, in general, it is also possible that we simply missed the emission lines because of sky lines. Indeed, based on the calculations in \rsec{SEC:sensitivity}, we can statistically expect this to happen in two of these $12$ galaxies.

Nevertheless, and statistically excluding two galaxies for which lines are not detectable because of sky lines, we could confirm that their $K$-band (and, for a few, also $H$-band) photometry is not significantly contaminated by emission lines (see \rsec{SEC:sensitivity}). As per the above, this implies their photometric redshifts and derived \uvj colors should not suffer from systematic errors, hence that most of these unconfirmed galaxies should be reliable quiescent candidates. Consequently, ZF-COS-20032 and 3D-UDS-27939 should be the only two galaxies with catastrophic redshift failure, resulting in a failure rate of $8\%$ (or $9\%$ if we account for galaxies with potentially missed emission lines).

\subsection{Stacked spectrum}

\begin{figure*}
\begin{center}
\includegraphics[width=\textwidth]{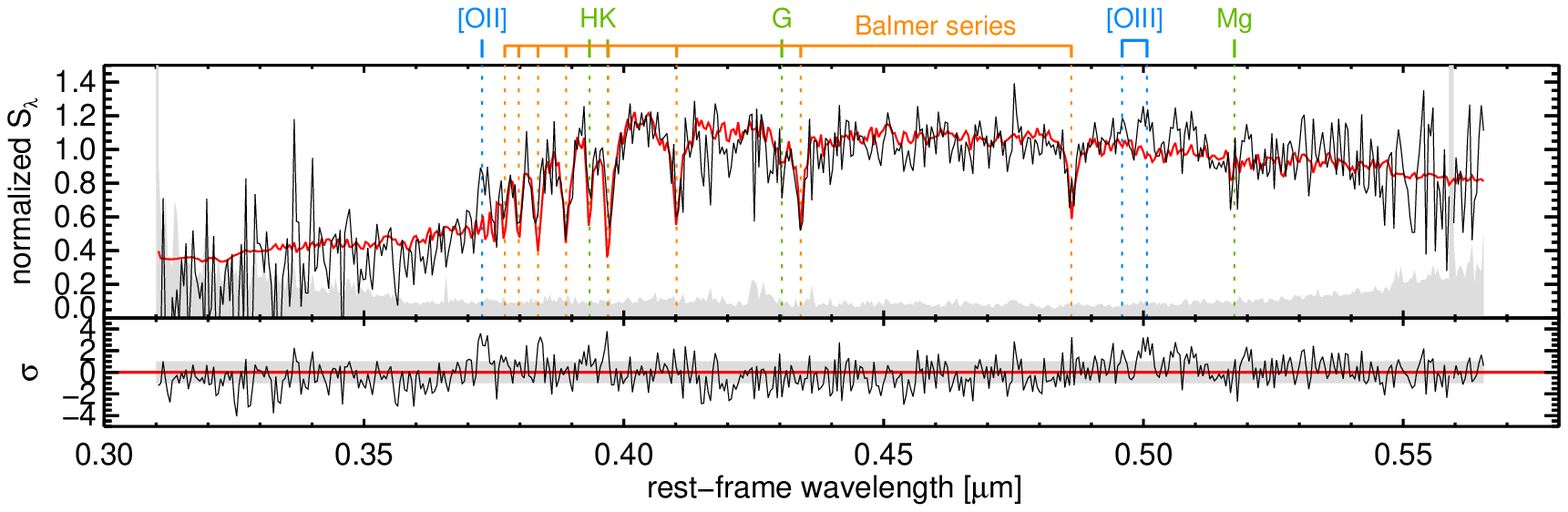}
\end{center}
\caption{Stacked rest-frame optical spectrum of the eight spectroscopically confirmed $z>3$ galaxies with no strong emission lines (i.e., all the galaxies with robust or uncertain redshifts at $z>3$, except ZF-COS-20133 and ZF-UDS-8197). The stacked MOSFIRE spectrum is shown in black, normalized to unit flux density at $\lambda\sim0.48\,\um$, and the error spectrum is shown in shaded gray at the bottom. We overlay the stacked model spectrum of all the galaxies in red (as obtained in \rsec{SEC:fast_fit}), and we indicate the main absorption and emission lines with colored lines (blue: emission, green: absorption, orange: Balmer series) with labels at the top of the figure. The residuals of the spectrum, after subtracting the stacked model and normalizing by the uncertainty, are displayed at the bottom of the figure. \label{FIG:stacked}}
\end{figure*}

We show in \rfig{FIG:stacked} the stack of all the eight $z>3$ galaxies with robust or uncertain redshifts. This stack was obtained as the inverse-variance-weighted average flux, after each spectrum was re-normalized to a unit flux at rest-frame $0.48\,\um$. The galaxies that entered the stack have different rest-wavelength coverage (as shown in \rfig{FIG:spec2d}), such that only the wavelengths from $0.475$ to $0.495\,\um$ (which includes $\hbeta$) were covered for all galaxies. The $\oiii$ and $\oii$ emission lines were covered in all but one galaxy, so their stacked amplitude should be representative, but $\hgamma$ and $\hdelta$ were only covered in half of the sample. We simultaneously stacked the best-fitting stellar continuum models of the galaxies (derived below in \rsec{SEC:fast_fit}), using the same weighting.

In this stacked spectrum, the Balmer absorption series can be readily identified, with $\hbeta$, $\hgamma$, $\hdelta$, $\hepsilon$, $\hzeta$, $\heta$ and H10. We also identified the calcium $H$ absorption feature (calcium $K$ is blended with $\hepsilon$), and tentatively the $G$-band and $\mgi$ absorption. In emission, we only find $\oii$ and $\oiii_{5007}$ to be significantly detected in the residual spectrum.

\subsection{Emission lines ratios and equivalent widths \label{SEC:lines}}

The measured emission line properties for all galaxies in the robust and uncertain categories are listed in \rtab{TAB:galaxies_em}. The most commonly detected line ($>2\sigma$) is the $\oiii$ doublet, which was detected in five galaxies with $\restew$ ranging from $14$ to $282\,\AA$ (median $49\,\AA$), while $\hbeta$ emission was detected in three galaxies, with $\restew$ ranging from $8$ to $34\,\AA$. For one galaxy, 3D-EGS-18996, we find $\oiii$ in emission and $\hbeta$ in absorption, as shown in \rfig{FIG:allspecs}. We also formally detected $\oii$ in two galaxies, 3D-EGS-26047 and 3D-EGS-40032, with equivalent widths of $43$ and $23\,\AA$, respectively.

Among the galaxies with $\oiii$ detections, the $\log_{10}(\oiii/\hbeta)$ ratio (corrected for Balmer absorption, see \rsec{SEC:redshifts_method}) ranges from $0.17^{+0.26}_{-0.30}$ (3D-EGS-26047) to $1.41^{+0.07}_{-0.26}$ (ZF-UDS-8197), with a median of $0.91$. Using the stellar masses derived in the next section (or the ones initially derived at $z=\zphot$), the mass-excitation diagram \citep{juneau2011} classifies all the $\oiii$-detected galaxies as ``AGN'', and this remains true even if we use the stricter criterion derived for $z>2$ galaxies in \cite{coil2015}. Recent results suggest this criterion should be made even stricter, shifting the $z\sim0$ critetion of \cite{juneau2011} by $1\,\dex$ in mass \citep{strom2017}; this would reduce the fraction of AGNs among our $\oiii$ emitters to $40\%$, which remains substantial. The $\oiii$ luminosity ranges from $1.5\times10^{8}$ to $1.8\times10^{9}\,\lsun$ ($0.7$ to $7.8\times10^{42}\,{\rm erg/s}$). The line velocity profile are unresolved ($\sigma_{v} < 60\,\kms$) for two galaxies, ZF-COS-20133 and ZF-COS-17779, and particularly broad for all other galaxies, with $\sigma_{v} = 530$ to $582\,\kms$. While the narrow $\oiii$ in ZF-COS-20133 may be powered by an AGN, the broad $\oiii$ of ZF-UDS-8197 should instead reflect shocked gas in the galaxy's gravitational potential, since $\oiii$ is not produced in AGN broad line regions \citep[e.g.,][]{baldwin1975}. We defer further analysis of these line kinematics and links to AGN activity to a future paper.

Lastly, for the two redshift interlopers we detected the $\halpha$ line with an $\restew$ of $84\pm31$ for ZF-COS-20032 and $109\pm9\,\AA$ for 3D-UDS-27939. The $\nii$ doublet was weakly detected in the former, and more clearly in the latter; the resulting $\log_{10}(\nii/\halpha)$ are $-0.33^{+0.27}_{-0.30}$ and $-0.40^{+0.06}_{-0.07}$, respectively, which are both inconclusive as they might correspond to any category in the Baldwin-Phillips-Terlevich (BPT; \citealt{baldwin1981}) diagram \citep{kauffmann2003-a}. The $\sii$ doublet was also covered and only detected for 3D-UDS-27939, leading to $\log_{10}(\sii/\halpha) = -0.24^{+0.06}_{-0.07}$, which is similarly inconclusive.

Over the entire sample, ``high-EW'' emission line complexes with a summed ${\rm EW_{rest}} > 100\,\AA$ were observed in four galaxies. This implies that \uvj-selected samples are contaminated by high-EW lines at the rate of $17\%$ (or $18\%$ if we account for galaxies with potentially missed emission lines), half of these being redshift interlopers. Even if all of these high-EW galaxies happened to not be truly quiescent (which is a question we address later in \rsecs{SEC:new_colors} and \ref{SEC:results_sfh}), this would not affect the number densities \citep[e.g.,][]{straatman2014} in a significant way.

\subsection{Subtracting emission lines from the broadband photometry \label{SEC:decontam}}

To go forward (see \rsec{SEC:results_sfh}) we needed to analyze the continuum emission, using both the spectra and the broadband photometry. Since some of our galaxies displayed particularly high EW emission lines, we had to correct the $H$ and $K$ broadband photometry for this contamination. For each NIR broadband, we selected the lines with $\obsew/\Delta\obsew>3$ and computed the corrected flux densities $S_{\rm BB}^{\rm cor}$. With a similar reasoning as in \rsec{SEC:sensitivity}, we can derive
\begin{align}
\frac{S_{\rm BB}}{S_{\rm BB}^{\rm cor}} = 1 + \sum_{\ell} \obsew^\ell \frac{R(\lambda_\ell)}{\int \dd\lambda\, R(\lambda)}\,
\end{align}
where $S_{\rm BB}$ is the original flux density, $R(\lambda)$ is the response curve of the corresponding filter, and where $\obsew^\ell$ and $\lambda_\ell$ are the observer-frame equivalent width and central wavelength of the line $\ell$, respectively. The above equation assumes a constant continuum flux density within the filter, and a constant filter response over the spectral extent of the line. The flux uncertainties were updated to account for the uncertainty associated with this correction, using Monte Carlo simulations where the original flux and all the EWs were randomly perturbed within their respective uncertainties. Because it is based only on the measured equivalent widths, this correction is by construction not affected by systematic errors in flux calibration.

In the end, this had a significant impact only in the \Ks band, and only for the galaxies ZF-COS-20133 ($30\%$ of the flux removed), ZF-UDS-8197 ($19\%$) and 3D-UDS-27939 ($16\%$). The fluxes of the other galaxies were affected by less than $5\%$, and the corrected photometry is displayed in \rfig{FIG:allseds}.

\subsection{Updated rest-frame colors \label{SEC:new_colors}}

\begin{figure}
\begin{center}
\includegraphics[width=0.5\textwidth]{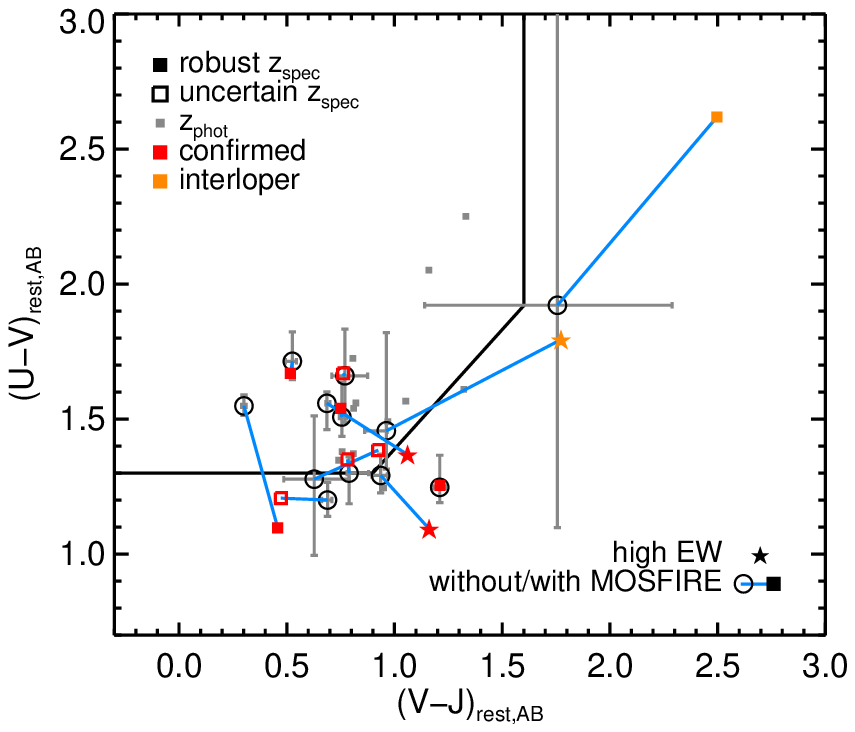}
\end{center}
\caption{\uvj colors of our candidate quiescent galaxies targeted with MOSFIRE. The legend is the same as in \rfig{FIG:zphotzspec}. The black solid line is the quiescent-or-star-forming dividing line used in \citetalias{straatman2014}. For galaxies with no spectroscopic redshift, shown in small gray symbols, the colors shown were computed at $z=\zphot$. For galaxies with a spectroscopic redshift, the colors shown were computed at $z=\zspec$; the colors initially computed at $z=\zphot$ and without the correction for emission lines are shown with empty squares, and blue lines connect the old (empty black circle) and new (colored star or square) colors of a given galaxy. These values are listed in \rtab{TAB:final_props}. \label{FIG:uvj}}
\end{figure}

Using the photometry corrected for emission lines, as described in the previous section, and assuming $z=\zspec$, we then re-computed the \uvj colors with EAzY. We show in \rfig{FIG:uvj} the change in colors resulting from the knowledge of the spectroscopic redshifts and the line-subtracted photometry.

The most striking change naturally occurs for the two redshift interlopers, which are now clearly located in the ``dusty star-forming'' region of the \uvj diagram. The reason for this change is different for the two galaxies. For ZF-COS-20032, the observed colors are redder than presently allowed by the SED template set of EAzY, which may explain why its redshift was incorrect in the first place. Indeed we show later in \rsec{SEC:fast_fit} that this galaxy suffers exceptional obscuration by dust, with $A_{\rm V}\sim4\,{\rm mag}$, while the dustiest template provided with EAzY has $A_{\rm V}=2\,{\rm mag}$; this suggests that such redshift outliers could be avoided if redder templates were included in the $\zphot$ determination, but demonstrating this goes beyond the scope of this paper. For 3D-UDS-27939, the rest-frame colors are still within range of the EAzY template set; the main reason for the $\zphot$ failure was that the \Ks-band flux was significantly contaminated by $\halpha$, mimicking the Balmer break at $z\sim3$.

Otherwise, the colors also changed significantly for the two confirmed $z>3$ galaxies with high equivalent-width $\oiii$ emission, namely ZF-COS-20133 and ZF-UDS-8197. These two galaxies are now outside the fiducial \uvj ``quiescent'' region, although ZF-COS-20133 still lies within $0.05\,{\rm mag}$ of the dividing line. The situation for these two objects is similar to that of 3D-UDS-27939, in that the apparent strength of the Balmer break was reduced once the emission line was subtracted from the photometry.

For the rest of the sample, the only significant change was for 3D-EGS-18996 which saw its $U-V$ color reduced by about $0.5\,{\rm mag}$, moving it outside of the quiescent region. This change was caused by a revision of the redshift, as the $\zphot$ was significantly underestimated. Interestingly, this galaxy nevertheless displays strong Balmer absorption and no $\hbeta$ or $\oii$ emission, which demonstrates the absence of current star formation. In fact, this region of the \uvj diagram was shown to be mainly populated by post-starburst galaxies \citep[e.g.,][]{whitaker2012,wild2014}. Albeit not satisfying the fiducial \uvj color cut owing to their too recent quenching, such galaxies still host little to no current star-formation activity \citep[e.g,][]{merlin2018}, and can thus still be considered quiescent.

In the end, out of the $24$ galaxies that were observed with MOSFIRE, two turned out to be redshift interlopers, two had bright emission lines contaminating their rest $V$ magnitude, and one saw its redshift sufficiently revised to change its \uvj colors and move it out of the quiescent region (albeit in the post-starburst area). Combined, this implies that $21\%$ of the galaxies initially classified as \uvj-quiescent were spurious. We thus concluded that the initial \uvj colors of our galaxies were robust for the majority of the sample, but that the number of quiescent galaxies estimated at $z>3$ from \uvj-selected samples is overestimated by $20\%$ because of contaminants. In the following, we take this figure into account to correct the observed number densities.

\section{Star formation rates and histories \label{SEC:results_sfh}}

In this section we take advantage of the knowledge derived from the MOSFIRE spectra, namely the redshifts and absence of strong emission lines, as well as the spectra themselves, to model the star formation histories of our quiescent galaxies candidates. The goal of this section is to investigate if these galaxies are truly quiescent, and if so, to provide the first constraints on their star formation histories.

\subsection{Modeling \label{SEC:fast_fit}}

Using the updated photometry and redshifts, we re-ran FAST++ to update the stellar masses and other physical properties of the quiescent galaxies. In addition, for each galaxy with a measured $\zspec$, we used the MOSFIRE spectrum to constrain the fit further, masking emission lines detected at more than $2\sigma$ significance since FAST++ does not model them, and only using spectra for which the synthetic broadband flux (in the $H$ or $K$ passband) was detected at more than $5\sigma$. In the fit, the spectrum was renormalized independently of the broadband photometry to account for mis-corrections of the slit losses and residual aperture systematics (\texttt{AUTO\_SCALE=1}). The spectrum was still included in the $\chi^2$, but the independent rescaling ensured that only the shape of the spectrum constrained the fit (i.e., absorption lines and spectral breaks, or the absence thereof) and not its absolute normalization. Lastly, if a galaxy was detected by ALMA, we computed its $\lir$ using the dust templates from \cite{schreiber2018-a}, assuming the average dust temperature at the redshift of each galaxy ($\tdust(z)$, Eq.~15 in \citealt{schreiber2018-a}; e.g, $\tdust\sim40\,\kelvin$ at $z\sim3.5$), and used this value to constrain the attenuation in FAST++. For ZF-COS-20115 we used the $\lir$ non-detection derived in \cite{schreiber2018}.

Similarly to \cite{schreiber2018}, we post-processed each model SFH generated by FAST++ and identified the two main phases of the galaxy's history, as illustrated in \rfig{FIG:sfh}. Firstly, we located the time of peak $\sfr$ and determined the smallest contiguous time period surrounding it where $68\%$ of the integrated $\sfr$ took place. We considered this as the ``main'' formation phase, and defined its length as $T_{\rm SF}$ and its mean $\sfr$ as $\mean{\sfr}_{\rm main}$. To locate this formation phase in time, we computed the time at which half of the mass had formed, $t_{\rm form}$, ignoring mass loss. Having identified the main formation phase, we then looked for the longest contiguous time period, starting from the epoch of observation and running backward, where the $\sfr$ was less than $10\%$ of $\mean{\sfr}_{\rm main}$. If such a time period existed, we considered it as the ``quenched'' phase, and defined its starting time as $t_{\rm quench}$. Knowing the redshift of the galaxy, we then inferred the formation and quenching redshifts, $z_{\rm form}$ and $z_{\rm quench}$, respectively.

For each galaxy and for all model parameters, we defined the best-fit values from the model with $\chi^2 = {\rm min}(\chi^2)$, and defined the range of allowed values as the range spanned by models with $\chi^2 - {\rm min}(\chi^2) < 2.71$. This corresponds to a $90\%$ confidence interval \citep{avni1976}. The resulting galaxy properties for the entire sample can be found in \rtab{TAB:final_props}. To study these two quantities in more detail, we also computed the probability distribution functions for $t_{\rm quench}$ and $t_{\rm form}$. We defined this probability on a one-dimensional grid of values $t_i$ with fixed step $\Delta t = 50\,\Myr$ such that $p(t_i)\propto\exp(-\chi_i^2)$, where $\chi_i^2$ is the minimum $\chi^2$ of all models with $|t-t_i| < \Delta t/2$ (where $t$ is either $t_{\rm quench}$ or $t_{\rm form}$). This same approach was used in CIGALE (\citealt{noll2009}; see in particular their Fig.~6 for an illustration).

To check that our modeling provided a good description of the data, we also computed the reduced $\chi^2$ ($\chi^2_{\rm red}$) for each galaxy. For this exercise, we excluded the spectra from the $\chi^2$ since we already showed they have $\chi^2_{\rm red}\sim1$ (see \rtab{TAB:galaxies_zspecs}). We find a median $\chi^2_{\rm red}$ of $1.13$, which indicates an overall good fit to the photometry, however three galaxies have values larger than $2.0$ which deserved further inspection. The largest values is $\chi^2_{\rm red}=2.34$ for ZF-UDS-6496, and is mainly caused by a flux excess in the $B$ band blueward of the Lyman limit (rest $860$-$1070\,\AA$). Excluding this band brings the $\chi^2_{\rm red}$ down to $1.39$. The second highest value is obtained for 3D-EGS-18996, with $\chi_{\rm red}^2=2.28$, and this is caused by the WirCAM $J$ band, which is inconsistent with the well-measured \hst F125W flux at the $2.4\sigma$ level. Without this band, we find $\chi_{\rm red}^2=1.46$. The last case is ZF-COS-10559, the faintest galaxy of our sample, with $\chi_{\rm red}^2=2.22$. There we could not find a single band causing the poor $\chi^2$, rather a number of bands with inconsistent fluxes (see \rfig{FIG:allseds}). The only trend we found was for the CFHT photometry to be lower than that from Subaru (the CFHT $g$ band in particular is lower than the overlapping Subaru $B$ and $V$ bands at $3.8$ and $4.2\sigma$, respectively). This may indicate variability. All the other galaxies have $\chi^2_{\rm red} < 1.5$, indicating that our models were able to capture all the significant features of the observed photometry.

\subsection{Impact of star-formation history parametrization \label{SEC:two_sfh}}

\begin{figure}
\begin{center}
\includegraphics[width=0.5\textwidth]{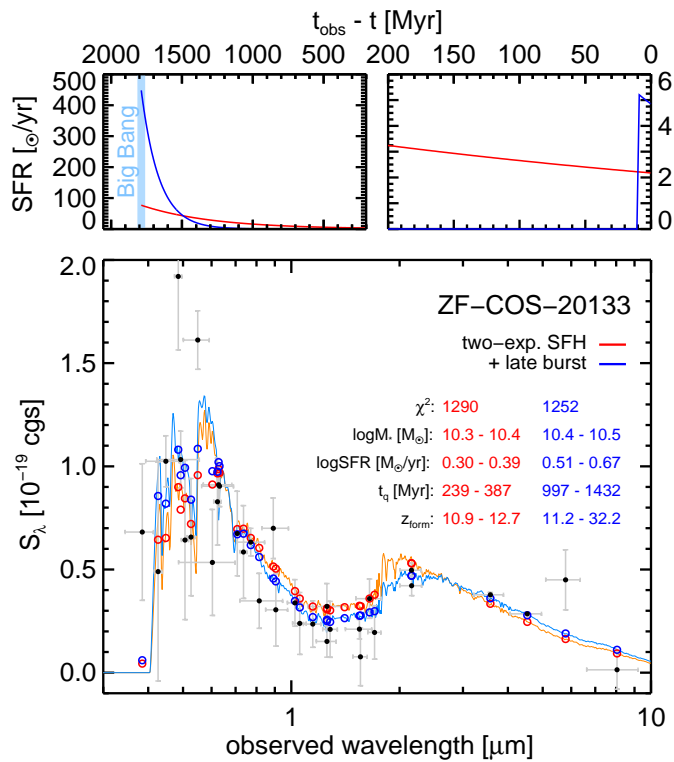}
\end{center}
\caption{Photometry and models of ZF-COS-20133. The observed photometry is shown as black circles with gray error bars. Our model spectrum and photometry using the fiducial SFH parametrization (\req{EQ:sfh}) is shown in blue, and the model without the additional late burst (\req{EQ:sfh_noburst}) is shown in red/orange. The fit parameters are given in inset. At the top of the figure, we show the star formation histories of both models. The time axis was split in half to better display the behavior during the last $200\,\Myr$. \label{FIG:20133-SED}}
\end{figure}

As discussed in \rsec{SEC:zphot}, in this work we considered a more involved SFH parametrization than the traditional models (e.g., delayed exponentially declining, or constant truncated SFH). In this section we discuss the impact of adding an additional degree of freedom regarding the recent $\sfr$ (\req{EQ:sfh}). For this purpose, we ran FAST++ again with a simplified SFH where we forced $R_{\rm SFR}=1$, and therefore where this freedom was removed.

Compared to the fits with $R_{\rm SFR}=1$, the best-fit values obtained with the SFH of \req{EQ:sfh} are mostly the same except for one galaxy, ZF-COS-20133, which is discussed below. For the rest of the sample, the stellar masses have a median ratio of unity and a scatter of only $0.07\,\dex$. The quiescence and formation times have a scatter of $115$ and $175\,\Myr$, respectively. The most important difference is found for the $10\,\Myr$-averaged $\sfr$s, which are increased on average by $35\%$. Still, for $80\%$ of the sample these changes are contained within the error bars, and are thus not significant.

The case of ZF-COS-20133 clearly stands apart from the rest of the sample. As illustrated in \rfig{FIG:20133-SED}, this galaxy is the only one for which setting $R_{\rm SFR}$ free provided a significant improvement of the fit ($\Delta\chi^2 = 39$). With $R_{\rm SFR}=1$, the adopted models consisted of an $\sfr$ that slowly declined since the Big Bang ($t_{\rm SF} \sim 600\,\Myr$), with a high formation redshift ($\zform = 11.1$ to $12.7$) and a low current $\sfr$ ($2.0$ to $2.4\,\msun/\yr$). With $R_{\rm SFR}$ free, the main formation episode of the galaxy was shortened ($t_{\rm SF} < 400\,\Myr$) and pushed to even higher redshifts, and a recent short burst was used to reproduce the current $\sfr$ ($3.2$ to $4.7\,\msun/\yr$). The galaxy was thus modeled as a maximally old stellar population with a small rejuvenation event.

Such a large age would imply an extreme star-formation efficiency within the first few hundred million years after the Big Bang. However, this galaxy is one of the few for which we observed strong $\oiii$ emission ($\log(\oiii/\hbeta) = 0.84$). This line is  unresolved, which suggests it may originate from the narrow-line region of an AGN. Since it also has a remarkably high EW of $\sim300\,\AA$, and since the $\oiii$ EW is known to correlate with AGN obscuration \citep[e.g.,][]{caccianiga2011}, an alternative and more plausible scenario is that the photometry contains some continuum emission from an obscured AGN (see, e.g., \citealt{marsan2017}). This would impact particularly the IRAC bands, increasing the flux there and thus faking the presence of an old population (the galaxy is not detected at $24\,\um$ however, so this putative AGN cannot be very luminous). If this is true, then the SFH of ZF-COS-20133 cannot be constrained without accurately accounting for the presence of the AGN, which we cannot do here. In the following, we therefore do not use the inferred SFH for this galaxy.

\subsection{Comparison of SFR estimates \label{SEC:sfr_comp}}

\begin{figure}
\begin{center}
\includegraphics[width=0.5\textwidth]{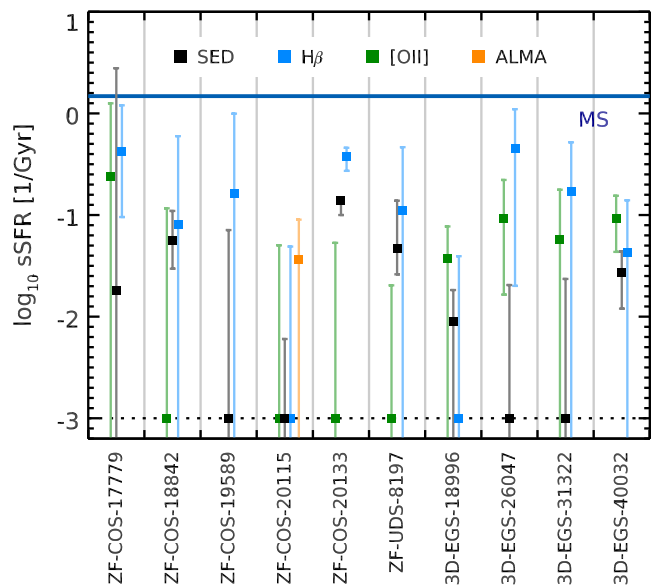}
\end{center}
\caption{Comparison of sSFR estimates from several sources: SED modeling with FAST++ (black), $\hbeta$ luminosity (blue), $\oii$ luminosity (green), and, when available, $\lir$ from ALMA (orange). The error bars for all quantities indicate the $90\%$ confidence interval. The line luminosities were corrected for dust attenuation using the $A_V$ from the SED modeling. Each galaxy with a spectroscopic redshift at $z>3$ is shown in a different column. For visualization purposes we limit the minimum sSFR to $10^{-3}\,\Gyr^{-1}$, which is indicated with the dotted line; galaxies on this line are actually at lower $\ssfr$. The blue horizontal line gives the average sSFR of main-sequence galaxies (see \rsec{SEC:ssfr}). \label{FIG:sfrcomp}}
\end{figure}

While we have explored a large variety of star formation histories, our $\sfr$ estimates may still be biased low if we underestimated the attenuation by dust, since red colors can be produced both by old stellar populations and dust obscuration \citep[e.g.,][]{dunlop2007}. The few galaxies in our sample covered by ALMA show no detection in the sub-millimeter (save for the two redshift outliers). Stacking the ALMA-derived $\lir$ for these galaxies leads to $\sfr=11\pm17\,\msun/\yr$, while the deep upper limit for ZF-COS-20115 is $\sfr < 13\,\msun/\yr$ \citep{schreiber2018}. The non-detections with ALMA therefore rules out strong starbursts, but still allows for moderate amounts of star formation.

To double check our $\sfr$s, we therefore obtained alternative estimates using the measured emission line luminosities (\rtab{TAB:galaxies_em}), empirical conversion factors from the literature, and the dust attenuation estimated by FAST++ from the broadband photometry. While it is usually assumed that emission lines suffer more attenuation than the stellar continuum \cite[e.g.,][]{calzetti2000}, recent results suggest this reddening excess becomes negligible at high redshifts \citep[e.g.,][]{pannella2015,reddy2015}. Here we therefore assume the same $A_{\rm V}$ for the lines as for the continuum. Even though the assumed reddening is the same as for the SED-based $\sfr$s, these line-based estimates are still independent. Indeed, the emission lines are located at visible rather than far-UV wavelengths, and are therefore affected by dust differently than the continuum flux of young OB stars (which drives the SED-based $\sfr$ estimates). The comparison can thus allow us to reveal systematic issues with the dust correction (with the caveat that part of the star-forming regions may be optically thick; this can only be tackled with deep FIR imaging). The results of this analysis are summarized in \rtab{TAB:sfr} and \rfig{FIG:sfrcomp}.

Given the wavelength coverage of our MOSFIRE observations, the best emission line at our disposal for $\sfr$ estimates is $\hbeta$, which is covered for all our spectroscopically-confirmed galaxies. We recall that our line flux measurement procedure automatically corrects for the underlying stellar absorption, see \rsec{SEC:redshifts_method}. To translate the $\hbeta$ luminosities to $\sfr$, we assumed case-B recombination ($L_{\halpha}/L_{\hbeta} = 2.86$) and the \cite{kennicutt1998-a} conversion factor (converted to a Chabrier IMF):
\begin{align}
\sfr_{\rm \hbeta}^{\rm nodust}/L_{\rm \hbeta} = 5.46\times10^{-8}\,\msun/\yr/\lsun.
\end{align}
Using this relation and without dust correction, the $1\sigma$ $\sfr$ sensitivity from the MOSFIRE spectra is of the order of $1$ to $3\,\msun/\yr$ (see \rtab{TAB:sfr}). We also considered $\sfr$s estimated from the $\oii$ line luminosity, which is covered in all but two galaxies. We used the \cite{kewley2004} calibration (converted to a Chabrier IMF):
\begin{align}
\sfr_{\rm \oii}^{\rm nodust}/L_{\rm \oii} = 1.59\times10^{-8}\,\msun/\yr/\lsun.
\end{align}
Without correction for dust, the $\oii$ emission in our MOSFIRE spectra provided comparable $\sfr$ sensitivity to $\hbeta$. We considered the $\oii$-based $\sfr$s less reliable than that based on $\hbeta$ because the physical connection between star-formation and $\oii$ emission is less immediate. In fact, given that our galaxies have (a priori) low $\ssfr$, both emission lines may be significantly contaminated by energy sources other than star-formation; an AGN, or low-ionization emission from old/intermediate-age stars (see, e.g., \citealt{cidfernandes2010,cidfernandes2011} for Balmer lines, and \citealt{yan2006} or \citealt{lemaux2010} for $\oii$). This will tend to bias the line-based $\sfr$s toward higher values.

From the broadband modeling (\rtab{TAB:final_props}), we find only modest attenuation in the confirmed $z>3$ galaxies, with $A_{V} < 0.5$ for all but two galaxies. At the wavelength of $\hbeta$, and assuming the \cite{calzetti2000} attenuation curve, this implies attenuation factors of at most $1.7$ for $\hbeta$, and $2.1$ for $\oii$. In all that follows, we considered the entire range of allowed $A_{V}$ for each galaxy to correct the line $\sfr$s for attenuation.

As illustrated in \rfig{FIG:sfrcomp}, in most cases (8/10) the $\hbeta$-based $\sfr$s are consistent with the range of allowed $10\,\Myr$-averaged $\sfr$ derived using FAST++. Only one galaxy is clearly discrepant, with larger $\hbeta$-based $\sfr$s than expected from the SED: ZF-COS-20133. This is probably caused by it hosting an AGN (see \rsec{SEC:two_sfh}), so we did not consider it further.

The $\oii$-based $\sfr$s are consistent with the FAST++ estimates in most cases (7/9), and the most striking inconsistency is again for ZF-COS-20133, which we thus ignored. The other inconsistent galaxy is for ZF-UDS-8197, for which the $\oii$-based $\sfr$ is actually lower than that inferred from the SED modeling. In fact, overall we find the $\oii$-based $\sfr$s are systematically lower than that derived from $\hbeta$. This could imply that we underestimated the attenuation by dust, and that star-forming regions are more attenuated than the older stars. Before applying any dust correction, the median $\sfr^{\rm nodust}_{\oii}/\sfr^{\rm nodust}_{\hbeta}$ is $0.3$, and reproducing this ratio with the \cite{calzetti2000} attenuation would require an average $A_V\sim3\,{\rm mag}$, which is substantially larger than the $A_V < 0.5$ we obtain from the SED modeling. However, with $A_V\sim3$ the observed $\hbeta$ luminosities would then translate to a median $\sfr=100\,\msun/\yr$, and violate the upper limit from ALMA ($<62\,\msun/\yr$ at $3\sigma$) or \herschel ($<85\,\msun/\yr$ at $3\sigma$, \citealt{straatman2014}).

Therefore, either the $z\sim0$ calibration of the $L_{\oii}$-to-$\sfr$ conversion factor does not apply to $z>3$, or the luminosities of these lines is not related to star formation. Providing a definitive answer to this question would require observing the $\halpha$ line to infer the Balmer decrement, which will only be possible with the upcoming {\it James Webb Space Telescope} (\jwst), although given the expected change in gas-phase metallicity and ionization parameters in high redshift galaxies \citep[e.g.,][]{erb2010,steidel2016,shapley2017}, an evolution of the $\oii$ $\sfr$ calibration would not be surprising. In any case, the fact that the $\hbeta$-based $\sfr$s agree with our SED-based estimates suggests that our SED modeling is not affected by significant biases regarding the dust attenuation, and that the derived star-formation histories should thus be reliable.

\subsection{Specific star-formation rates \label{SEC:ssfr}}

\begin{figure}
\begin{center}
\includegraphics[width=0.5\textwidth]{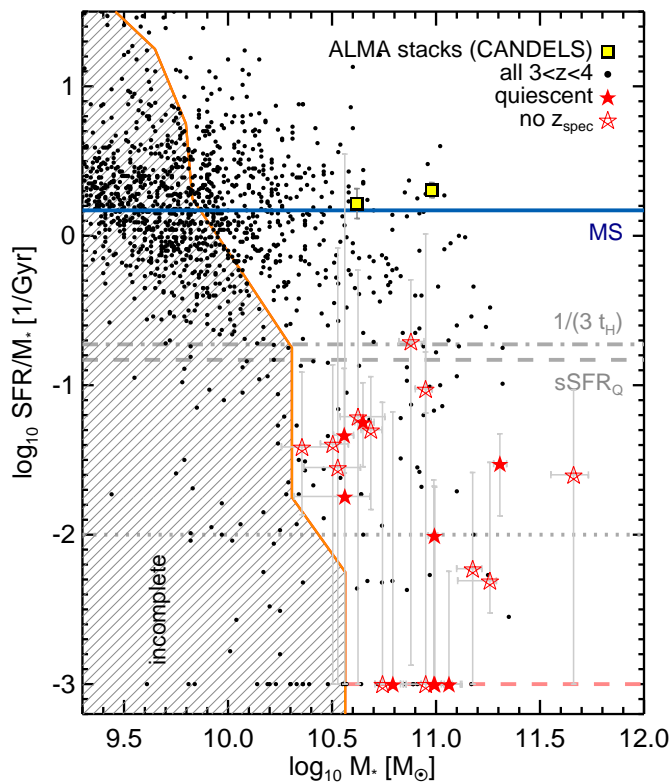}
\end{center}
\caption{Best fit specific $\sfr$ ($\ssfr$) as a function of stellar mass for galaxies at $3<z<4$, as derived from their UV-to-NIR SEDs. The whole ZFOURGE sample (in COSMOS and UDS) is shown as small dots, and the $z>3$ quiescent galaxies observed with MOSFIRE are shown as red stars with gray error bars. Filled and open symbols refer to galaxies with and without a $\zspec$, respectively (values are listed in \rtab{TAB:final_props}). The locus of the main sequence for the ZFOURGE galaxies is shown with a blue line. Different definitions for ``quiescence'' are shown in gray: the one we adopted in this paper, $\ssfr = \ssfr_Q = 0.15\,\Gyr^{-1}$ (dashed), $\ssfr = 1/(3\,t_H)$ (dot-dashed, $t_H$ being the age of the Universe), and $\ssfr = 0.01\,\Gyr^{-1}$ (dotted). The $\ssfr$ measured in stacks of ALMA imaging at $z\sim4$ is shown with yellow squares \citep{schreiber2017}, and was corrected down by a factor $1.3$ to account for redshift evolution of the $\ssfr$ between $z=3.5$ and $z=4$. As in \rfig{FIG:sfrcomp}, for display purposes we place a limit on the minimum $\ssfr$ of $10^{-3}\,\Gyr^{-1}$, which is indicated with the dashed pink line. The hashed region on the bottom left shows the region of this parameter space where the ZFOURGE catalogs are incomplete because of the \Ks magnitude limit.
\label{FIG:ms}}
\end{figure}

We show in \rfig{FIG:ms} the distribution of the specific $\sfr$ ($\ssfr$) of our $z>3$ quiescent galaxies, and compare it to that of ``normal'' main-sequence galaxies. Here and in all that follows, we excluded the two $z<3$ interlopers and ZF-COS-20133 from the quiescent sample. We defined the locus of the main sequence as the average $\ssfr$ of the ZFOURGE galaxies at $3<\zphot<4$ and $3\times10^{9}<\mstar/\msun<10^{10}$, using the values derived in the present paper with the same SED modeling as for the quiescent galaxies (\rsec{SEC:zphot}). We find $\ssfr_{\rm MS} = 1.5\,\Gyr^{-1}$, which is comparable to dust-based estimates \citep[e.g.,][]{schreiber2017}. We chose to use the same approach to derive $\sfr$s for both the main-sequence and quiescent galaxies (i.e., rather than using stacked far-IR measurements to define the main sequence) in order to make the comparison between the two populations as straightforward as possible.

Looking only at the best-fitting models to begin with, we find our quiescent candidates are located more than a factor of ten below the $z\sim3.5$ main sequence, with the exception of ZF-UDS-3651 which is a factor of seven below. When using the upper limits on the $\ssfr$, only five galaxies (24\%) can actually be located within a factor of three of the main sequence, and thus not be truly quiescent; these are the galaxies ZF-COS-17779, ZF-UDS-3651, 3D-UDS-35168, 3D-UDS-39102, and 3D-EGS-34322. ZF-COS-17779, in particular, is the only one which may be located on or above the main sequence. The vast majority of the MOSFIRE sample thus remains at least a factor of ten below the main sequence, within the uncertainties, which confirms that these galaxies must be genuinely quiescent.

In the absolute sense, although the recovered $\ssfr$ are low, the best-fit values still span a range up to $\ssfr = 0.1\,\Gyr^{-1}$ (and ZF-UDS-3651 at $0.2\,\Gyr^{-1}$). In particular, $57\%$ of our galaxies are above the $\ssfr = 0.01\,\Gyr^{-1}$ threshold used in other studies to isolate ``red-and-dead'' galaxies \citep[shown in \rfig{FIG:ms}; see, e.g.,][]{fontana2009,merlin2018}, and only one has an upper limit below this threshold (ZF-COS-20115). This shows that the \uvj selection does not only select galaxies with zero on-going star-formation, but can also include galaxies with low residual $\sfr$ (see also the discussion in \citealt{merlin2018}). While here we consider this more of a feature than a fundamental flaw (these galaxies are still an order of magnitude below the main sequence), it is clearly an important distinction to keep in mind when comparing observed galaxy number counts to models, which we do in \rsec{SEC:rhogal}.

\subsection{Inferred star-formation histories \label{SEC:sfh}}

\begin{figure}
\begin{center}
\includegraphics[width=0.5\textwidth]{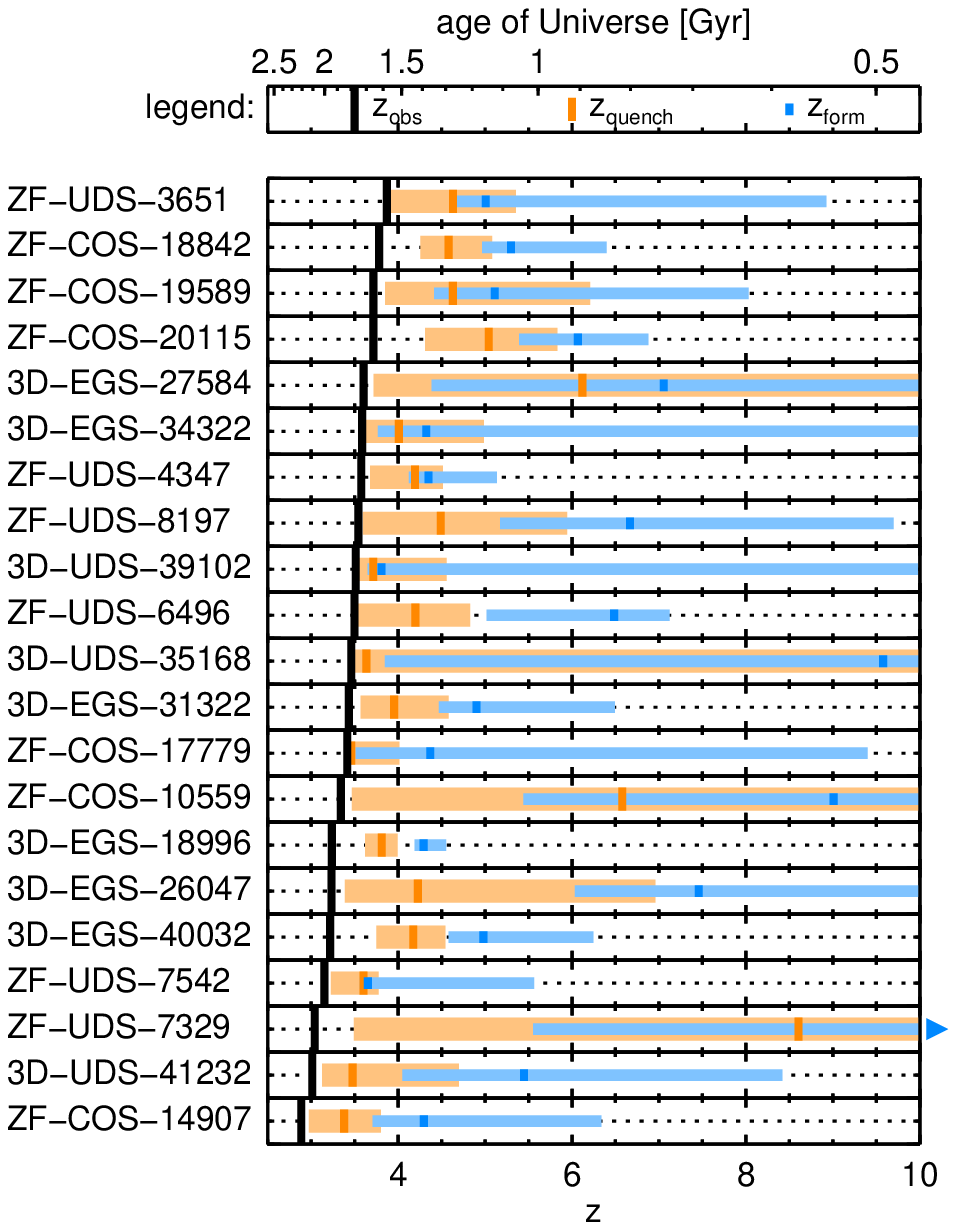}
\end{center}
\caption{Summarized star-formation histories of our $z>3$ quiescent galaxies. Galaxies are sorted by descending redshift, and are each displayed on a separate line. The black vertical bar indicates the redshift at which the galaxy is observed. The orange and blue bands show the $90\%$ confidence range for the quenching and formation redshifts, respectively. Bars of darker colors indicate the corresponding best fit values. These values are listed in \rtab{TAB:final_props}. \label{FIG:zz}}
\end{figure}

\begin{figure}
\begin{center}
\includegraphics[width=0.5\textwidth]{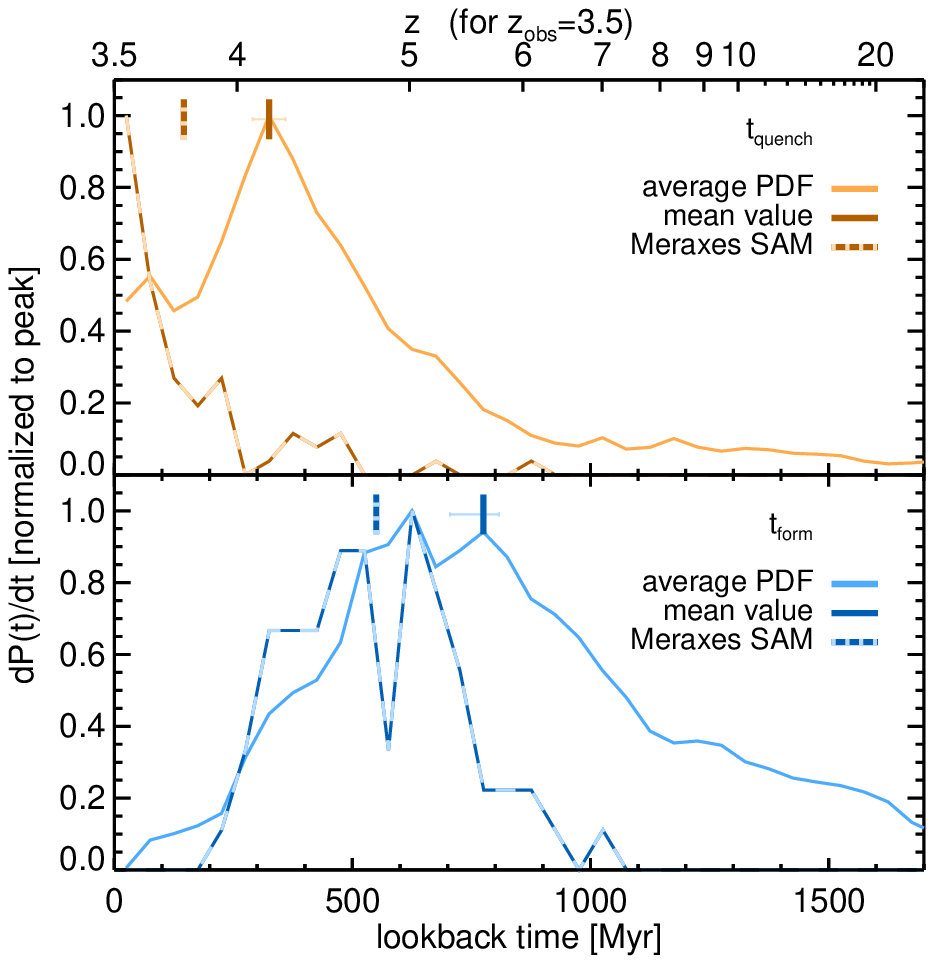}
\end{center}
\caption{Probability distributions functions (PDF) of the quenching (top) and formation (bottom) lookback times for our $z>3$ quiescent galaxies observed with MOSFIRE. For each quantity we show the average of the individual PDFs with a solid line, the mean of the population with a darker vertical line (error bars are the error on the mean), as well as the distribution of values measured in a $z=3.5$ snapshot of the \textsc{Meraxes} semi-analytic model (SAM) with a dashed line (the dashed vertical bar indicates the mean value for the \textsc{Meraxes} galaxies). \label{FIG:tprob}}
\end{figure}

In this section we discuss the inferred star-formation histories of the quiescent galaxies. In \rfig{FIG:zz}, we show the values for $z_{\rm form}$ --- the point in time when half of the mass had formed, and $z_{\rm quench}$ --- the point in time when the galaxy's $\sfr$ dropped below $10\%$ of their respective past average (see \rsec{SEC:fast_fit} for the precise definitions of these quantities). The values are listed in \rtab{TAB:final_props}. In addition, we show in \rfig{FIG:tprob} the probability distribution functions (PDFs) of $t_{\rm form}$ and $t_{\rm quench}$. We show the mean of all PDFs, to illustrate of how well the quantity is constrained for individual objects, as well as the sample's average value. We note that, because of outshining from the youngest stars, our modeling can tend to underestimate the mass of old stellar populations, and the actual age of formation and quenching (see, e.g., \citealt{papovich2001} and the discussion in the Appendix of \citealt{schreiber2018}). In this context, our estimates of $z_{\rm quench}$ and $z_{\rm form}$ could be considered as lower limits. These values are later compared to models in \rsec{SEC:model_sfh}.

\begin{table*}
\begin{center}
\caption{Final properties of the galaxies observed with MOSFIRE ($90\%$ confidence error bars, unless otherwise stated).\label{TAB:final_props}}
\begin{tabular}{lcccccccccc}
\hline\hline \\[-0.3cm]
ID       & $z$ $^a$ & $U-V$ $^a$ & $V-J$ $^a$ & $\mstar$         & $A_{\rm V}$ & $\sfr_{10}$ & $t_{\rm quench}$ & $z_{\rm form}$ & $t_{\rm SF}$ & $\mean{\sfr}_{\rm main}$ \\
         &          & rest, AB   & rest, AB   & $10^{11}\,\msun$ & mag.        & $\msun/\yr$ & $\Gyr$           &                & $\Gyr$       & $\log(\msun/\yr)$ \\
\hline \\[-0.3cm]
\multicolumn{2}{l}{Robust $\zspec$} \\
ZF-COS-20032 & $2.474$ & $2.62^{+0.71}_{-1.61}$ & $2.50^{+0.05}_{-1.40}$ & $1.12^{+0.19}_{-0.47}$ & $3.7^{+0.3}_{-0.2}$ & $139^{+75}_{-139}$ & $0.28^{+0.17}_{-0.28}$ & $2.9^{+0.3}_{-0.3}$ & $0.06^{+0.56}_{-0.05}$ & $3.26^{+0.58}_{-0.98}$ \\[0.1cm]
ZF-COS-20115 & $3.715$ & $1.67^{+0.16}_{-0.02}$ & $0.52^{+0.03}_{-0.03}$ & $1.15^{+0.16}_{-0.09}$ & $0.3^{+0.1}_{-0.1}$ & $0.0^{+0.7}_{-0.0}$ & $0.51^{+0.19}_{-0.24}$ & $6.1^{+0.8}_{-0.7}$ & $0.15^{+0.58}_{-0.13}$ & $2.93^{+0.83}_{-0.66}$ \\[0.1cm]
ZF-COS-20133 & $3.481$ & $1.37^{+0.32}_{-0.01}$ & $1.06^{+0.03}_{-0.01}$ & $0.33^{+0.01}_{-0.06}$ & $0.1^{+0.1}_{-0.1}$ & $4.6^{+0.1}_{-1.3}$ & $1.43^{+0.00}_{-0.44}$ & $32.2^{+0.0}_{-21.1}$ & $0.14^{+0.28}_{-0.02}$ & $2.43^{+0.00}_{-0.54}$ \\[0.1cm]
3D-EGS-18996 & $3.239$ & $1.10^{+0.47}_{-0.01}$ & $0.46^{+0.01}_{-0.15}$ & $0.98^{+0.04}_{-0.06}$ & $0.0^{+0.1}_{-0.0}$ & $1.0^{+1.0}_{-0.9}$ & $0.33^{+0.09}_{-0.10}$ & $4.3^{+0.3}_{-0.1}$ & $0.27^{+0.14}_{-0.20}$ & $2.58^{+0.53}_{-0.16}$ \\[0.1cm]
3D-EGS-26047 & $3.234$ & $1.25^{+0.11}_{-0.06}$ & $1.21^{+0.03}_{-0.03}$ & $0.99^{+0.21}_{-0.14}$ & $0.2^{+0.2}_{-0.2}$ & $0.1^{+1.9}_{-0.0}$ & $0.52^{+0.66}_{-0.42}$ & $7.5^{+7.2}_{-1.4}$ & $0.42^{+0.53}_{-0.32}$ & $2.42^{+0.64}_{-0.34}$ \\[0.1cm]
3D-EGS-40032 & $3.219$ & $1.54^{+0.04}_{-0.09}$ & $0.75^{+0.02}_{-0.01}$ & $2.03^{+0.16}_{-0.14}$ & $0.4^{+0.1}_{-0.1}$ & $6.1^{+3.7}_{-3.4}$ & $0.51^{+0.14}_{-0.20}$ & $5.0^{+1.3}_{-0.4}$ & $0.23^{+0.76}_{-0.17}$ & $2.99^{+0.60}_{-0.61}$ \\[0.1cm]
ZF-UDS-8197 & $3.543$ & $1.09^{+0.25}_{-0.01}$ & $1.16^{+0.02}_{-0.18}$ & $0.36^{+0.04}_{-0.04}$ & $0.0^{+0.3}_{-0.0}$ & $1.7^{+3.3}_{-0.7}$ & $0.43^{+0.39}_{-0.40}$ & $6.7^{+3.0}_{-1.5}$ & $0.69^{+0.21}_{-0.64}$ & $1.76^{+1.03}_{-0.11}$ \\[0.1cm]
3D-UDS-27939 & $2.210$ & $1.79^{+1.27}_{-1.20}$ & $1.77^{+0.87}_{-0.81}$ & $0.46^{+0.05}_{-0.17}$ & $1.0^{+0.6}_{-0.2}$ & $1.8^{+8.0}_{-1.2}$ & $2.36^{+0.00}_{-2.36}$ & $23.6^{+0.0}_{-20.1}$ & $0.23^{+1.14}_{-0.10}$ & $2.39^{+0.17}_{-0.90}$ \\[0.1cm]
\hline \\[-0.3cm]
\multicolumn{2}{l}{Uncertain $\zspec$} \\
ZF-COS-17779 & $3.415$ & $1.39^{+0.13}_{-0.45}$ & $0.92^{+0.01}_{-0.43}$ & $0.36^{+0.12}_{-0.16}$ & $1.4^{+0.6}_{-0.8}$ & $0.7^{+100}_{-0.7}$ & $0.03^{+0.29}_{-0.03}$ & $4.4^{+5.0}_{-0.9}$ & $0.85^{+0.66}_{-0.84}$ & $1.65^{+1.76}_{-0.21}$ \\[0.1cm]
ZF-COS-18842 & $3.782$ & $1.21^{+0.19}_{-0.17}$ & $0.47^{+0.26}_{-0.01}$ & $0.45^{+0.06}_{-0.04}$ & $0.0^{+0.2}_{-0.0}$ & $2.5^{+2.4}_{-1.2}$ & $0.33^{+0.16}_{-0.12}$ & $5.3^{+1.1}_{-0.3}$ & $0.27^{+0.50}_{-0.22}$ & $2.24^{+0.79}_{-0.40}$ \\[0.1cm]
ZF-COS-19589 & $3.715$ & $1.67^{+0.16}_{-0.18}$ & $0.76^{+0.12}_{-0.06}$ & $0.62^{+0.10}_{-0.08}$ & $0.9^{+0.4}_{-0.5}$ & $0.0^{+4.4}_{-0.0}$ & $0.38^{+0.39}_{-0.32}$ & $5.1^{+2.9}_{-0.7}$ & $0.09^{+1.12}_{-0.08}$ & $2.87^{+0.86}_{-1.11}$ \\[0.1cm]
3D-EGS-31322 & $3.434$ & $1.35^{+0.02}_{-0.16}$ & $0.78^{+0.10}_{-0.01}$ & $0.98^{+0.12}_{-0.08}$ & $0.3^{+0.2}_{-0.2}$ & $0.0^{+2.3}_{-0.0}$ & $0.28^{+0.25}_{-0.20}$ & $4.9^{+1.6}_{-0.4}$ & $0.28^{+0.87}_{-0.24}$ & $2.57^{+0.83}_{-0.58}$ \\[0.1cm]
\hline \\[-0.3cm]
\multicolumn{2}{l}{$\zphot$} \\
ZF-COS-10559 & $3.34^{+0.30}_{-1.04}$ & $1.57^{+0.75}_{-0.88}$ & $1.05^{+0.89}_{-0.32}$ & $0.23^{+0.08}_{-0.05}$ & $0.0^{+0.5}_{-0.0}$ & $0.9^{+2.2}_{-0.5}$ & $1.06^{+0.46}_{-0.98}$ & $9.0^{+23.3}_{-3.6}$ & $0.16^{+0.74}_{-0.07}$ & $2.22^{+0.29}_{-0.75}$ \\[0.1cm]
ZF-COS-14907 & $2.89^{+0.06}_{-0.06}$ & $1.38^{+0.08}_{-0.07}$ & $0.76^{+0.04}_{-0.03}$ & $0.49^{+0.06}_{-0.05}$ & $0.2^{+0.3}_{-0.2}$ & $2.4^{+3.2}_{-1.8}$ & $0.36^{+0.24}_{-0.29}$ & $4.3^{+2.0}_{-0.6}$ & $0.68^{+0.55}_{-0.63}$ & $1.90^{+1.13}_{-0.25}$ \\[0.1cm]
3D-EGS-27584 & $3.60^{+0.18}_{-0.23}$ & $2.25^{+0.29}_{-0.23}$ & $1.33^{+0.16}_{-0.14}$ & $4.59^{+0.82}_{-1.02}$ & $1.3^{+0.7}_{-0.3}$ & $11.5^{+29.7}_{-11.5}$ & $0.82^{+0.61}_{-0.76}$ & $7.1^{+30.7}_{-2.7}$ & $0.09^{+1.23}_{-0.07}$ & $3.73^{+0.60}_{-1.16}$ \\[0.1cm]
3D-EGS-34322 & $3.59^{+0.33}_{-0.32}$ & $1.25^{+0.51}_{-0.16}$ & $0.95^{+0.05}_{-0.23}$ & $0.42^{+0.15}_{-0.08}$ & $0.9^{+0.7}_{-0.8}$ & $2.6^{+25.0}_{-2.6}$ & $0.21^{+0.35}_{-0.21}$ & $4.3^{+8.5}_{-0.6}$ & $0.09^{+1.28}_{-0.08}$ & $2.67^{+0.92}_{-1.09}$ \\[0.1cm]
ZF-UDS-3651 & $3.87^{+0.12}_{-0.12}$ & $1.35^{+0.11}_{-0.25}$ & $0.74^{+0.01}_{-0.01}$ & $0.76^{+0.12}_{-0.09}$ & $0.7^{+0.2}_{-0.5}$ & $14.8^{+25.1}_{-14.7}$ & $0.31^{+0.21}_{-0.31}$ & $5.0^{+3.9}_{-0.4}$ & $0.13^{+0.93}_{-0.09}$ & $2.79^{+0.50}_{-0.89}$ \\[0.1cm]
ZF-UDS-4347 & $3.58^{+0.04}_{-0.05}$ & $1.54^{+0.06}_{-0.09}$ & $0.81^{+0.08}_{-0.04}$ & $0.32^{+0.06}_{-0.04}$ & $0.4^{+0.2}_{-0.2}$ & $1.3^{+3.7}_{-1.3}$ & $0.30^{+0.12}_{-0.24}$ & $4.4^{+0.8}_{-0.2}$ & $0.06^{+0.75}_{-0.05}$ & $2.75^{+0.71}_{-1.09}$ \\[0.1cm]
ZF-UDS-6496 & $3.50^{+0.04}_{-0.04}$ & $1.73^{+0.04}_{-0.02}$ & $0.81^{+0.01}_{-0.02}$ & $0.89^{+0.07}_{-0.10}$ & $0.3^{+0.2}_{-0.1}$ & $8.4^{+6.2}_{-2.7}$ & $0.34^{+0.22}_{-0.34}$ & $6.5^{+0.6}_{-1.5}$ & $0.81^{+0.10}_{-0.73}$ & $2.08^{+0.99}_{-0.08}$ \\[0.1cm]
ZF-UDS-7329 & $3.04^{+0.17}_{-0.17}$ & $2.05^{+0.28}_{-0.32}$ & $1.16^{+0.13}_{-0.10}$ & $1.82^{+0.14}_{-0.55}$ & $0.5^{+0.3}_{-0.1}$ & $0.9^{+3.3}_{-0.4}$ & $1.51^{+0.22}_{-1.21}$ & $15.0^{+17.2}_{-9.5}$ & $0.27^{+0.54}_{-0.19}$ & $2.90^{+0.36}_{-0.53}$ \\[0.1cm]
ZF-UDS-7542 & $3.15^{+0.06}_{-0.06}$ & $1.49^{+0.08}_{-0.05}$ & $0.97^{+0.02}_{-0.01}$ & $0.55^{+0.14}_{-0.04}$ & $1.1^{+0.2}_{-0.5}$ & $0.0^{+4.7}_{-0.0}$ & $0.28^{+0.09}_{-0.23}$ & $3.7^{+1.9}_{-0.1}$ & $0.02^{+1.23}_{-0.01}$ & $3.43^{+0.29}_{-1.69}$ \\[0.1cm]
3D-UDS-35168 & $3.46^{+0.32}_{-0.29}$ & $1.37^{+0.35}_{-0.28}$ & $0.81^{+0.11}_{-0.12}$ & $0.34^{+0.10}_{-0.11}$ & $0.3^{+1.1}_{-0.3}$ & $0.9^{+24.6}_{-0.9}$ & $0.10^{+1.41}_{-0.10}$ & $9.6^{+27.4}_{-5.7}$ & $0.73^{+0.77}_{-0.71}$ & $1.72^{+1.56}_{-0.37}$ \\[0.1cm]
3D-UDS-39102 & $3.51^{+0.42}_{-0.36}$ & $1.61^{+0.51}_{-0.50}$ & $1.32^{+0.11}_{-0.24}$ & $0.89^{+0.44}_{-0.22}$ & $1.9^{+0.4}_{-0.8}$ & $0.0^{+100}_{-0.0}$ & $0.11^{+0.36}_{-0.11}$ & $3.8^{+6.2}_{-0.2}$ & $0.07^{+1.39}_{-0.06}$ & $3.11^{+0.85}_{-1.20}$ \\[0.1cm]
3D-UDS-41232 & $3.01^{+0.07}_{-0.08}$ & $1.56^{+0.16}_{-0.12}$ & $0.82^{+0.04}_{-0.03}$ & $1.50^{+0.16}_{-0.24}$ & $0.3^{+0.3}_{-0.2}$ & $0.9^{+2.9}_{-0.9}$ & $0.32^{+0.54}_{-0.23}$ & $5.4^{+3.0}_{-1.4}$ & $0.98^{+0.18}_{-0.92}$ & $2.24^{+1.20}_{-0.08}$ \\[0.1cm]
\hline
\end{tabular}
\end{center}
{\footnotesize This table is available in electronic format at the CDS. $^a$ $68\%$ confidence intervals.}
\end{table*}

\subsubsection{When did star-formation stop?}

All the galaxies in our sample have best-fit solutions which have quenched (as per our adopted definition for $t_{\rm quench}$). Quantitatively, the $16$ and $84$th percentiles of $t_{\rm quench}$ are $210$ and $510\,\Myr$, $71\%$ of the galaxies have an upper bound on $t_{\rm quench}$ larger than $500\,\Myr$, and $t_{\rm quench}=0$ is ruled for $76\%$ of the sample. Stacking the probability distributions, as shown in \rfig{FIG:tprob}, we find that the average $t_{\rm quench}$ is $330\pm30\,\Myr$. In other words, $76\%$ of our quiescent galaxies have quenched with certainty, on average $\sim330\,\Myr$ before being observed. For reference, at $z=3$, $3.5$, and $4$, $t_{\rm quench}=330\,\Myr$ implies a quenching at $z_{\rm quench}=3.5$, $4.2$ and $4.9$, respectively.

Looking at individual objects, $t_{\rm quench} < 200\,\Myr$ is excluded for five of our quiescent galaxies ($25\%$). These are ZF-COS-18842, ZF-COS-20115, ZF-UDS-7329, 3D-EGS-18996, and 3D-EGS-40032 (all but ZF-UDS-7329 are spectroscopically confirmed). ZF-COS-20115 has the highest redshift of these five galaxies, and its case was already discussed at length in our previous works (\citealt{glazebrook2017,schreiber2018}). Here we confirm once more that the galaxy must have quenched between $270$ and $700\,\Myr$ prior to observation ($4.3<z_{\rm quench}<5.8$).

Otherwise, ZF-UDS-7329, albeit not spectroscopically confirmed, is particularly interesting: its photometry requires a quenching as recently as $300\,\Myr$ prior to observation ($z_{\rm quench} > 3.5$), but it also allows much older solutions up to $1.7\,\Gyr$ prior to observation, equivalent to $z_{\rm quench}=12$, while the other galaxies are limited to $z_{\rm quench} < 6$. The reason why is apparent in \rfig{FIG:allseds}, where it is clear that ZF-UDS-7329 has the SED least resembling that of an A star (as opposed, e.g., to ZF-COS-20115). An older stellar population may also explain why no significant absorption line was found in its deep $K$-band spectrum (at its $\zphot$, we would expect to see only $\hbeta$). If it indeed stopped forming stars at such extremely high redshifts, this galaxy may prove difficult to explain in our current understanding of cosmology \citep[e.g.,][]{behroozi2018}. Follow up observations are under way to confirm its redshift and constrain further its star formation history.

Two other galaxies in our sample may have had such an early quenching, ZF-COS-10559 and 3D-EGS-27584. However, their photometry also allows a more recent quenching, as late as $70\,\Myr$ prior to observation. While ZF-COS-10559 is simply too faint to offer reliable constraints on the quenching, 3D-EGS-27584 has a bright and clean SED. The main source of uncertainty for this galaxy lies in dust obscuration, which could be as high as $A_{\rm V}=2$, in which case the quenching might be quite recent, while the earliest quenching scenarios correspond to the fits with $A_{\rm V}<1.2$. Deep sub-mm imaging could break this degeneracy.

On the other hand, five galaxies have a lower bound of $t_{\rm quench}=0\,\Myr$, that is, their photometry is also compatible with SFHs that have not fully quenched. These are ZF-COS-17779, ZF-UDS-3651, 3D-UDS-35168, 3D-UDS-39102, and 3D-EGS-34322. All of them still have best-fit solutions which have quenched however, and the possibility of non-quenched solutions can be partly explained by their simply being fainter (median $K$ magnitude fainter than the rest of the sample by $0.5\,{\rm mag}$), so their photometry is less constraining. This is apparent in \rfig{FIG:allseds}. In fact, only one of these has a spectroscopic redshift. Based on the brighter galaxies, we expect in most cases that deeper NIR spectra or photometry would lift this ambiguity and eventually rule out these non-quenched solutions.

\subsubsection{When and how did these galaxies form?}

From \rfig{FIG:zz}, the formation epoch of these galaxies appears less well constrained, but this is partly an effect of the non-linearity of the redshift. The $16$ and $84$th percentiles of $t_{\rm form}$ are $360$ and $1070\,\Myr$, $71\%$ of the galaxies have an upper bound on $t_{\rm form}$ larger than $1\,\Gyr$, and values less than $400\,\Myr$ are excluded for $57\%$ of the sample. From the stacked probability distributions, we find an average of $t_{\rm form}=780^{+30}_{-70}\,\Myr$. At $z=3$, $3.5$, and $4$, $t_{\rm form}=780\,\Myr$ implies a formation at $z_{\rm form}=4.4$, $5.6$ and $7.1$, respectively.

During this main formation epoch, the average $\sfr$ of the galaxies must have been high: the $16$ and $84$th percentiles of $\mean{\sfr}_{\rm main}$ are $80$ and $850\,\msun/\yr$, and the stacked probability distributions lead to an average of $340^{+150}_{-30}\,\msun/\yr$. We recall that this value is the average $\sfr$ during the formation phase; the peak $\sfr$ would be even higher. The duration of this phase, $T_{\rm SF}$, has a $16$ and $84$th percentile range of $90$ to $700\,\Myr$, with an average of $280^{+40}_{-134}\,\Myr$. This implies that substantial star-formation must have occurred at $z\sim5$ in brief and intense episodes in order to form the bulk of the $3<z<4$ quiescent population, and supports the results of \citetalias{glazebrook2017}.

Considering individual galaxies, $z_{\rm form} < 5$ is excluded for six galaxies ($28\%$), namely ZF-COS-10559, ZF-COS-20115, ZF-UDS-6496, ZF-UDS-8197, ZF-UDS-7329, and 3D-EGS-26047. The latter is the only galaxy with a lower limit of $z_{\rm form} > 6$, while ZF-COS-20115 has the highest implied past $\sfr$, with a lower limit of $\mean{\sfr}_{\rm main} > 190\,\msun/\yr$.

\section{Number density \label{SEC:rhogal}}

In the previous sections, we have demonstrated that the \uvj colors can be used to select genuinely quiescent galaxies at $z>3$ with a high purity ($80\%$). Using this knowledge, we came back to the full ZFOURGE sample and revised the observed number density of quiescent galaxies. This is discussed in the following subsections.

By coming back to the entire ZFOURGE catalogs, rather than only working with those galaxies observed with MOSFIRE, we ensured that the derived number densities are unaffected by the non-trivial selection function of our masks (\rsec{SEC:masks}). Therefore the only relevant factor limiting our completeness is the $K$-band flux, which we already addressed in \rsec{SEC:zphot}.

\subsection{Number density of \uvj-quiescent galaxies}

Combined, the three ZFOURGE fields contain $130$ galaxies at $3<z<4$ and $\mstar > 3\times10^{10},\msun$, where our catalogs are complete (see \rsec{SEC:zphot}). The \uvj colors classify $23$ galaxies as quiescent, eight of which were observed with MOSFIRE and confirmed to be robust candidates from the absence of strong emission lines. For the quiescent galaxies without MOSFIRE coverage, following our results from \rsec{SEC:new_colors} we assumed that $20\%$ are not truly quiescent (i.e., they are either dusty interlopers or high equivalent-width $\oiii$ emitters). This lead to a statistically corrected number of $20$ galaxies, corresponding to a number density of $(1.4\pm0.3)\times10^{-5}\,\Mpc^{-3}$ and a stellar mass density of $(1.0\pm0.2)\times10^{6}\,\msun/\Mpc^{3}$ (error bars are only Poisson noise). Following \cite{moster2011}, we estimate the amplitude of cosmic variance on these numbers to be $22\%$. This number density is lower by $30\%$ compared to the value first reported in \cite{straatman2014}, even though their sample was built at slightly higher redshift and masses ($3.4<z<4.2$ and $\mstar > 4\times10^{10}\,\msun$). This is mostly the result of the de-contamination from redshift interlopers, and strong $\oiii$ emitters. Given the moderate number of objects at play, this change is nevertheless contained within the Poisson error bars.

\subsection{Completeness of the \uvj selection and link to $\ssfr$}

\begin{figure*}
\begin{center}
\includegraphics[width=\textwidth]{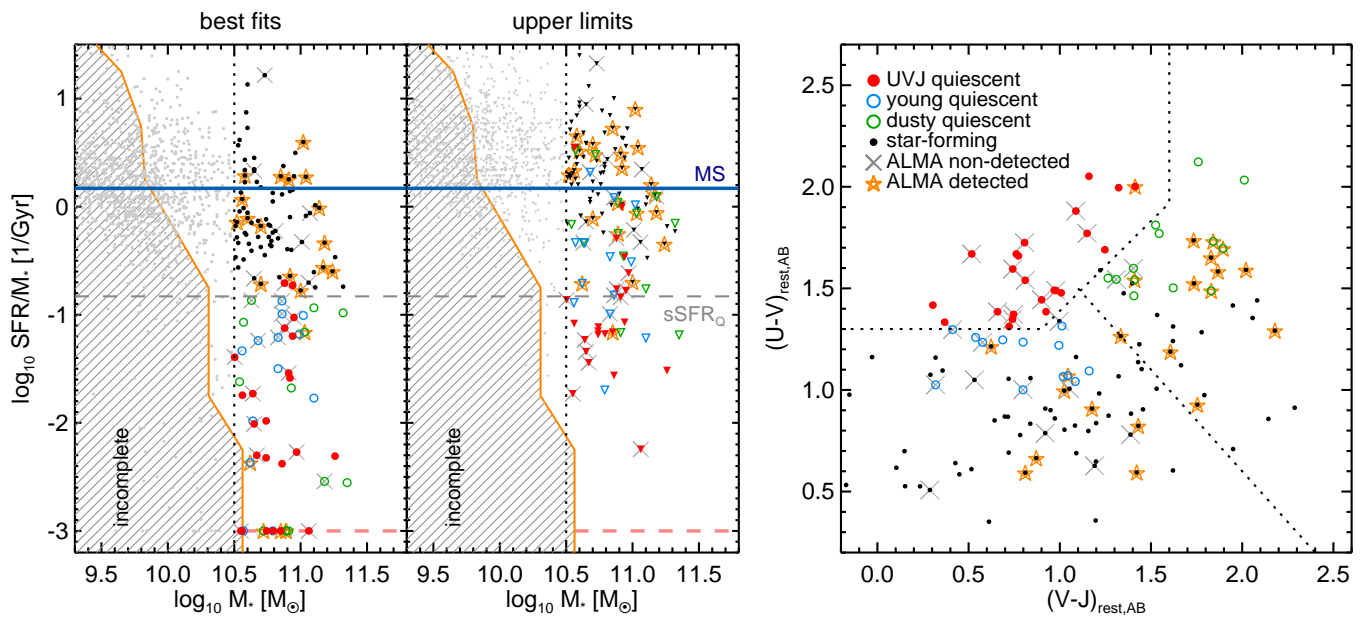}
\end{center}
\caption{Comparison of $\ssfr$ and \uvj classification for the $3<z<4$ galaxies with $\mstar > 3\times10^{10}\,\msun$ in ZFOURGE (COSMOS and UDS). The $\ssfr$ are displayed on the left (best fits on the left, and upper limits on the right), while \uvj colors are shown on the right. \uvj-quiescent galaxies are highlighted in red. Galaxies outside of the quiescent region but with $\ssfr<\ssfr_{\rm Q}=0.15\,\Gyr^{-1}$ are labeled with green and blue open symbols, depending on whether they are in the ``dusty'' or ``non-dusty'' part of the \uvj diagram (respectively), as delimited with the diagonal dotted line ($U-J = 2.6$). \label{FIG:uvj_comp}}
\end{figure*}

While the MOSFIRE observations we present in this paper demonstrate that the \uvj selection has a high purity, its completeness still remains to be addressed. Indeed, the galaxy 3D-EGS-18996 studied in the present paper (see \rsec{SEC:new_colors}) and the analysis of \cite{merlin2018} suggest that genuinely quiescent galaxies can be found outside of the fiducial \uvj boundaries. Merlin et al.~showed that this can be the case either when a quiescent galaxy is both old and significantly obscured by dust (it then fails the $V-J$ cut), or if it is devoid of dust and quenched abruptly less than a few million years prior observation (in which case it fails the $U-V$ cut, as is the case for 3D-EGS-18996). The latter could be a common occurrence, especially at high redshifts where galaxies had little available time to evolve passively after quenching. If true, \uvj-selected sample may underestimate the actual number densities of quiescent galaxies.

We illustrate this issue in \rfig{FIG:uvj_comp} by showing the $\ssfr$s of all the ZFOURGE galaxies, in relation to their position on the \uvj diagram. As discussed in the previous section, \uvj-quiescent galaxies in our sample are predominantly located at $\ssfr < \ssfr_{\rm MS}/10$, which we adopt as a threshold for quiescence: $\ssfr_{\rm Q}=0.15\,\Gyr^{-1}$. We note that this threshold is valid only for $z>3$ galaxies, and corresponds to a ``relative'' quiescence; galaxies below this threshold at $z>3$ are forming stars with rates an order of magnitude lower than the bulk of the star-forming population. Galaxies thus labeled ``quiescent'' may still form stars at low rates of up to $10\,\msun/\yr$, which would be high for $z\sim0$ galaxies of similar masses, but is nonetheless far enough from the average at $z>3$ to require an abnormal event (e.g., quenching) in the galaxy's history (see discussion in the introduction).

As shown in \rfig{FIG:uvj_comp}, while the vast majority of the \uvj-quiescent galaxies are found below $\ssfr = \ssfr_{\rm Q}$, $55\%$ of the galaxies below this threshold are not classified as \uvj-quiescent. This would imply indeed that the \uvj selection is incomplete.

We find $26$ such galaxies with a best-fit $\ssfr < \ssfr_Q$ but \uvj-``star-forming'' colors, and note that their colors remain nonetheless within $0.4\,{\rm mag}$ of the \uvj dividing line. Based on the above discussion, we split this sample in two parts according to their \uvj colors (see \rfig{FIG:uvj_comp}): ``young-quiescent'' galaxies with ``blue'' colors, $(U-J) < 2.6$, and ``dusty-quiescent'' galaxies with ``red'' colors, $(U-J) > 2.6$. Each sample contains $13$ galaxies. In the following, we quantify what fraction of these are genuinely quiescent, that is, with $\ssfr < \ssfr_{\rm Q}$ and no dust-obscured star formation.

\subsection{Young or dusty quiescent galaxies \label{SEC:qq}}

In this section we analyze the available data for these ``young'' or ``dusty-quiescent'' galaxies to prune the sample, and determine which of them are most likely to be genuinely quiescent. We first use archival ALMA data to identify dust-obscured star-forming galaxies and, for those galaxies without ALMA coverage, to statistically infer the rate of contamination from dusty SFGs. We then use the probability distribution function of the $\ssfr$ as obtained from the SED modeling to estimate how many are truly at $\ssfr < \ssfr_{\rm Q}$. Using these data, we finally estimate their contribution to the number density of quiescent galaxies.

\subsubsection{Removing dusty SFGs}

We cross-matched all the ZFOURGE galaxies at $3<z<4$ and $\mstar > 3\times10^{10},\msun$ with the ALMA archive, as in \rsec{SEC:alma} for the MOSFIRE sample, and looked for detections\footnote{For this exercise we used data from the same programs listed in \rsec{SEC:alma}, adding Band 7 data from 2015.1.01495.S (PI: Wang), 2012.1.00307.S (PI: Hodge), 2012.1.00869.S (PI: Mullaney), as well as Band 6 observations from 2013.1.00151.S and 2015.1.00379.S (PI: Schinnerer), and 2013.1.00118.S (PI: Aravena).}. Of the $130$ massive galaxies in ZFOURGE, $46$ were observed with ALMA, and $22$ were detected at more than $3\sigma$ significance. The detection limit is typically $0.8$ and $0.2\,\mJy$ in Band 7 and 6, respectively, which is equivalent to an $\lir$ detection limit of about $10^{12}\,\lsun$ \citep{schreiber2018-a}, or $\sfr \gtrsim 100\,\msun/\yr$ \citep{kennicutt1998-a}. An ALMA detection can therefore be used to rule out quiescence (see however \citealt{schreiber2018}). The coverage and detection rates are summarized in \rtab{TAB:alma}. There, and in what follows, we only consider as star-forming the galaxies that have \uvj-star-forming colors and $\ssfr > \ssfr_Q$.

\begin{table}
\begin{center}
\caption{ALMA coverage and detection rates. \label{TAB:alma}}
\begin{tabular}{lllll}
\hline\hline\\[-0.3cm]
Category$^a$ & Num.$^b$ & Cov.$^c$ & Det.$^d$ & Frac.$^e$ \\
\hline \\[-0.3cm]
All               & 130 & 46 & 22 & 48\% \\
Star-forming      & 81  & 23 & 16 & 70\% \\
$\dots$ dusty     & 24  & 8  & 8  & 100\% \\
$\dots$ non-dusty & 57  & 15 & 8  & 53\% \\
Dusty-quiescent   & 13  & 6  & 4  & 67\% \\
Young-quiescent   & 13  & 5  & 1  & 20\% \\
\uvj-quiescent    & 23  & 12 & 1  & 8\% \\
\hline
\end{tabular}
\end{center}
{\footnotesize $^a$ Star-forming: \uvj star-forming and $\ssfr > \ssfr_Q$, Dusty-quiescent: \uvj star-forming, $(U-J) > 2.6$, and $\ssfr < \ssfr_Q$, Young-quiescent: \uvj star-forming, $(U-J) < 2.6$, and $\ssfr < \ssfr_Q$. For star-forming galaxies, ``dusty'' and ``non-dusty'' refer to the location of galaxies on the \uvj diagram as per the ``dusty'' and ``young'' quiescent samples. $^b$ Total number of galaxies in this category. $^c$ Number of galaxies located inside the full-width-half-power area covered by archival ALMA observations. $^d$ Number of detections at more than $3\sigma$ significance. $^e$ Fraction of detections among the covered galaxies.}
\end{table}

Among galaxies covered by ALMA, dusty-quiescent galaxies have a detection rate of $67\%$ (4/6), while the young-quiescent have a detection rate of only $20\%$ (1/5). Albeit to a lesser extent, a similar trend can be observed for star-forming galaxies, which have detection rates of $100\%$ (8/8) for the dusty galaxies and $53\%$ (8/15) for the non-dusty galaxies. We also note that the detection rate for \uvj-quiescent galaxies is particularly low ($8\%$, 1/12).

From this we conclude that the vast majority of the dusty-quiescent galaxies are not robustly quiescent. The possibility of their dust emission being powered by old or intermediate age stars should be explored in more detail (e.g., \citealt{fumagalli2014,schreiber2018}), but to be conservative we plainly discarded them as unreliable and focused exclusively on the young-quiescent galaxies.

\begin{figure*}
\begin{center}
\includegraphics[width=\textwidth]{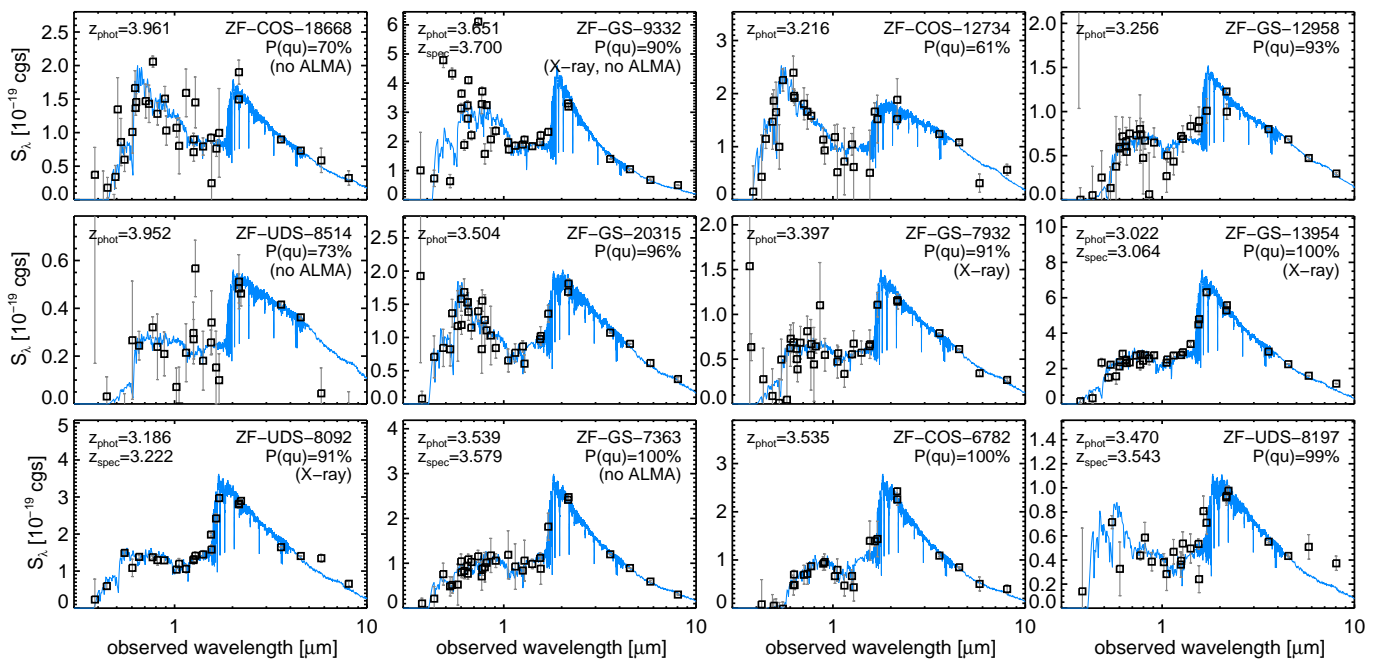}
\end{center}
\caption{Same as \rfig{FIG:allseds}, but for the ``young-quiescent'' galaxies, namely galaxies in ZFOURGE that lie outside of the \uvj-quiescent region but are probably still quiescent. Beside the name of each galaxy, we show the probability that its $\ssfr$ is lower than $\ssfr_{\rm Q}$, our adopted threshold for quiescence, and we also indicate which galaxies have X-ray detections (``X-ray'') and those which are covered by ALMA but not detected (``no ALMA''). \label{FIG:seds_psb} }
\end{figure*}

Of the $13$ young-quiescent galaxies, five were observed with ALMA, and only one is detected (ZF-GS-8706), which we excluded in the following. This implies that, although a priori non-dusty from their \uvj colors, a small fraction are nevertheless dusty contaminants, of order $20\%$. We statistically take this into account later for the eight galaxies lacking ALMA coverage. The SEDs of the $12$ remaining galaxies, excluding ZF-GS-8706, are shown in \rfig{FIG:seds_psb}. These are the galaxies we considered as genuine quiescent candidates complementing the \uvj-selected quiescent galaxies.

\subsubsection{Spectroscopic redshifts and high AGN fraction \label{SEC:qqz}}

Four of the young-quiescent galaxies are X-ray detected, either by Chandra (ZF-GS-7932, 9332, and 13954) or XMM-Newton (ZF-UDS-8092), resulting in an X-ray detection rate of $33\%$, twice higher than for \uvj-quiescent galaxies ($18\%$). This suggests that AGNs are particularly common in this population, similarly to ``green-valley'' galaxies at lower redshifts \citep[e.g.,][]{wang2017}.

One galaxy, ZF-UDS-8197, had its redshift confirmed to $\zspec=3.543$ in our MOSFIRE masks (see \rsec{SEC:redshifts}). Otherwise, four galaxies have spectroscopic redshifts from the literature, all in excellent agreement with the photometric redshifts: ZF-GS-7363 at $\zspec=3.579$ \citep{tasca2017}, ZF-GS-9332 at $\zspec=3.700$ \citep{hernan-caballero2011}, ZF-GS-13954 at $\zspec=3.064$ \citep{silverman2010}, and ZF-UDS-8092 at $\zspec=3.222$ \citep{akiyama2015}. This suggests that redshift outliers are not a major concern for these galaxies, especially given that three of these galaxies with confirmed redshifts are among the X-ray detected AGNs, for which we could expect the photometric redshifts to be the most uncertain. This can be explained by their characteristic SEDs (\rfig{FIG:seds_psb}), in which both the Lyman and Balmer breaks are present and unambiguously constrain the redshift. We thus expect this population to not be significantly affected by redshift outliers, and assumed in the following that their $\zphot$ are accurate.

However, a fraction of this sample could be affected by high equivalent-width emission lines, particularly those hosting an AGN. Two young-quiescent galaxies were observed in our MOSFIRE runs, ZF-UDS-8092 and ZF-UDS-8197 (the former was not included in our analysis so far because it failed to fulfill our UVJ color cuts, see \rsec{SEC:sample_final}). Both galaxies indeed show significant $\oiii$ emission. ZF-UDS-8197 was already discussed in \rsec{SEC:decontam} as one of the ``high-EW'' galaxy, and its photometry was subsequently corrected for the $\oiii$ flux. ZF-UDS-8092 has a much weaker equivalent-width ($35\,\AA$) such that its impact on the photometry is marginal. A more systematic near-IR spectroscopic follow-up would be required to definitively address this question, but this suggests the implied SFH for these galaxies should be interpreted with caution, as ages (or stellar masses) might be biased high if $\oiii$ contributes significantly to the $K$ band flux\footnote{To some extent, $\oii$ could counter-balance this bias by increasing the flux in the $H$ band, however none of the galaxies we observed with MOSFIRE here had $\oii$ contributing more than a few percent of the $H$-band photometry. We therefore only expect $\oiii$ to be a source of concern in case of bright emission lines.}. We show however in \rsec{SEC:qq_sfh} that these galaxies have measurably younger stellar populations than the other quiescent galaxies, which suggests this should not be a major issue.

\subsubsection{Contribution to the number density}

While these young-quiescent galaxies have best-fit $\ssfr$ lower than $\ssfr_{\rm Q}$, $67\%$ actually have an upper limit on their $\ssfr$ larger than $\ssfr_{\rm Q}$, compared to only $30\%$ for the \uvj-quiescent galaxies. To account for this, we used the probability distribution of the $\ssfr$ derived from the SED modeling and weighted each galaxy by the probability that its $\ssfr$ is indeed lower than $\ssfr_{\rm Q}$. We find that $10.6$ galaxies should be truly below this threshold. Assuming a $20\%$ ALMA detection rate for the galaxies without ALMA coverage, this number was further reduced to a total of $9.2$ genuinely quiescent galaxies. Applying the same weighting schemed to the \uvj-quiescent galaxies, we find $18.6$ \uvj-quiescent galaxies below $\ssfr_{\rm Q}$.

Combined, the two samples produce a total of $27.8$ galaxies with $\ssfr < \ssfr_{\rm Q}$, or a space density of $(2.0\pm0.3)\times10^{-5}\,\Mpc^{-3}$. This is $40\%$ larger than the space density of \uvj-quiescent galaxies alone: young-quiescent galaxies therefore form a sizable fraction of the quiescent population at $z>3$.

\subsubsection{Star formation histories \label{SEC:qq_sfh}}

To understand what sets these young-quiescent galaxies apart from the rest of the \uvj-quiescent population, we compared their estimated star-formation histories from the SED modeling, keeping in mind that high-EW $\oiii$ emission might tend to artificially increase the strength of their Balmer break (see \rsec{SEC:qqz}), hence drive models toward older ages.

As expected based on the analysis of \cite{merlin2018}, we find that young-quiescent galaxies have experienced a more recent quenching: the 16th and 84th percentiles of $t_{\rm quench}$ are $15$ and $330\,\Myr$, respectively, about $200\,\Myr$ later than the other quiescent galaxies. There is evidence that they also entered their main formation phase on average $300\,\Myr$ later. This would indicate at first order that their SFHs are not intrinsically different from that of other quiescent galaxies, but are simply shifted to later times.

The existence of such objects is natural if the quiescent population was assembled gradually, with new galaxies becoming quiescent at all epochs. In this context, their relative importance to the overall quiescent population can be expected to increase toward earlier times, to the point where they must be dominant at the epoch of formation of the very first quiescent galaxies. As showed in the previous section, \uvj-quiescent galaxies were still dominant at $3<z<4$ ($66\%$ of the population), but not by a large margin. This suggests that the \uvj selection may not be adequate in finding $z>4$ quiescent galaxies, and that a revised color selection or an $\ssfr$ criterion will have to be used instead.

One way to achieve this would be to adjust the bottom edge of the \uvj selection. For example, at $3<z<4$, 6 out of 13 of the young-quiescent galaxies would be included in the \uvj selection if we were to remove the constraint $(U-V)>1.3$. This constraint was initially introduced in \cite{williams2009} to prevent blue galaxies from entering this region by random scatter, which is critical when blue galaxies dominate the parent sample. For galaxies as massive as those considered here, this is not so much an issue, both because the photometry is quite accurate so random scatter is limited, but also because relatively few massive galaxies are blue or un-obscured. One could therefore adopt this as a revised \uvj selection, fine-tuned for more massive samples. However, if we were to remove this constraint we would also include $7$ galaxies with $\ssfr > \ssfr_{\rm Q}$, which would reduce the purity of the sample. This could be alleviated by adding more colors, for example sensitive to more recent star-formation (far-UV).

\section{Discussion \label{SEC:discussion}}

Our observations clearly demonstrate the existence of quiescent galaxies at $z>3$, and confirm the high number density estimated from photometric samples. With these facts solidly established, the main question still remains open: do we understand how these objects formed? In this section we describe if and how the $z>3$ quiescent population is reproduced in current galaxy formation models. Following \citetalias{glazebrook2017}, we looked in particular at how well the number densities are reproduced, and, when possible, how the simulated star formation histories compare to that inferred from our sample.

We based the comparison of number densities on the observed number of galaxies with low $\ssfr$, since this selection is immediately applicable to models, and we therefore included the young-quiescent galaxies in this analysis (i.e., the galaxies with low $\ssfr$ located outside of the \uvj quiescent region, see \rsec{SEC:qq}). The following discussion would not change significantly if we were to only consider the \uvj quiescent galaxies.

\subsection{Number densities predicted by semi-analytic models}

We first looked at semi-analytic models (SAMs). We downloaded the first $3.14{\rm deg}^2$ light cone of the \cite{henriques2015} SAM\footnote{\url{http://galformod.mpa-garching.mpg.de/public/LGalaxies/downloads.php}} and searched for galaxies in the same range of mass and redshift as ours. Looking at quiescent galaxies with $\ssfr < \ssfr_{\rm Q}$ we find only four objects. The corresponding number density, $1.1\times10^{-7}\,\Mpc^{-3}$, is smaller than our observed value by a factor $\sim$$200$. Simulating uncertainties in stellar masses as a log-normal error of $0.07\,\dex$ (the median uncertainty of our galaxies) increased the predicted number density by $20\%$, which is insufficient to match the observed value. The overall number density of massive galaxies this model predicts, regardless of $\ssfr$, is also too small by a factor of about four: $2.4\times10^{-5}\,\Mpc^{-3}$, compared to $9.3\times10^{-5}\,\Mpc^{-3}$ in the ZFOURGE catalogs. This suggests that both the formation of massive galaxies and their subsequent quenching happens too late in this model.

More recently, \cite{rong2017} used an earlier version of this SAM \citep{guo2011} to study analogs of ZF-COS-20115, the $z=3.7$ quiescent galaxy discussed in \citetalias{glazebrook2017}. While they were indeed able to find quiescent galaxies at these redshifts, their predicted space density was comparably low: $7.5\times10^{-8}\,\Mpc^{-3}$ for galaxies with $\ssfr < \ssfr_Q$ and $\mstar > 10^{11}\,\msun$, a factor $30$ lower than we observe for this same mass threshold (i.e., $(2.2\pm1.1)\times10^{-6}\,\Mpc^{-3}$).

In both these SAMs, quenching of massive galaxies is explicitly caused by AGNs in the so-called radio mode, that is, when the black hole slowly accretes hot gas from the host galaxy. Energy is artificially deposited into the galaxy's halo, supposedly from the action of the AGN radio jets, and inhibits cooling. The more intense quasar mode, which happens only during galaxy mergers, does not generate explicit feedback but is accompanied by stellar feedback from the resulting starburst, which \cite{guo2011} found is sufficient to remove all the gas from the galaxy. In the Henriques et al.~SAM, the parameters of the quenching model were tuned to match additional observables at $0<z<3$, chiefly the fraction of color-selected (\uvj) quiescent galaxies as function of mass and redshift. The fact that, despite a calibration up to $z=3$, this model still cannot reproduce the number densities of QGs at $3<z<4$ is surprising. We suspect this could be caused by two effects. First, an over-estimation of the stellar mass uncertainties; Henriques et al.~assumed $0.36\,\dex$ at $z\sim3.5$, which is a factor two larger than reported in observations \citep[e.g.,][]{ilbert2013}. Second, the mismatch of definition for ``quiescent'' galaxies between their model and observations. Indeed, the dividing line between quiescent and star-forming in their \uvj diagram is different than in observations, and is shifted toward dusty galaxies;  their predicted quiescent sample used in the model calibration phase will therefore include a number of red but nonetheless star-forming contaminants. In the model calibration, this will make it possible to match the observed number of ``red'' galaxies while under-predicting the actual number of galaxies with low $\ssfr$.

The \textsc{Meraxes} SAM \citep{qin2017-a,qin2017} was recently shown to predict more accurate number densities, both for the quiescent and the overall massive population. After correcting their masses from the \cite{salpeter1955} to the \cite{chabrier2003} IMF we used here, their model predicts a number densities of quiescent galaxies of $1.1\times10^{-5}\,\Mpc^{-3}$ at $z=3.5$ (Y.~Qin, private communication), which is only a factor two lower than our observed value. The prediction for the number density of massive galaxies, regardless of $\ssfr$, is within 10\% of the observed value ($8.5\times10^{-5}\,\Mpc^{-3}$). Therefore massive galaxies in this model are formed in the right numbers, but quench slightly too late (see below).

The ingredients for quenching in \textsc{Meraxes} are essentially the same as in the Henriques and Guo SAMs, but the calibration strategy and implementation differ. Perhaps the most immediate difference is that this model, introduced to study reionization, was calibrated to match the observed stellar mass functions at $0.6<z<7$ with an emphasis on $z>5$. Most other models, like the Henriques SAM, are instead calibrated with data at $0<z<3$, usually with a special emphasis on the $z=0$ mass function. This different focus may allow the \textsc{Meraxes} model to better reproduce the $3<z<4$ galaxy population, but possibly at the expanse of a poorer description of the $z\sim0$ Universe (their merger trees stop at $z=0.56$). \textsc{Meraxes} also includes a calibration of the black hole masses (anchored at $z\sim0$), and an indirect calibration on $\sfr$s through the matching of reionization. Although the latter should only constrain star formation in low-mass galaxies (which dominate reionization), this may impact the star-formation efficiency in the progenitors of massive $z\sim3.5$ quiescent galaxies as well. Finally, a last notable difference is that this SAM was implemented on a dark matter merger tree generated with a finer time step ($11\,\Myr$ at $z>5$, $30\,\Myr$ at $z\sim4$), which allows a better temporal sampling of star-formation and accretion histories.

The fact that this model manages to reproduce the number density of quiescent galaxies at $3<z<4$ is encouraging and suggests that, with adequate calibration, the AGN-based quenching model is indeed able to explain the existence of such galaxies. However, it remains to be shown that this agreement is not reached at the expense of other observable properties which were not the focus of the \textsc{Meraxes} model, in particular at lower redshifts.

\subsection{Number densities predicted by cosmological hydrodynamic simulations}

The strength of SAMs is their ability to be anchored to different observables while exploring a wide space of parameters and models. This freedom is also a weakness: since the calibration of these models plays such an important role, the predictive power of SAMs, albeit not null, is limited. Cosmological hydrodynamic simulations come closer to a ``first-principle'' approach, and although their limited spatial resolution still requires the use of empirical sub-grid recipes and regularization schemes, their predictive power is enhanced compared to SAMs. The downside is that the studied volume is usually smaller, but modern simulations are now executed on large enough volumes that a comparison to our results can be attempted.

In Illustris (\citealt{wellons2015}, Fig.~6), only five galaxies have $\mstar > 3\times10^{10}\,\msun$ and $\ssfr<\ssfr_Q$ at $z=3$, and none exist at $z=4$ (as first noticed in \citetalias{glazebrook2017}). Given the box size of Illustris, this translates into an average number density at $3<z<4$ of $2.1\times10^{-6}\,\Mpc^{-3}$, an order of magnitude too low. The \textsc{mufasa} simulation \citep{dave2016} has a three times smaller volume, $(50/h)^3\,\Mpc^3$, and will therefore offer more limited statistics, but a comparison is still possible. In this simulation, only two galaxies match the required properties at $z=3$, and none at $z=4$ (R.~Dav\'e, private communication). The average number density at $3<z<4$ is therefore similar to that of Illustris, $2.5\times10^{-6}\,\Mpc^{-3}$.

While Illustris implemented AGN feedback in a fashion similar to that discussed above for SAMs (i.e., including both quasar- and radio-mode feedback), \textsc{mufasa} adopted a simpler phenomenological model. There, AGN feedback was not invoked explicitly; instead, a ``halo quenching'' mass was introduced above which all the gas which was not self-shielded was artificially heated, hence prevented from forming stars. The source of this heating is unspecified in the model, but AGNs would be obvious candidates. Despite its simplicity, this model was still able to offer a good match to the $z<4$ stellar mass functions, but with a lack of massive galaxies at $z>4$ \citep{dave2016}.

These two simulations therefore fall short of the observed number density by about an order of magnitude despite implementing different quenching mechanisms. Given that the \textsc{Meraxes} SAM was able to match our observations, a side-to-side comparison of galaxy properties with these different models could reveal what feature might be currently missing in the hydrodynamic simulations, or if the feedback model implemented in \textsc{Meraxes} can be implemented at all in a more physically-motivated context.

\subsection{Star formation histories in \textsc{Meraxes} \label{SEC:model_sfh}}

Since, among all the models we explored, \textsc{Meraxes} was the only one that came close to reproducing the observed number density of quiescent galaxies, we proceeded to investigate whether the star formation histories of the quiescent galaxies in this model match the ones we inferred in \rsec{SEC:sfh}. For this analysis we used the 70 quiescent galaxies in the $z=3.5$ \textsc{Meraxes} snapshot kindly provided by Y.~Qin (private communication). We applied the same procedure as for the observed galaxies to compute $t_{\rm quench}$ and $t_{\rm form}$, using the tabulated SFHs produced by \textsc{Meraxes}. The results are illustrated on \rfig{FIG:tprob}.

We find that, on average, quiescent galaxies in \textsc{Meraxes} have been quiescent for $150\,\Myr$, a duration about twice shorter than that measured for the observed galaxies, $330\pm30\,\Myr$. The 16th and 84th percentiles of $t_{\rm quench}$ are $0$ and $210\,\Myr$ in \textsc{Meraxes}, versus $210$ and $510\,\Myr$ in observations. In this simulation, the quenching therefore happened at later times, which is consistent with the slight under-prediction of the number density. However we find that this time shift also affects the formation epoch: in the simulation, the quiescent galaxies had assembled half of their observed mass on average $550\,\Myr$ prior to observation (percentiles: $370$ and $700\,\Myr$), compared to $780^{+30}_{-70}\,\Myr$ for the observed galaxies (percentiles: $360$, $1070\,\Myr$). While these numbers are comparable at first order, the simulated SFHs are systematically shifted to later times by about $200\,\Myr$ ($10\%$ of the Hubble time at $z=3.5$). This would be consistent with the predicted space density of $z\sim3.5$ quiescent galaxies being lower than the observed value.

In addition, we find that the simulated SFHs are more extended in time than the observed ones. The average $\sfr$ of these galaxies during their formation phase was $160\,\msun/\yr$ in the simulation, while we inferred $340^{+150}_{-30}\,\msun/\yr$ for the observed galaxies. Consequently, the duration of the star-forming phase was longer in the simulation: $480\,\Myr$ versus $280^{+40}_{-134}\,\Myr$ for our observed sample.

Our observations therefore suggest that the formation of $3<z<4$ quiescent galaxies happened earlier, in shorter, more intense bursts than what this model predicts. This inevitably requires a higher star formation efficiency, as already argued in \citetalias{glazebrook2017}, and a corresponding increase in quenching efficiency to avoid over-producing star-forming galaxies. The possibility of achieving this with an AGN-driven quenching model goes beyond the scope of this paper, but should definitely be explored.

We recall, however, that the galaxies studied in \rsec{SEC:sfh} are only those for which we obtained a MOSFIRE spectrum, and these are mostly \uvj-quiescent. Young-quiescent galaxies are under-represented in this sample, and including them would tend to shift the observed population toward overall younger ages (\rsec{SEC:qq_sfh}). While this would improve the agreement with \textsc{Meraxes}, we do not expect the discrepancy to disappear entirely. Indeed, we quantified the fraction of young-quiescent galaxies in \textsc{Meraxes} by producing synthetic \uvj colors with the \cite{bruzual2003} model and assuming $A_V = 0.3\,{\rm mag}$ of attenuation (the median of our sample, see \rtab{TAB:final_props}). In this model, young-quiescent galaxies make up $73\%$ of the quiescent population, while they remain a minority in ZFOURGE ($33\%$). It is therefore clear that quiescent galaxies in \textsc{Meraxes} are overall younger.

\section{Conclusions \label{SEC:conclusion}}

We have obtained NIR spectra for a sample of $3<z<4$ massive, \uvj-selected quiescent galaxies with MOSFIRE. These spectra allow us to measure a spectroscopic redshift for $40\%$ of our targets, revealing that their photometric redshifts have an excellent accuracy of $1.2\%$, with a catastrophic failure rate of $10\%$ coming from dusty galaxies at lower redshifts. An additional $10\%$ are found at the correct redshift, but with high-equivalent-width $\oiii$ emission contributing significantly to the $K$-band flux, hence for which the steepness of the Balmer break was overestimated. The rest of the sample shows no strong emission lines. Combined, this demonstrates that the \uvj selection of quiescent galaxies suffers from a contamination rate of only $20\%$ at $3<z<4$.

Balmer absorption lines are detected in four galaxies (among the brightest of the sample), a clear signpost of a recent quenching, and in agreement with expectations from their broadband photometry. The star-formation histories inferred for the entire sample show that all but one of the galaxies have quenched with certainty, on average $330\,\Myr$ before being observed, sometime between $z=3.5$ and $z=5$. Half of their stars were formed by $z=4.4$ to $z=7.1$ in brief ($280\,\Myr$) star-formation episodes with $\sfr$s of $80$ to $850\,\msun/\yr$.

These results therefore confirm that the \uvj selection is efficient in selecting genuinely quiescent galaxies. Building on this result and our estimated contamination rate, we came back to the parent catalogs and updated the number density of \uvj-quiescent galaxies at $3<z<4$. At $\mstar > 3\times10^{10}\,\msun$, we find $(1.4\pm0.3)\times10^{-5}\,\Mpc^{-3}$. This number is slightly lower than previous estimates, but the overall picture remains unchanged: a substantial population of quiescent galaxies did exist at these early epochs.

To compare this number to models, we then investigated the completeness of the \uvj selection in terms of $\ssfr$. We find a sizable population of galaxies with $\ssfr < 0.15\,\Gyr^{-1}$ (a factor of ten below the main sequence) and no dust-obscured star-formation, some of which were missing from our initial \uvj-selected sample. We dubbed these ``young-quiescent'' galaxies. Their modeling suggest they quenched later than the other quiescent galaxies, such that their colors are not yet red enough to enter the \uvj selection. We find they have a number density comparable to that of the \uvj-quiescent galaxies. The combined number density of galaxies with low $\ssfr$ is therefore larger than that of \uvj-quiescent galaxies alone, namely, $(2.0\pm0.3)\times10^{-5}\,\Mpc^{-3}$.

We finally compared our results to recent models of galaxy formation. All the models we explored predict number densities from one to two orders of magnitude too low. The only exception is the \textsc{Meraxes} semi-analytic model, which was tuned with a particular emphasis on high redshift observations, and which falls short by only a factor two. This shows that the AGN-based quenching model, adopted in \textsc{Meraxes}, can produce quiescent galaxies in almost the right numbers, albeit only with some specific tuning to high-redshift galaxies. Yet the predicted star formation histories in \textsc{Meraxes} do not match our observations. The formation and quenching of massive $z\sim3.5$ galaxies in this model happens $200\,\Myr$ too late, and their star-forming phase is twice longer than observed, with past $\sfr$s too low by a similar factor. This suggests that quenching of high-redshift galaxies is not yet a fully understood process.

From here, a number of follow-up studies can be undertaken. In a future work, we will attempt to predict the properties of the star-forming progenitors of our quiescent galaxies, and compare their inferred $\sfr$s and space densities to known SFGs at high redshifts. A more detailed analysis of the ALMA-detected quiescent candidates is also in order, to investigate whether the sub-mm emission is always powered by obscured star-formation, or if it may originate from heating by older stars \citep[e.g.,][]{schreiber2018}. In addition, for those which are truly star-forming, it would be worthwhile to understand why the color classification failed in the first place.

Follow-up observations would also be beneficial to better understand this population. Deeper spectra would allow us to place tighter constraints on the SFHs, pin-pointing with greater accuracy the time of quenching and the duration of the star-forming episode. Perhaps more importantly, this would also allows us to start constraining the chemical enrichment histories, which can further constrain the SFH and the physics of stellar evolution in the early, metal-poor Universe. Velocity dispersion measurements would constrain dynamical masses and the stellar initial mass function, which may or may not be as universal as thought. There is only so much one can do from ground-based $8\,{\rm m}$ class telescopes however, and these are questions that will undoubtedly require the help of the next generation of instruments, chiefly the \jwst and its on-board spectrograph, NIRSpec.

The origin of the low level residual star-formation in these otherwise ``dead'' galaxies is also intriguing, and could be linked to different processes. New gas can be brought in from the generous infall expected at these high redshifts, or from gas recycling. Alternatively, this residual activity may just be the declining ``tail'' of a past starburst, if quenching is not an instantaneous event. These different scenarios could be investigated with higher resolution NIR imaging to locate these star-forming regions, and with deep spectroscopy to reveal the physical condition in these small reservoirs of star-forming gas (i.e., metallicity). \jwst will again be a key instrument to answer these questions.

Finally, while it was thought for a long time that galaxies must fully deplete (or expel) their gas reservoirs before quenching, evidence is building that this is not always true and that sizable reservoirs exist in quenched or post-starburst galaxies. This question could be addressed by looking at those galaxies in our sample which have spent the smallest amount of time since quenching, and study their gas or dust content, for example through systematic and deep follow-ups with ALMA. At these high redshifts, ALMA can cover a number of interesting emission lines, chiefly $[\ion{C}{ii}]$, $[\ion{N}{ii}]$, and the CO ladder, while providing a reasonable sampling of the dust SED to derive infrared luminosities and dust (or gas) masses.

\begin{acknowledgements}
The authors want to thank the anonymous referee for his/her comments and suggestions, which improved the consistency and quality of this paper.

Most of the analysis for this paper was done using {\tt phy++}, a free and open source C++ library for fast and robust numerical astrophysics (\hlink{cschreib.github.io/phypp/}).

K.G. acknowledges support from Australian Research Council (ARC) Discovery Program grants DP130101460 and DP160102235.

This work is based on observations taken by the CANDELS Multi-Cycle Treasury Program with the NASA/ESA \hst, which is operated by the Association of Universities for Research in Astronomy, Inc., under NASA contract NAS5-26555.

This paper makes use of the following ALMA data: ADS/JAO.ALMA\#2012.1.00307.S, 2012.1.00869.S, 2013.1.00118.S, 2013.1.00151.S, 2013.1.01292.S, 2015.A.00026.S, 2015.1.00379.S, 2015.1.01074.S, 2015.1.01495.S, 2015.1.01528.S. ALMA is a partnership of ESO (representing its member states), NSF (USA) and NINS (Japan), together with NRC (Canada) and NSC and ASIAA (Taiwan) and KASI (Republic of Korea), in cooperation with the Republic of Chile. The Joint ALMA Observatory is operated by ESO, AUI/NRAO and NAOJ.

Data in this paper were obtained at the W.M.~Keck Observatory, made possible by the generous financial support of the W.M.~Keck Foundation, and operated as a scientific partnership among Caltech, the University of California and NASA. We recognize and acknowledge the very significant cultural role and reverence that the summit of Mauna Kea has always had within the indigenous Hawaiian community.
\end{acknowledgements}

\bibliographystyle{aa}
\bibliography{../bbib/full}

\begin{thebibliography}{149}
\expandafter\ifx\csname natexlab\endcsname\relax\def\natexlab#1{#1}\fi

\bibitem[{Akiyama {et~al.}(2015)Akiyama, Ueda, Watson, Furusawa, Takata,
  Simpson, Morokuma, Yamada, Ohta, Iwamuro, Yabe, Tamura, Moritani, Takato,
  Kimura, Maihara, Dalton, Lewis, Lee, {Curtis-Lake}, Macaulay, Clarke,
  Silverman, Croom, Ouchi, Hanami, Díaz~Tello, Yoshikawa, Fujishiro, \&
  Sekiguchi}]{akiyama2015}
Akiyama, M., Ueda, Y., Watson, M.~G., {et~al.} 2015, \pasj, 67, 82

\bibitem[{Alatalo {et~al.}(2014)Alatalo, Appleton, Lisenfeld, Bitsakis,
  Guillard, Charmandaris, Cluver, Dopita, Freeland, Jarrett, Kewley, Ogle,
  Rasmussen, Rich, {Verdes-Montenegro}, Xu, \& Yun}]{alatalo2014}
Alatalo, K., Appleton, P.~N., Lisenfeld, U., {et~al.} 2014, \apj, 795, 159

\bibitem[{Alatalo {et~al.}(2015)Alatalo, Lacy, Lanz, Bitsakis, Appleton,
  Nyland, Cales, Chang, Davis, de~Zeeuw, Lonsdale, Martín, Meier, \&
  Ogle}]{alatalo2015}
Alatalo, K., Lacy, M., Lanz, L., {et~al.} 2015, \apj, 798, 31

\bibitem[{Audouze \& Tinsley(1976)}]{audouze1976}
Audouze, J. \& Tinsley, B.~M. 1976, \araa, 14, 43

\bibitem[{Avni(1976)}]{avni1976}
Avni, Y. 1976, \apj, 210, 642

\bibitem[{Baldry {et~al.}(2004)Baldry, Glazebrook, Brinkmann, Ivezić, Lupton,
  Nichol, \& Szalay}]{baldry2004}
Baldry, I.~K., Glazebrook, K., Brinkmann, J., {et~al.} 2004, \apj, 600, 681

\bibitem[{Baldwin(1975)}]{baldwin1975}
Baldwin, J.~A. 1975, \apj, 201, 26

\bibitem[{Baldwin {et~al.}(1981)Baldwin, Phillips, \& Terlevich}]{baldwin1981}
Baldwin, J.~A., Phillips, M.~M., \& Terlevich, R. 1981, \pasp, 93, 5

\bibitem[{Barro {et~al.}(2013)Barro, Faber, {P\'erez-González}, Koo, Williams,
  Kocevski, Trump, Mozena, {McGrath}, van~der Wel, Wuyts, Bell, Croton,
  Ceverino, Dekel, Ashby, Cheung, Ferguson, Fontana, Fang, Giavalisco, Grogin,
  Guo, Hathi, Hopkins, Huang, Koekemoer, Kartaltepe, Lee, Newman, Porter,
  Primack, Ryan, Rosario, Somerville, Salvato, \& Hsu}]{barro2013}
Barro, G., Faber, S.~M., {P\'erez-González}, P.~G., {et~al.} 2013, \apj, 765,
  104

\bibitem[{Barro {et~al.}(2016)Barro, Kriek, {P\'erez-González}, Trump, Koo,
  Faber, Dekel, Primack, Guo, Kocevski, {Muñoz-Mateos}, Rujoparkarn, \&
  Seth}]{barro2016-a}
Barro, G., Kriek, M., {P\'erez-González}, P.~G., {et~al.} 2016, \apjl, 827,
  L32

\bibitem[{Behroozi \& Silk(2018)}]{behroozi2018}
Behroozi, P. \& Silk, J. 2018, \mnras, 477, 5382

\bibitem[{Belli {et~al.}(2017{\natexlab{a}})Belli, Genzel, F\"orster~Schreiber,
  Wisnioski, Wilman, Wuyts, Mendel, Beifiori, Bender, Brammer, Burkert, Chan,
  Davies, Davies, Fabricius, Fossati, Galametz, Lang, Lutz, Momcheva, Nelson,
  Saglia, Tacconi, Tadaki, Übler, \& van Dokkum}]{belli2017}
Belli, S., Genzel, R., F\"orster~Schreiber, N.~M., {et~al.} 2017{\natexlab{a}},
  \apjl, 841, L6

\bibitem[{Belli {et~al.}(2017{\natexlab{b}})Belli, Newman, \&
  Ellis}]{belli2017-a}
Belli, S., Newman, A.~B., \& Ellis, R.~S. 2017{\natexlab{b}}, \apj, 834, 18

\bibitem[{Belli {et~al.}(2014)Belli, Newman, Ellis, \& Konidaris}]{belli2014}
Belli, S., Newman, A.~B., Ellis, R.~S., \& Konidaris, N.~P. 2014, \apjl, 788,
  L29

\bibitem[{Benson {et~al.}(2003)Benson, Bower, Frenk, Lacey, Baugh, \&
  Cole}]{benson2003}
Benson, A.~J., Bower, R.~G., Frenk, C.~S., {et~al.} 2003, \apj, 599, 38

\bibitem[{Benítez(2000)}]{benitez2000}
Benítez, N. 2000, \apj, 536, 571

\bibitem[{Birnboim \& Dekel(2003)}]{birnboim2003}
Birnboim, Y. \& Dekel, A. 2003, \mnras, 345, 349

\bibitem[{Blanton {et~al.}(2001)Blanton, Sarazin, {McNamara}, \&
  Wise}]{blanton2001}
Blanton, E.~L., Sarazin, C.~L., {McNamara}, B.~R., \& Wise, M.~W. 2001, \apjl,
  558, L15

\bibitem[{Bower {et~al.}(2006)Bower, Benson, Malbon, Helly, Frenk, Baugh, Cole,
  \& Lacey}]{bower2006}
Bower, R.~G., Benson, A.~J., Malbon, R., {et~al.} 2006, \mnras, 370, 645

\bibitem[{Brammer {et~al.}(2008)Brammer, van Dokkum, \& Coppi}]{brammer2008}
Brammer, G.~B., van Dokkum, P.~G., \& Coppi, P. 2008, \apj, 686, 1503

\bibitem[{Brammer {et~al.}(2011)Brammer, Whitaker, van Dokkum, Marchesini,
  Franx, Kriek, Labb\'e, Lee, Muzzin, Quadri, Rudnick, \&
  Williams}]{brammer2011}
Brammer, G.~B., Whitaker, K.~E., van Dokkum, P.~G., {et~al.} 2011, \apj, 739,
  24

\bibitem[{Bruzual \& Charlot(2003)}]{bruzual2003}
Bruzual, G. \& Charlot, S. 2003, \mnras, 344, 1000

\bibitem[{Butcher \& Oemler(1978)}]{butcher1978}
Butcher, H. \& Oemler, A., J. 1978, \apj, 219, 18

\bibitem[{Caccianiga \& Severgnini(2011)}]{caccianiga2011}
Caccianiga, A. \& Severgnini, P. 2011, \mnras, 415, 1928

\bibitem[{Calzetti {et~al.}(2000)Calzetti, Armus, Bohlin, Kinney, Koornneef, \&
  {Storchi-Bergmann}}]{calzetti2000}
Calzetti, D., Armus, L., Bohlin, R.~C., {et~al.} 2000, \apj, 533, 682

\bibitem[{Cappellari \& Emsellem(2004)}]{cappellari2004}
Cappellari, M. \& Emsellem, E. 2004, \pasp, 116, 138

\bibitem[{Cattaneo {et~al.}(2009)Cattaneo, Faber, Binney, Dekel, Kormendy,
  Mushotzky, Babul, Best, Br\"uggen, Fabian, Frenk, Khalatyan, Netzer, Mahdavi,
  Silk, Steinmetz, \& Wisotzki}]{cattaneo2009}
Cattaneo, A., Faber, S.~M., Binney, J., {et~al.} 2009, \nat, 460, 213

\bibitem[{Chabrier(2003)}]{chabrier2003}
Chabrier, G. 2003, \pasp, 115, 763

\bibitem[{Cid~Fernandes {et~al.}(2011)Cid~Fernandes, Stasińska, Mateus, \&
  Vale~Asari}]{cidfernandes2011}
Cid~Fernandes, R., Stasińska, G., Mateus, A., \& Vale~Asari, N. 2011, \mnras,
  413, 1687

\bibitem[{Cid~Fernandes {et~al.}(2010)Cid~Fernandes, Stasińska, Schlickmann,
  Mateus, Vale~Asari, Schoenell, \& Sodr\'e}]{cidfernandes2010}
Cid~Fernandes, R., Stasińska, G., Schlickmann, M.~S., {et~al.} 2010, \mnras,
  403, 1036

\bibitem[{Ciesla {et~al.}(2016)Ciesla, Boselli, Elbaz, Boissier, Buat,
  Charmandaris, Schreiber, B\'ethermin, Baes, Boquien, De~Looze,
  {Fernández-Ontiveros}, Pappalardo, Spinoglio, \& Viaene}]{ciesla2016}
Ciesla, L., Boselli, A., Elbaz, D., {et~al.} 2016, \aap, 585, A43

\bibitem[{Ciesla {et~al.}(2017)Ciesla, Elbaz, \& Fensch}]{ciesla2017}
Ciesla, L., Elbaz, D., \& Fensch, J. 2017, \aap, 608, A41

\bibitem[{Cimatti {et~al.}(2004)Cimatti, Daddi, Renzini, Cassata, Vanzella,
  Pozzetti, Cristiani, Fontana, Rodighiero, Mignoli, \& Zamorani}]{cimatti2004}
Cimatti, A., Daddi, E., Renzini, A., {et~al.} 2004, \nat, 430, 184

\bibitem[{Cimatti {et~al.}(2002)Cimatti, Pozzetti, Mignoli, Daddi, Menci, Poli,
  Fontana, Renzini, Zamorani, Broadhurst, Cristiani, {D'Odorico}, Giallongo, \&
  Gilmozzi}]{cimatti2002}
Cimatti, A., Pozzetti, L., Mignoli, M., {et~al.} 2002, \aap, 391, L1

\bibitem[{Coil {et~al.}(2015)Coil, Aird, Reddy, Shapley, Kriek, Siana,
  Mobasher, Freeman, Price, \& Shivaei}]{coil2015}
Coil, A.~L., Aird, J., Reddy, N., {et~al.} 2015, \apj, 801, 35

\bibitem[{Croton {et~al.}(2006)Croton, Springel, White, De~Lucia, Frenk, Gao,
  Jenkins, Kauffmann, Navarro, \& Yoshida}]{croton2006}
Croton, D.~J., Springel, V., White, S. D.~M., {et~al.} 2006, \mnras, 365, 11

\bibitem[{Daddi {et~al.}(2000)Daddi, Cimatti, \& Renzini}]{daddi2000}
Daddi, E., Cimatti, A., \& Renzini, A. 2000, \aap, 362, L45

\bibitem[{Daddi {et~al.}(2004)Daddi, Cimatti, Renzini, Fontana, Mignoli,
  Pozzetti, Tozzi, \& Zamorani}]{daddi2004-a}
Daddi, E., Cimatti, A., Renzini, A., {et~al.} 2004, \apj, 617, 746

\bibitem[{Dav\'e {et~al.}(2017)Dav\'e, Rafieferantsoa, \& Thompson}]{dave2017}
Dav\'e, R., Rafieferantsoa, M.~H., \& Thompson, R.~J. 2017, \mnras, 471, 1671

\bibitem[{Dav\'e {et~al.}(2016)Dav\'e, Thompson, \& Hopkins}]{dave2016}
Dav\'e, R., Thompson, R., \& Hopkins, P.~F. 2016, \mnras, 462, 3265

\bibitem[{Davis {et~al.}(2014)Davis, Young, Crocker, Bureau, Blitz, Alatalo,
  Emsellem, Naab, Bayet, Bois, Bournaud, Cappellari, Davies, de~Zeeuw, Duc,
  Khochfar, Krajnović, Kuntschner, {McDermid}, Morganti, Oosterloo, Sarzi,
  Scott, Serra, \& Weijmans}]{davis2014}
Davis, T.~A., Young, L.~M., Crocker, A.~F., {et~al.} 2014, \mnras, 444, 3427

\bibitem[{de~Vaucouleurs(1948)}]{devaucouleurs1948}
de~Vaucouleurs, G. 1948, Annales {d'Astrophysique}, 11, 247

\bibitem[{Dekel \& Birnboim(2008)}]{dekel2008}
Dekel, A. \& Birnboim, Y. 2008, \mnras, 383, 119

\bibitem[{Dunlop {et~al.}(2007)Dunlop, Cirasuolo, \& {McLure}}]{dunlop2007}
Dunlop, J.~S., Cirasuolo, M., \& {McLure}, R.~J. 2007, \mnras, 376, 1054

\bibitem[{Elbaz {et~al.}(2007)Elbaz, Daddi, Le~Borgne, Dickinson, Alexander,
  Chary, Starck, Brandt, Kitzbichler, {MacDonald}, Nonino, Popesso, Stern, \&
  Vanzella}]{elbaz2007}
Elbaz, D., Daddi, E., Le~Borgne, D., {et~al.} 2007, \aap, 468, 33

\bibitem[{Elbaz {et~al.}(2017)Elbaz, Leiton, Nagar, Okumura, Franco, Schreiber,
  Pannella, Wang, Dickinson, {Diaz-Santos}, Ciesla, Daddi, Bournaud, Magdis,
  Zhou, \& Rujopakarn}]{elbaz2017}
Elbaz, D., Leiton, R., Nagar, N., {et~al.} 2017, {ArXiv} e-prints,
  1711.arXiv:1711.10047

\bibitem[{Erb {et~al.}(2010)Erb, Pettini, Shapley, Steidel, Law, \&
  Reddy}]{erb2010}
Erb, D.~K., Pettini, M., Shapley, A.~E., {et~al.} 2010, \apj, 719, 1168

\bibitem[{Faber {et~al.}(2007)Faber, Willmer, Wolf, Koo, Weiner, Newman, Im,
  Coil, Conroy, Cooper, Davis, Finkbeiner, Gerke, Gebhardt, Groth,
  Guhathakurta, Harker, Kaiser, Kassin, Kleinheinrich, Konidaris, Kron, Lin,
  Luppino, Madgwick, Meisenheimer, Noeske, Phillips, Sarajedini, Schiavon,
  Simard, Szalay, Vogt, \& Yan}]{faber2007}
Faber, S.~M., Willmer, C. N.~A., Wolf, C., {et~al.} 2007, \apj, 665, 265

\bibitem[{Fabian(1994)}]{fabian1994}
Fabian, A.~C. 1994, \araa, 32, 277

\bibitem[{Fontana {et~al.}(2009)Fontana, Santini, Grazian, Pentericci, Fiore,
  Castellano, Giallongo, Menci, Salimbeni, Cristiani, Nonino, \&
  Vanzella}]{fontana2009}
Fontana, A., Santini, P., Grazian, A., {et~al.} 2009, \aap, 501, 15

\bibitem[{Franx {et~al.}(2003)Franx, Labb\'e, Rudnick, van Dokkum, Daddi,
  F\"orster~Schreiber, Moorwood, Rix, R\"ottgering, van~der Wel, van~der Werf,
  \& van Starkenburg}]{franx2003}
Franx, M., Labb\'e, I., Rudnick, G., {et~al.} 2003, \apjl, 587, L79

\bibitem[{French {et~al.}(2015)French, Yang, Zabludoff, Narayanan, Shirley,
  Walter, Smith, \& Tremonti}]{french2015}
French, K.~D., Yang, Y., Zabludoff, A., {et~al.} 2015, \apj, 801, 1

\bibitem[{Fumagalli {et~al.}(2014)Fumagalli, Labb\'e, Patel, Franx, van Dokkum,
  Brammer, da~Cunha, F\"orster~Schreiber, Kriek, Quadri, Rix, Wake, Whitaker,
  Lundgren, Marchesini, Maseda, Momcheva, Nelson, Pacifici, \&
  Skelton}]{fumagalli2014}
Fumagalli, M., Labb\'e, I., Patel, S.~G., {et~al.} 2014, \apj, 796, 35

\bibitem[{Gabor \& Dav\'e(2012)}]{gabor2012}
Gabor, J.~M. \& Dav\'e, R. 2012, \mnras, 427, 1816

\bibitem[{Gatz \& Smith(1995)}]{gatz1995}
Gatz, D.~F. \& Smith, L. 1995, Atmospheric Environment, 29, 1185

\bibitem[{Glazebrook {et~al.}(2004)Glazebrook, Abraham, {McCarthy}, Savaglio,
  Chen, Crampton, Murowinski, Jørgensen, Roth, Hook, Marzke, \&
  Carlberg}]{glazebrook2004}
Glazebrook, K., Abraham, R.~G., {McCarthy}, P.~J., {et~al.} 2004, \nat, 430,
  181

\bibitem[{Glazebrook {et~al.}(2017)Glazebrook, Schreiber, Labb\'e, Nanayakkara,
  Kacprzak, Oesch, Papovich, Spitler, Straatman, Tran, \&
  Yuan}]{glazebrook2017}
Glazebrook, K., Schreiber, C., Labb\'e, I., {et~al.} 2017, \nat, 544, 71

\bibitem[{Gobat {et~al.}(2018)Gobat, Daddi, Magdis, Bournaud, Sargent, Martig,
  Jin, Finoguenov, B\'ethermin, Hwang, Renzini, Wilson, Aretxaga, Yun,
  Strazzullo, \& Valentino}]{gobat2018}
Gobat, R., Daddi, E., Magdis, G., {et~al.} 2018, Nature Astronomy, 2, 239

\bibitem[{Gobat {et~al.}(2012)Gobat, Strazzullo, Daddi, Onodera, Renzini,
  B\'ethermin, Dickinson, Carollo, \& Cimatti}]{gobat2012}
Gobat, R., Strazzullo, V., Daddi, E., {et~al.} 2012, \apjl, 759, L44

\bibitem[{Grogin {et~al.}(2011)Grogin, Kocevski, Faber, Ferguson, Koekemoer,
  Riess, Acquaviva, Alexander, Almaini, Ashby, Barden, Bell, Bournaud, Brown,
  Caputi, Casertano, Cassata, Castellano, Challis, Chary, Cheung, Cirasuolo,
  Conselice, Roshan~Cooray, Croton, Daddi, Dahlen, Dav\'e, de~Mello, Dekel,
  Dickinson, Dolch, Donley, Dunlop, Dutton, Elbaz, Fazio, Filippenko,
  Finkelstein, Fontana, Gardner, Garnavich, Gawiser, Giavalisco, Grazian, Guo,
  Hathi, H\"aussler, Hopkins, Huang, Huang, Jha, Kartaltepe, Kirshner, Koo,
  Lai, Lee, Li, Lotz, Lucas, Madau, {McCarthy}, {McGrath}, {McIntosh},
  {McLure}, Mobasher, Moustakas, Mozena, Nandra, Newman, Niemi, Noeske,
  Papovich, Pentericci, Pope, Primack, Rajan, Ravindranath, Reddy, Renzini,
  Rix, Robaina, Rodney, Rosario, Rosati, Salimbeni, Scarlata, Siana, Simard,
  Smidt, Somerville, Spinrad, Straughn, Strolger, Telford, Teplitz, Trump,
  van~der Wel, Villforth, Wechsler, Weiner, Wiklind, Wild, Wilson, Wuyts, Yan,
  \& Yun}]{grogin2011}
Grogin, N.~A., Kocevski, D.~D., Faber, S.~M., {et~al.} 2011, \apjs, 197, 35

\bibitem[{Guo {et~al.}(2011)Guo, White, {Boylan-Kolchin}, De~Lucia, Kauffmann,
  Lemson, Li, Springel, \& Weinmann}]{guo2011}
Guo, Q., White, S., {Boylan-Kolchin}, M., {et~al.} 2011, \mnras, 413, 101

\bibitem[{Henriques {et~al.}(2015)Henriques, White, Thomas, Angulo, Guo,
  Lemson, Springel, \& Overzier}]{henriques2015}
Henriques, B. M.~B., White, S. D.~M., Thomas, P.~A., {et~al.} 2015, \mnras,
  451, 2663

\bibitem[{{Hernán-Caballero} \& Hatziminaoglou(2011)}]{hernan-caballero2011}
{Hernán-Caballero}, A. \& Hatziminaoglou, E. 2011, \mnras, 414, 500

\bibitem[{Hill {et~al.}(2016)Hill, Muzzin, Franx, \& van~de Sande}]{hill2016}
Hill, A.~R., Muzzin, A., Franx, M., \& van~de Sande, J. 2016, \apj, 819, 74

\bibitem[{Hopkins {et~al.}(2008)Hopkins, Hernquist, Cox, \&
  Kereš}]{hopkins2008}
Hopkins, P.~F., Hernquist, L., Cox, T.~J., \& Kereš, D. 2008, \apjs, 175, 356

\bibitem[{Husser {et~al.}(2013)Husser, Wende-von Berg, Dreizler, Homeier,
  Reiners, Barman, \& Hauschildt}]{husser2013}
Husser, T., Wende-von Berg, S., Dreizler, S., {et~al.} 2013, \aap, 553, A6

\bibitem[{Ilbert {et~al.}(2013)Ilbert, {McCracken}, Le~F\`evre, Capak, Dunlop,
  Karim, Renzini, Caputi, Boissier, Arnouts, Aussel, Comparat, Guo, Hudelot,
  Kartaltepe, Kneib, Krogager, Le~Floc'h, Lilly, Mellier, {Milvang-Jensen},
  Moutard, Onodera, Richard, Salvato, Sanders, Scoville, Silverman, Taniguchi,
  Tasca, Thomas, Toft, Tresse, Vergani, Wolk, \& Zirm}]{ilbert2013}
Ilbert, O., {McCracken}, H.~J., Le~F\`evre, O., {et~al.} 2013, \aap, 556, 55

\bibitem[{Ilbert {et~al.}(2010)Ilbert, Salvato, Le~Floc'h, Aussel, Capak,
  {McCracken}, Mobasher, Kartaltepe, Scoville, Sanders, Arnouts, Bundy,
  Cassata, Kneib, Koekemoer, Le~F\`evre, Lilly, Surace, Taniguchi, Tasca,
  Thompson, Tresse, Zamojski, Zamorani, \& Zucca}]{ilbert2010}
Ilbert, O., Salvato, M., Le~Floc'h, E., {et~al.} 2010, \apj, 709, 644

\bibitem[{Inami {et~al.}(2017)Inami, Bacon, Brinchmann, Richard, Contini,
  Conseil, Hamer, Akhlaghi, Bouch\'e, Cl\'ement, Desprez, Drake, Hashimoto,
  Leclercq, Maseda, {Michel-Dansac}, Paalvast, Tresse, Ventou, Kollatschny,
  Boogaard, Finley, Marino, Schaye, \& Wisotzki}]{inami2017}
Inami, H., Bacon, R., Brinchmann, J., {et~al.} 2017, \aap, 608, A2

\bibitem[{Juneau {et~al.}(2011)Juneau, Dickinson, Alexander, \&
  Salim}]{juneau2011}
Juneau, S., Dickinson, M., Alexander, D.~M., \& Salim, S. 2011, \apj, 736, 104

\bibitem[{{Kado-Fong} {et~al.}(2017){Kado-Fong}, Marchesini, Marsan, Muzzin,
  Quadri, Brammer, Bezanson, Labb\'e, Lundgren, Rudnick, Stefanon, Tal, Wake,
  Williams, Whitaker, \& van Dokkum}]{kado-fong2017}
{Kado-Fong}, E., Marchesini, D., Marsan, Z.~C., {et~al.} 2017, \apj, 838, 57

\bibitem[{Kauffmann {et~al.}(2003{\natexlab{a}})Kauffmann, Heckman, Tremonti,
  Brinchmann, Charlot, White, Ridgway, Brinkmann, Fukugita, Hall, Ivezić,
  Richards, \& Schneider}]{kauffmann2003-a}
Kauffmann, G., Heckman, T.~M., Tremonti, C., {et~al.} 2003{\natexlab{a}},
  \mnras, 346, 1055

\bibitem[{Kauffmann {et~al.}(2003{\natexlab{b}})Kauffmann, Heckman, White,
  Charlot, Tremonti, Brinchmann, Bruzual, Peng, Seibert, Bernardi, Blanton,
  Brinkmann, Castander, Csábai, Fukugita, Ivezic, Munn, Nichol, Padmanabhan,
  Thakar, Weinberg, \& York}]{kauffmann2003}
Kauffmann, G., Heckman, T.~M., White, S. D.~M., {et~al.} 2003{\natexlab{b}},
  \mnras, 341, 33

\bibitem[{Kennicutt(1998)}]{kennicutt1998-a}
Kennicutt, Robert~C., J. 1998, \araa, 36, 189

\bibitem[{Kewley {et~al.}(2004)Kewley, Geller, \& Jansen}]{kewley2004}
Kewley, L.~J., Geller, M.~J., \& Jansen, R.~A. 2004, \aj, 127, 2002

\bibitem[{Koekemoer {et~al.}(2011)Koekemoer, Faber, Ferguson, Grogin, Kocevski,
  Koo, Lai, Lotz, Lucas, {McGrath}, Ogaz, Rajan, Riess, Rodney, Strolger,
  Casertano, Castellano, Dahlen, Dickinson, Dolch, Fontana, Giavalisco,
  Grazian, Guo, Hathi, Huang, van~der Wel, Yan, Acquaviva, Alexander, Almaini,
  Ashby, Barden, Bell, Bournaud, Brown, Caputi, Cassata, Challis, Chary,
  Cheung, Cirasuolo, Conselice, Roshan~Cooray, Croton, Daddi, Dav\'e, de~Mello,
  de~Ravel, Dekel, Donley, Dunlop, Dutton, Elbaz, Fazio, Filippenko,
  Finkelstein, Frazer, Gardner, Garnavich, Gawiser, Gruetzbauch, Hartley,
  H\"aussler, Herrington, Hopkins, Huang, Jha, Johnson, Kartaltepe, Khostovan,
  Kirshner, Lani, Lee, Li, Madau, {McCarthy}, {McIntosh}, {McLure},
  {McPartland}, Mobasher, Moreira, Mortlock, Moustakas, Mozena, Nandra, Newman,
  Nielsen, Niemi, Noeske, Papovich, Pentericci, Pope, Primack, Ravindranath,
  Reddy, Renzini, Rix, Robaina, Rosario, Rosati, Salimbeni, Scarlata, Siana,
  Simard, Smidt, Snyder, Somerville, Spinrad, Straughn, Telford, Teplitz,
  Trump, Vargas, Villforth, Wagner, Wandro, Wechsler, Weiner, Wiklind, Wild,
  Wilson, Wuyts, \& Yun}]{koekemoer2011}
Koekemoer, A.~M., Faber, S.~M., Ferguson, H.~C., {et~al.} 2011, \apjs, 197, 36

\bibitem[{Kriek {et~al.}(2016)Kriek, Conroy, van Dokkum, Shapley, Choi, Reddy,
  Siana, van~de Voort, Coil, \& Mobasher}]{kriek2016}
Kriek, M., Conroy, C., van Dokkum, P.~G., {et~al.} 2016, \nat, 540, 248

\bibitem[{Kriek {et~al.}(2015)Kriek, Shapley, Reddy, Siana, Coil, Mobasher,
  Freeman, de~Groot, Price, Sanders, Shivaei, Brammer, Momcheva, Skelton, van
  Dokkum, Whitaker, Aird, Azadi, Kassis, Bullock, Conroy, Dav\'e, Kereš, \&
  Krumholz}]{kriek2015}
Kriek, M., Shapley, A.~E., Reddy, N.~A., {et~al.} 2015, \apjs, 218, 15

\bibitem[{Kriek {et~al.}(2006)Kriek, van Dokkum, Franx, F\"orster~Schreiber,
  Gawiser, Illingworth, Labb\'e, Marchesini, Quadri, Rix, Rudnick, Toft,
  van~der Werf, \& Wuyts}]{kriek2006}
Kriek, M., van Dokkum, P.~G., Franx, M., {et~al.} 2006, \apj, 645, 44

\bibitem[{Kriek {et~al.}(2009)Kriek, van Dokkum, Labb\'e, Franx, Illingworth,
  Marchesini, \& Quadri}]{kriek2009}
Kriek, M., van Dokkum, P.~G., Labb\'e, I., {et~al.} 2009, \apj, 700, 221

\bibitem[{Labb\'e {et~al.}(2005)Labb\'e, Huang, Franx, Rudnick, Barmby, Daddi,
  van Dokkum, Fazio, Schreiber, Moorwood, Rix, R\"ottgering, Trujillo, \&
  van~der Werf}]{labbe2005}
Labb\'e, I., Huang, J., Franx, M., {et~al.} 2005, \apjl, 624, L81

\bibitem[{Lacey \& Cole(1993)}]{lacey1993}
Lacey, C. \& Cole, S. 1993, \mnras, 262, 627

\bibitem[{Lemaux {et~al.}(2010)Lemaux, Lubin, Shapley, Kocevski, Gal, \&
  Squires}]{lemaux2010}
Lemaux, B.~C., Lubin, L.~M., Shapley, A., {et~al.} 2010, \apj, 716, 970

\bibitem[{Lin {et~al.}(2017)Lin, Belfiore, Pan, Bothwell, Hsieh, Huang, Xiao,
  Sánchez, Hsieh, Masters, Ramya, Lin, Hsu, Li, Maiolino, Bundy, Bizyaev,
  Drory, {Ibarra-Medel}, Lacerna, Haines, Smethurst, Stark, \&
  Thomas}]{lin2017}
Lin, L., Belfiore, F., Pan, H., {et~al.} 2017, \apj, 851, 18

\bibitem[{Marsan {et~al.}(2017)Marsan, Marchesini, Brammer, Geier, {Kado-Fong},
  Labb\'e, Muzzin, \& Stefanon}]{marsan2017}
Marsan, Z.~C., Marchesini, D., Brammer, G.~B., {et~al.} 2017, \apj, 842, 21

\bibitem[{Martig {et~al.}(2009)Martig, Bournaud, Teyssier, \&
  Dekel}]{martig2009}
Martig, M., Bournaud, F., Teyssier, R., \& Dekel, A. 2009, \apj, 707, 250

\bibitem[{Martis {et~al.}(2016)Martis, Marchesini, Brammer, Muzzin, Labb\'e,
  Momcheva, Skelton, Stefanon, van Dokkum, \& Whitaker}]{martis2016}
Martis, N.~S., Marchesini, D., Brammer, G.~B., {et~al.} 2016, \apjl, 827, L25

\bibitem[{Mawatari {et~al.}(2016)Mawatari, Yamada, Fazio, Huang, \&
  Ashby}]{mawatari2016}
Mawatari, K., Yamada, T., Fazio, G.~G., Huang, J., \& Ashby, M. L.~N. 2016,
  \pasj, 68, 46

\bibitem[{{McLean} {et~al.}(2012){McLean}, Steidel, Epps, Konidaris, Matthews,
  Adkins, Aliado, Brims, Canfield, Cromer, Fucik, Kulas, Mace, Magnone,
  Rodriguez, Rudie, Trainor, Wang, Weber, \& Weiss}]{mclean2012}
{McLean}, I.~S., Steidel, C.~C., Epps, H.~W., {et~al.} 2012, in , 84460J

\bibitem[{{McLure} {et~al.}(2017){McLure}, Pentericci, \& {{VANDELS}
  Team}}]{mclure2017}
{McLure}, R., Pentericci, L., \& {{VANDELS} Team}. 2017, The Messenger, 167, 31

\bibitem[{Merlin {et~al.}(2018)Merlin, Fontana, Castellano, Santini, Torelli,
  Boutsia, Wang, Grazian, Pentericci, Schreiber, Ciesla, {McLure}, Derriere,
  Dunlop, \& Elbaz}]{merlin2018}
Merlin, E., Fontana, A., Castellano, M., {et~al.} 2018, \mnras, 473, 2098

\bibitem[{Moster {et~al.}(2011)Moster, Somerville, Newman, \& Rix}]{moster2011}
Moster, B.~P., Somerville, R.~S., Newman, J.~A., \& Rix, H. 2011, \apj, 731,
  113

\bibitem[{Muzzin {et~al.}(2012)Muzzin, Labb\'e, Franx, van Dokkum, Holt,
  Szomoru, van~de Sande, Brammer, Marchesini, Stefanon, Buitrago, Caputi,
  Dunlop, Fynbo, Le~F\'evre, {McCracken}, \& {Milvang-Jensen}}]{muzzin2012}
Muzzin, A., Labb\'e, I., Franx, M., {et~al.} 2012, \apj, 761, 142

\bibitem[{Muzzin {et~al.}(2013)Muzzin, Marchesini, Stefanon, Franx,
  {Milvang-Jensen}, Dunlop, Fynbo, Brammer, Labb\'e, \& van
  Dokkum}]{muzzin2013}
Muzzin, A., Marchesini, D., Stefanon, M., {et~al.} 2013, \apjs, 206, 8

\bibitem[{Nanayakkara {et~al.}(2016)Nanayakkara, Glazebrook, Kacprzak, Yuan,
  Tran, Spitler, Kewley, Straatman, Cowley, Fisher, Labbe, Tomczak, Allen, \&
  Alcorn}]{nanayakkara2016}
Nanayakkara, T., Glazebrook, K., Kacprzak, G.~G., {et~al.} 2016, \apj, 828, 21

\bibitem[{Newman {et~al.}(2012)Newman, Ellis, Bundy, \& Treu}]{newman2012}
Newman, A.~B., Ellis, R.~S., Bundy, K., \& Treu, T. 2012, \apj, 746, 162

\bibitem[{Noeske {et~al.}(2007)Noeske, Faber, Weiner, Koo, Primack, Dekel,
  Papovich, Conselice, Le~Floc'h, Rieke, Coil, Lotz, Somerville, \&
  Bundy}]{noeske2007}
Noeske, K.~G., Faber, S.~M., Weiner, B.~J., {et~al.} 2007, \apjl, 660, L47

\bibitem[{Noll {et~al.}(2009)Noll, Burgarella, Giovannoli, Buat, Marcillac, \&
  {Muñoz-Mateos}}]{noll2009}
Noll, S., Burgarella, D., Giovannoli, E., {et~al.} 2009, \aap, 507, 1793

\bibitem[{Onodera {et~al.}(2016)Onodera, Carollo, Lilly, Renzini, Arimoto,
  Capak, Daddi, Scoville, Tacchella, Tatehora, \& Zamorani}]{onodera2016}
Onodera, M., Carollo, C.~M., Lilly, S., {et~al.} 2016, \apj, 822, 42

\bibitem[{Pannella {et~al.}(2015)Pannella, Elbaz, Daddi, Dickinson, Hwang,
  Schreiber, Strazzullo, Aussel, Bethermin, Buat, Charmandaris, Cibinel,
  Juneau, Ivison, Le~Borgne, Le~Floc’h, Leiton, Lin, Magdis, Morrison,
  Mullaney, Onodera, Renzini, Salim, Sargent, Scott, Shu, \&
  Wang}]{pannella2015}
Pannella, M., Elbaz, D., Daddi, E., {et~al.} 2015, \apj, 807, 141

\bibitem[{Papovich {et~al.}(2001)Papovich, Dickinson, \&
  Ferguson}]{papovich2001}
Papovich, C., Dickinson, M., \& Ferguson, H.~C. 2001, \apj, 559, 620

\bibitem[{Papovich {et~al.}(2011)Papovich, Finkelstein, Ferguson, Lotz, \&
  Giavalisco}]{papovich2011}
Papovich, C., Finkelstein, S.~L., Ferguson, H.~C., Lotz, J.~M., \& Giavalisco,
  M. 2011, \mnras, 412, 1123

\bibitem[{Papovich {et~al.}(2006)Papovich, Moustakas, Dickinson, Le~Floc'h,
  Rieke, Daddi, Alexander, Bauer, Brandt, Dahlen, Egami, Eisenhardt, Elbaz,
  Ferguson, Giavalisco, Lucas, Mobasher, {P\'erez-González}, Stutz, Rieke, \&
  Yan}]{papovich2006}
Papovich, C., Moustakas, L.~A., Dickinson, M., {et~al.} 2006, \apj, 640, 92

\bibitem[{Press \& Schechter(1974)}]{press1974}
Press, W.~H. \& Schechter, P. 1974, \apj, 187, 425

\bibitem[{Qin {et~al.}(2017{\natexlab{a}})Qin, Mutch, Duffy, Geil, Poole,
  Mesinger, \& Wyithe}]{qin2017-a}
Qin, Y., Mutch, S.~J., Duffy, A.~R., {et~al.} 2017{\natexlab{a}}, \mnras, 471,
  4345

\bibitem[{Qin {et~al.}(2017{\natexlab{b}})Qin, Mutch, Poole, Liu, Angel, Duffy,
  Geil, Mesinger, \& Wyithe}]{qin2017}
Qin, Y., Mutch, S.~J., Poole, G.~B., {et~al.} 2017{\natexlab{b}}, \mnras, 472,
  2009

\bibitem[{Reddy {et~al.}(2015)Reddy, Kriek, Shapley, Freeman, Siana, Coil,
  Mobasher, Price, Sanders, \& Shivaei}]{reddy2015}
Reddy, N.~A., Kriek, M., Shapley, A.~E., {et~al.} 2015, \apj, 806, 259

\bibitem[{Rees \& Ostriker(1977)}]{rees1977}
Rees, M.~J. \& Ostriker, J.~P. 1977, \mnras, 179, 541

\bibitem[{Rong {et~al.}(2017)Rong, Jing, Gao, Guo, Wang, Sun, Wang, \&
  Pan}]{rong2017}
Rong, Y., Jing, Y., Gao, L., {et~al.} 2017, \mnras, 471, L36

\bibitem[{Salpeter(1955)}]{salpeter1955}
Salpeter, E.~E. 1955, \apj, 121, 161

\bibitem[{Schreiber {et~al.}(2018{\natexlab{a}})Schreiber, Elbaz, Pannella,
  Ciesla, Wang, \& Franco}]{schreiber2018-a}
Schreiber, C., Elbaz, D., Pannella, M., {et~al.} 2018{\natexlab{a}}, \aap, 609,
  A30

\bibitem[{Schreiber {et~al.}(2016)Schreiber, Elbaz, Pannella, Ciesla, Wang,
  Koekemoer, Rafelski, \& Daddi}]{schreiber2016}
Schreiber, C., Elbaz, D., Pannella, M., {et~al.} 2016, \aap, 589, A35

\bibitem[{Schreiber {et~al.}(2018{\natexlab{b}})Schreiber, Labb\'e, Glazebrook,
  Bekiaris, Papovich, Costa, Elbaz, Kacprzak, Nanayakkara, Oesch, Pannella,
  Spitler, Straatman, Tran, \& Wang}]{schreiber2018}
Schreiber, C., Labb\'e, I., Glazebrook, K., {et~al.} 2018{\natexlab{b}}, \aap,
  611, A22

\bibitem[{Schreiber {et~al.}(2015)Schreiber, Pannella, Elbaz, B\'ethermin,
  Inami, Dickinson, Magnelli, Wang, Aussel, Daddi, Juneau, Shu, Sargent, Buat,
  Faber, Ferguson, Giavalisco, Koekemoer, Magdis, Morrison, Papovich, Santini,
  \& Scott}]{schreiber2015}
Schreiber, C., Pannella, M., Elbaz, D., {et~al.} 2015, \aap, 575, A74

\bibitem[{Schreiber {et~al.}(2017)Schreiber, Pannella, Leiton, Elbaz, Wang,
  Okumura, \& Labb\'e}]{schreiber2017}
Schreiber, C., Pannella, M., Leiton, R., {et~al.} 2017, \aap, 599, A134

\bibitem[{Shapley {et~al.}(2017)Shapley, Sanders, Reddy, Kriek, Freeman,
  Mobasher, Siana, Coil, Leung, {deGroot}, Shivaei, Price, Azadi, \&
  Aird}]{shapley2017}
Shapley, A.~E., Sanders, R.~L., Reddy, N.~A., {et~al.} 2017, \apjl, 846, L30

\bibitem[{Silk \& Rees(1998)}]{silk1998}
Silk, J. \& Rees, M.~J. 1998, \aap, 331, L1

\bibitem[{Silverman {et~al.}(2010)Silverman, Mainieri, Salvato, Hasinger,
  Bergeron, Capak, Szokoly, Finoguenov, Gilli, Rosati, Tozzi, Vignali,
  Alexander, Brandt, Lehmer, Luo, Rafferty, Xue, Balestra, Bauer, Brusa,
  Comastri, Kartaltepe, Koekemoer, Miyaji, Schneider, Treister, Wisotski, \&
  Schramm}]{silverman2010}
Silverman, J.~D., Mainieri, V., Salvato, M., {et~al.} 2010, \apjs, 191, 124

\bibitem[{Simpson {et~al.}(2017)Simpson, Smail, Swinbank, Ivison, Dunlop,
  Geach, Almaini, Arumugam, Bremer, Chen, Conselice, Coppin, Farrah, Ibar,
  Hartley, Ma, Michałowski, Scott, Spaans, Thomson, \& van~der
  Werf}]{simpson2017}
Simpson, J.~M., Smail, I., Swinbank, A.~M., {et~al.} 2017, \apj, 839, 58

\bibitem[{Skelton {et~al.}(2014)Skelton, Whitaker, Momcheva, Brammer, van
  Dokkum, Labb\'e, Franx, van~der Wel, Bezanson, Da~Cunha, Fumagalli,
  F\"orster~Schreiber, Kriek, Leja, Lundgren, Magee, Marchesini, Maseda,
  Nelson, Oesch, Pacifici, Patel, Price, Rix, Tal, Wake, \&
  Wuyts}]{skelton2014}
Skelton, R.~E., Whitaker, K.~E., Momcheva, I.~G., {et~al.} 2014, \apjs, 214, 24

\bibitem[{Smit {et~al.}(2012)Smit, Bouwens, Franx, Illingworth, Labb\'e, Oesch,
  \& van Dokkum}]{smit2012}
Smit, R., Bouwens, R.~J., Franx, M., {et~al.} 2012, \apj, 756, 14

\bibitem[{Smit {et~al.}(2016)Smit, Bouwens, Labb\'e, Franx, Wilkins, \&
  Oesch}]{smit2016}
Smit, R., Bouwens, R.~J., Labb\'e, I., {et~al.} 2016, \apj, 833, 254

\bibitem[{Sparre {et~al.}(2015)Sparre, Hayward, Springel, Vogelsberger, Genel,
  Torrey, Nelson, Sijacki, \& Hernquist}]{sparre2015}
Sparre, M., Hayward, C.~C., Springel, V., {et~al.} 2015, \mnras, 447, 3548

\bibitem[{Spitler {et~al.}(2014)Spitler, Straatman, Labb\'e, Glazebrook, Tran,
  Kacprzak, Quadri, Papovich, Persson, van Dokkum, Allen, Kawinwanichakij,
  Kelson, {McCarthy}, Mehrtens, Monson, Nanayakkara, Rees, Tilvi, \&
  Tomczak}]{spitler2014}
Spitler, L.~R., Straatman, C. M.~S., Labb\'e, I., {et~al.} 2014, \apjl, 787,
  L36

\bibitem[{Stefanon {et~al.}(2013)Stefanon, Marchesini, Rudnick, Brammer, \&
  Whitaker}]{stefanon2013}
Stefanon, M., Marchesini, D., Rudnick, G.~H., Brammer, G.~B., \& Whitaker,
  K.~E. 2013, \apj, 768, 92

\bibitem[{Steidel {et~al.}(2016)Steidel, Strom, Pettini, Rudie, Reddy, \&
  Trainor}]{steidel2016}
Steidel, C.~C., Strom, A.~L., Pettini, M., {et~al.} 2016, \apj, 826, 159

\bibitem[{Straatman {et~al.}(2014)Straatman, Labb\'e, Spitler, Allen, Altieri,
  Brammer, Dickinson, van Dokkum, Inami, Glazebrook, Kacprzak, Kawinwanichakij,
  Kelson, {McCarthy}, Mehrtens, Monson, Murphy, Papovich, Persson, Quadri,
  Rees, Tomczak, Tran, \& Tilvi}]{straatman2014}
Straatman, C. M.~S., Labb\'e, I., Spitler, L.~R., {et~al.} 2014, \apjl, 783,
  L14

\bibitem[{Straatman {et~al.}(2015)Straatman, Labb\'e, Spitler, Glazebrook,
  Tomczak, Allen, Brammer, Cowley, van Dokkum, Kacprzak, Kawinwanichakij,
  Mehrtens, Nanayakkara, Papovich, Persson, Quadri, Rees, Tilvi, Tran, \&
  Whitaker}]{straatman2015}
Straatman, C. M.~S., Labb\'e, I., Spitler, L.~R., {et~al.} 2015, \apjl, 808,
  L29

\bibitem[{Straatman {et~al.}(2016)Straatman, Spitler, Quadri, Labb\'e,
  Glazebrook, Persson, Papovich, Tran, Brammer, Cowley, Tomczak, Nanayakkara,
  Alcorn, Allen, Broussard, van Dokkum, Forrest, van Houdt, Kacprzak,
  Kawinwanichakij, Kelson, Lee, {McCarthy}, Mehrtens, Monson, Murphy, Rees,
  Tilvi, \& Whitaker}]{straatman2016}
Straatman, C. M.~S., Spitler, L.~R., Quadri, R.~F., {et~al.} 2016, \apj, 830,
  51

\bibitem[{Strom {et~al.}(2017)Strom, Steidel, Rudie, Trainor, Pettini, \&
  Reddy}]{strom2017}
Strom, A.~L., Steidel, C.~C., Rudie, G.~C., {et~al.} 2017, \apj, 836, 164

\bibitem[{Suess {et~al.}(2017)Suess, Bezanson, Spilker, Kriek, Greene,
  Feldmann, Hunt, \& Narayanan}]{suess2017}
Suess, K.~A., Bezanson, R., Spilker, J.~S., {et~al.} 2017, \apjl, 846, L14

\bibitem[{Tacconi {et~al.}(2010)Tacconi, Genzel, Neri, Cox, Cooper, Shapiro,
  Bolatto, Bouch\'e, Bournaud, Burkert, Combes, Comerford, Davis, Schreiber,
  {Garcia-Burillo}, {Gracia-Carpio}, Lutz, Naab, Omont, Shapley, Sternberg, \&
  Weiner}]{tacconi2010}
Tacconi, L.~J., Genzel, R., Neri, R., {et~al.} 2010, \nat, 463, 781

\bibitem[{Tasca {et~al.}(2017)Tasca, Le~F\`evre, Ribeiro, Thomas, Moreau,
  Cassata, Garilli, Le~Brun, Lemaux, Maccagni, Pentericci, Schaerer, Vanzella,
  Zamorani, Zucca, Amorin, Bardelli, Cassar\`a, Castellano, Cimatti, Cucciati,
  Durkalec, Fontana, Giavalisco, Grazian, Hathi, Ilbert, Paltani, Pforr,
  Scodeggio, Sommariva, Talia, Tresse, Vergani, Capak, Charlot, Contini, de~la
  Torre, Dunlop, Fotopoulou, Guaita, Koekemoer, {López-Sanjuan}, Mellier,
  Salvato, Scoville, Taniguchi, \& Wang}]{tasca2017}
Tasca, L. A.~M., Le~F\`evre, O., Ribeiro, B., {et~al.} 2017, \aap, 600, A110

\bibitem[{Tomczak {et~al.}(2014)Tomczak, Quadri, Tran, Labb\'e, Straatman,
  Papovich, Glazebrook, Allen, Brammer, Kacprzak, Kawinwanichakij, Kelson,
  {McCarthy}, Mehrtens, Monson, Persson, Spitler, Tilvi, \& van
  Dokkum}]{tomczak2014}
Tomczak, A.~R., Quadri, R.~F., Tran, K.~H., {et~al.} 2014, \apj, 783, 85

\bibitem[{van~de Sande {et~al.}(2013)van~de Sande, Kriek, Franx, van Dokkum,
  Bezanson, Bouwens, Quadri, Rix, \& Skelton}]{vandesande2013}
van~de Sande, J., Kriek, M., Franx, M., {et~al.} 2013, \apj, 771, 85

\bibitem[{van~der Wel {et~al.}(2014{\natexlab{a}})van~der Wel, Chang, Bell,
  Holden, Ferguson, Giavalisco, Rix, Skelton, Whitaker, Momcheva, Brammer,
  Kassin, Martig, Dekel, Ceverino, Koo, Mozena, van Dokkum, Franx, Faber, \&
  Primack}]{vanderwel2014}
van~der Wel, A., Chang, Y., Bell, E.~F., {et~al.} 2014{\natexlab{a}}, \apjl,
  792, L6

\bibitem[{van~der Wel {et~al.}(2014{\natexlab{b}})van~der Wel, Franx, van
  Dokkum, Skelton, Momcheva, Whitaker, Brammer, Bell, Rix, Wuyts, Ferguson,
  Holden, Barro, Koekemoer, Chang, {McGrath}, H\"aussler, Dekel, Behroozi,
  Fumagalli, Leja, Lundgren, Maseda, Nelson, Wake, Patel, Labb\'e, Faber,
  Grogin, \& Kocevski}]{vanderwel2014-a}
van~der Wel, A., Franx, M., van Dokkum, P.~G., {et~al.} 2014{\natexlab{b}},
  \apj, 788, 28

\bibitem[{van Dokkum {et~al.}(2008)van Dokkum, Franx, Kriek, Holden,
  Illingworth, Magee, Bouwens, Marchesini, Quadri, Rudnick, Taylor, \&
  Toft}]{vandokkum2008}
van Dokkum, P.~G., Franx, M., Kriek, M., {et~al.} 2008, \apjl, 677, L5

\bibitem[{van Dokkum {et~al.}(2015)van Dokkum, Nelson, Franx, Oesch, Momcheva,
  Brammer, F\"orster~Schreiber, Skelton, Whitaker, van~der Wel, Bezanson,
  Fumagalli, Illingworth, Kriek, Leja, \& Wuyts}]{vandokkum2015}
van Dokkum, P.~G., Nelson, E.~J., Franx, M., {et~al.} 2015, \apj, 813, 23

\bibitem[{van Dokkum {et~al.}(2010)van Dokkum, Whitaker, Brammer, Franx, Kriek,
  Labb\'e, Marchesini, Quadri, Bezanson, Illingworth, Muzzin, Rudnick, Tal, \&
  Wake}]{vandokkum2010}
van Dokkum, P.~G., Whitaker, K.~E., Brammer, G., {et~al.} 2010, \apj, 709, 1018

\bibitem[{Wang {et~al.}(2016)Wang, Elbaz, Daddi, Finoguenov, Liu, Schreiber,
  Martín, Strazzullo, Valentino, van~der Burg, Zanella, Ciesla, Gobat,
  Le~Brun, Pannella, Sargent, Shu, Tan, Cappelluti, \& Li}]{wang2016-a}
Wang, T., Elbaz, D., Daddi, E., {et~al.} 2016, \apj, 828, 56

\bibitem[{Wang {et~al.}(2017)Wang, Faber, Liu, Guo, Pacifici, Koo, Kassin, Mao,
  Fang, Chen, Koekemoer, Kocevski, \& Ashby}]{wang2017}
Wang, W., Faber, S.~M., Liu, F.~S., {et~al.} 2017, \mnras, 469, 4063

\bibitem[{Wellons {et~al.}(2015)Wellons, Torrey, Ma, {Rodriguez-Gomez},
  Vogelsberger, Kriek, van Dokkum, Nelson, Genel, Pillepich, Springel, Sijacki,
  Snyder, Nelson, Sales, \& Hernquist}]{wellons2015}
Wellons, S., Torrey, P., Ma, C., {et~al.} 2015, \mnras, 449, 361

\bibitem[{Whitaker {et~al.}(2012)Whitaker, Kriek, van Dokkum, Bezanson,
  Brammer, Franx, \& Labb\'e}]{whitaker2012}
Whitaker, K.~E., Kriek, M., van Dokkum, P.~G., {et~al.} 2012, \apj, 745, 179

\bibitem[{White \& Rees(1978)}]{white1978}
White, S. D.~M. \& Rees, M.~J. 1978, \mnras, 183, 341

\bibitem[{Wild {et~al.}(2014)Wild, Almaini, Cirasuolo, Dunlop, {McLure},
  Bowler, Ferreira, Bradshaw, Chuter, \& Hartley}]{wild2014}
Wild, V., Almaini, O., Cirasuolo, M., {et~al.} 2014, \mnras, 440, 1880

\bibitem[{Williams {et~al.}(2009)Williams, Quadri, Franx, van Dokkum, \&
  Labb\'e}]{williams2009}
Williams, R.~J., Quadri, R.~F., Franx, M., van Dokkum, P., \& Labb\'e, I. 2009,
  \apj, 691, 1879

\bibitem[{Wuyts {et~al.}(2011)Wuyts, F\"orster~Schreiber, van~der Wel,
  Magnelli, Guo, Genzel, Lutz, Aussel, Barro, Berta, Cava, {Graciá-Carpio},
  Hathi, Huang, Kocevski, Koekemoer, Lee, Le~Floc'h, {McGrath}, Nordon,
  Popesso, Pozzi, Riguccini, Rodighiero, Saintonge, \& Tacconi}]{wuyts2011-a}
Wuyts, S., F\"orster~Schreiber, N.~M., van~der Wel, A., {et~al.} 2011, \apj,
  742, 96

\bibitem[{Yan {et~al.}(2006)Yan, Newman, Faber, Konidaris, Koo, \&
  Davis}]{yan2006}
Yan, R., Newman, J.~A., Faber, S.~M., {et~al.} 2006, \apj, 648, 281

\end{thebibliography}

\appendix

\section{Tables}

\begin{table*}
\begin{center}
\caption{Galaxies targeted with MOSFIRE, photometry.\label{TAB:galaxies}}
\begin{tabular}{lccccccccc}
\hline\hline \\[-0.3cm]
ID $^a$     & R.A. & Dec. & \multicolumn{2}{c}{Observed mag. $^b$} & 870$\,\um$ \\
            & deg. & deg. & $H$ & $K$                              & $\mJy$ \\
\hline \\[-0.3cm]
ZF-COS-10559 & 150.07147 & 2.2911844 & $25.74\pm0.18$ & $24.15\pm0.08$ & $0.09 \pm 0.17$ \\[0.1cm]
ZF-COS-14907 & 150.12422 & 2.3374486 & $23.44\pm0.02$ & $22.65\pm0.02$ & -- \\[0.1cm]
ZF-COS-17779 & 150.04651 & 2.3673911 & $24.94\pm0.15$ & $23.82\pm0.06$ & -- \\[0.1cm]
ZF-COS-18842 & 150.08728 & 2.3960431 & $24.45\pm0.08$ & $23.03\pm0.04$ & -- \\[0.1cm]
ZF-COS-19589 & 150.06671 & 2.3823645 & $25.41\pm0.15$ & $23.54\pm0.06$ & -- \\[0.1cm]
ZF-COS-20032 & 150.11256 & 2.3765266 & $25.14\pm0.13$ & $23.62\pm0.06$ & $1.64 \pm 0.28$ \\[0.1cm]
ZF-COS-20115 & 150.06149 & 2.3787093 & $24.39\pm0.03$ & $22.43\pm0.02$ & $0.09 \pm 0.06$ $^c$ \\[0.1cm]
ZF-COS-20133 & 150.12173 & 2.3745940 & $25.75\pm0.19$ & $23.84\pm0.06$ & -- \\[0.1cm]
3D-EGS-18996 & 214.89563 & 52.856556 & $22.75\pm0.01$ & $21.60\pm0.02$ & -- \\[0.1cm]
3D-EGS-26047 & 214.90512 & 52.891621 & $23.83\pm0.03$ & $22.55\pm0.05$ & -- \\[0.1cm]
3D-EGS-27584 & 214.85387 & 52.861393 & $24.62\pm0.12$ & $22.25\pm0.07$ & -- \\[0.1cm]
3D-EGS-31322 & 214.86606 & 52.884312 & $23.72\pm0.03$ & $22.20\pm0.04$ & -- \\[0.1cm]
3D-EGS-34322 & 214.81316 & 52.858986 & $25.02\pm0.12$ & $23.52\pm0.21$ & -- \\[0.1cm]
3D-EGS-40032 & 214.76062 & 52.845383 & $22.92\pm0.02$ & $21.59\pm0.03$ & -- \\[0.1cm]
ZF-UDS-3651 & 34.289452 & -5.2698030 & $24.54\pm0.05$ & $22.95\pm0.03$ & $0.64 \pm 0.42$ \\[0.1cm]
ZF-UDS-4347 & 34.290428 & -5.2620687 & $24.90\pm0.07$ & $23.17\pm0.03$ & $-0.44 \pm 0.45$ \\[0.1cm]
ZF-UDS-6496 & 34.340358 & -5.2412550 & $24.27\pm0.05$ & $22.62\pm0.02$ & $-0.04 \pm 0.24$ \\[0.1cm]
ZF-UDS-7329 & 34.255872 & -5.2338210 & $23.90\pm0.04$ & $22.36\pm0.01$ & -- \\[0.1cm]
ZF-UDS-7542 & 34.258888 & -5.2322803 & $24.16\pm0.04$ & $23.02\pm0.02$ & -- \\[0.1cm]
ZF-UDS-8197 & 34.293755 & -5.2269478 & $24.77\pm0.07$ & $23.22\pm0.03$ & -- \\[0.1cm]
3D-UDS-27939 & 34.487273 & -5.1821730 & $24.69\pm0.09$ & $23.26\pm0.07$ & $1.57 \pm 0.57$ \\[0.1cm]
3D-UDS-35168 & 34.485131 & -5.1578340 & $25.40\pm0.13$ & $23.89\pm0.09$ & -- \\[0.1cm]
3D-UDS-39102 & 34.526210 & -5.1438100 & $25.10\pm0.18$ & $23.28\pm0.15$ & $0.11 \pm 0.56$ \\[0.1cm]
3D-UDS-41232 & 34.526589 & -5.1360390 & $22.90\pm0.02$ & $21.74\pm0.01$ & $0.09 \pm 0.26$ \\[0.1cm]
\hline
\end{tabular}
\end{center}
{\footnotesize
$^a$ IDs starting with ``ZF'' are from the ZFOURGE DR1 catalogs, and IDs starting with ``3D'' are from the 3DHST v4.1 catalogs.
$^b$ Observed AB magnitudes.
$^c$ Obtained at $\lambda=744\,\um$ (Band 9).}
\end{table*}

\clearpage

\begin{table*}
\begin{center}
\caption{Galaxies targeted with MOSFIRE, properties from photometric redshift.\label{TAB:galaxies_zphot}}
\begin{tabular}{lcccc}
\hline\hline \\[-0.3cm]
ID $^a$     & $\zphot$ & $\mstar$         & $U-V$    & $V-J$      \\
            &          & $10^{11}\,\msun$ & rest, AB & rest, AB   \\
\hline \\[-0.3cm]
ZF-COS-10559 & $3.34^{+0.30}_{-1.04}$ & 0.23 & $1.57^{+0.75}_{-0.88}$ & $1.05^{+0.89}_{-0.32}$ \\[0.1cm]
ZF-COS-14907 & $2.89^{+0.06}_{-0.06}$ & 0.47 & $1.38^{+0.08}_{-0.07}$ & $0.76^{+0.04}_{-0.03}$ \\[0.1cm]
ZF-COS-17779 & $3.91^{+0.35}_{-0.33}$ & 0.43 & $1.28^{+0.23}_{-0.28}$ & $0.63^{+0.16}_{-0.14}$ \\[0.1cm]
ZF-COS-18842 & $3.47^{+0.07}_{-0.07}$ & 0.40 & $1.20^{+0.07}_{-0.06}$ & $0.69^{+0.02}_{-0.03}$ \\[0.1cm]
ZF-COS-19589 & $3.73^{+0.16}_{-0.15}$ & 0.64 & $1.66^{+0.17}_{-0.17}$ & $0.77^{+0.11}_{-0.06}$ \\[0.1cm]
ZF-COS-20032 & $3.55^{+0.69}_{-0.68}$ & 2.54 & $1.92^{+1.13}_{-0.82}$ & $1.76^{+0.53}_{-0.62}$ \\[0.1cm]
ZF-COS-20115 & $3.64^{+0.08}_{-0.09}$ & 1.24 & $1.71^{+0.11}_{-0.07}$ & $0.53^{+0.02}_{-0.03}$ \\[0.1cm]
ZF-COS-20133 & $3.51^{+0.04}_{-0.04}$ & 0.24 & $1.56^{+0.04}_{-0.10}$ & $0.69^{+0.01}_{-0.01}$ \\[0.1cm]
3D-EGS-18996 & $2.99^{+0.03}_{-0.03}$ & 0.86 & $1.55^{+0.04}_{-0.04}$ & $0.30^{+0.01}_{-0.01}$ \\[0.1cm]
3D-EGS-26047 & $3.24^{+0.08}_{-0.08}$ & 1.03 & $1.25^{+0.12}_{-0.06}$ & $1.21^{+0.03}_{-0.03}$ \\[0.1cm]
3D-EGS-27584 & $3.60^{+0.18}_{-0.23}$ & 4.54 & $2.25^{+0.29}_{-0.23}$ & $1.33^{+0.16}_{-0.14}$ \\[0.1cm]
3D-EGS-31322 & $3.47^{+0.07}_{-0.07}$ & 0.98 & $1.30^{+0.06}_{-0.11}$ & $0.79^{+0.09}_{-0.01}$ \\[0.1cm]
3D-EGS-34322 & $3.59^{+0.33}_{-0.32}$ & 0.43 & $1.25^{+0.51}_{-0.16}$ & $0.95^{+0.05}_{-0.23}$ \\[0.1cm]
3D-EGS-40032 & $3.22^{+0.08}_{-0.09}$ & 1.98 & $1.51^{+0.13}_{-0.07}$ & $0.76^{+0.02}_{-0.01}$ \\[0.1cm]
ZF-UDS-3651 & $3.87^{+0.12}_{-0.12}$ & 0.77 & $1.35^{+0.11}_{-0.25}$ & $0.74^{+0.01}_{-0.01}$ \\[0.1cm]
ZF-UDS-4347 & $3.58^{+0.04}_{-0.05}$ & 0.31 & $1.54^{+0.06}_{-0.09}$ & $0.81^{+0.08}_{-0.04}$ \\[0.1cm]
ZF-UDS-6496 & $3.50^{+0.04}_{-0.04}$ & 0.79 & $1.73^{+0.04}_{-0.02}$ & $0.81^{+0.01}_{-0.02}$ \\[0.1cm]
ZF-UDS-7329 & $3.04^{+0.17}_{-0.17}$ & 1.67 & $2.05^{+0.28}_{-0.32}$ & $1.16^{+0.13}_{-0.10}$ \\[0.1cm]
ZF-UDS-7542 & $3.15^{+0.06}_{-0.06}$ & 0.70 & $1.49^{+0.08}_{-0.05}$ & $0.97^{+0.02}_{-0.01}$ \\[0.1cm]
ZF-UDS-8197 & $3.47^{+0.06}_{-0.06}$ & 0.36 & $1.29^{+0.08}_{-0.06}$ & $0.94^{+0.03}_{-0.04}$ \\[0.1cm]
3D-UDS-27939 & $3.22^{+0.20}_{-0.22}$ & 0.54 & $1.46^{+0.36}_{-0.14}$ & $0.96^{+0.01}_{-0.10}$ \\[0.1cm]
3D-UDS-35168 & $3.46^{+0.32}_{-0.29}$ & 0.32 & $1.37^{+0.35}_{-0.28}$ & $0.81^{+0.11}_{-0.12}$ \\[0.1cm]
3D-UDS-39102 & $3.51^{+0.42}_{-0.36}$ & 0.86 & $1.61^{+0.51}_{-0.50}$ & $1.32^{+0.11}_{-0.24}$ \\[0.1cm]
3D-UDS-41232 & $3.01^{+0.07}_{-0.08}$ & 1.52 & $1.56^{+0.16}_{-0.12}$ & $0.82^{+0.04}_{-0.03}$ \\[0.1cm]
\hline
\end{tabular}
\end{center}
\end{table*}

\clearpage

\begin{table*}
\begin{center}
\caption{MOSFIRE observations of our targets.\label{TAB:galaxies_limits}}
\begin{tabular}{lR{0.8cm}R{0.8cm}R{1.0cm}R{1.0cm}R{0.7cm}R{0.7cm}R{0.7cm}R{0.7cm}l}
\hline\hline \\[-0.3cm]
ID       & \multicolumn{2}{c}{Integration time} & \multicolumn{2}{c}{Uncertainty $^{a,b}$} & \multicolumn{2}{c}{Median $S/N$ $^b$} & \multicolumn{2}{c}{Maximum $S/N$ $^b$} & Masks \\
         & \multicolumn{2}{c}{hours} & \multicolumn{2}{c}{$10^{-19}\,{\rm erg/s/cm^2/\AA}$} &
         & & & & \\[0.1cm]
         & $H$ & $K$ & $H$ & $K$ & $H$ & $K$ & $H$ & $K$ \\
\hline \\[-0.3cm]
ZF-COS-10559 & -- & 1.6 & -- & 0.50 & -- & 0.7 & -- & 4.3 & {\scriptsize Z245} \\
ZF-COS-14907 & -- & 3.3 & -- & 0.54 & -- & 3.6 & -- & 9.9 & {\scriptsize Y259-A} \\
ZF-COS-17779 & 3.9 & 7.2 & 0.40 & 0.24 & 1.1 & 3.1 & 5.1 & 6.7 & {\scriptsize W182} \\
ZF-COS-18842 & 0.3 & 3.6 & 0.89 & 0.36 & 0.7 & 3.9 & 3.5 & 8.7 & {\scriptsize U069} \\
ZF-COS-19589 & 3.9 & 7.2 & 0.42 & 0.25 & 0.8 & 3.7 & 3.2 & 7.9 & {\scriptsize W182} \\
ZF-COS-20032 & 3.9 & 7.2 & 0.84 & 0.31 & 1.5 & 1.5 & 3.4 & 6.8 & {\scriptsize W182} \\
ZF-COS-20115 & 4.2 & 14.4 & 0.47 & 0.21 & 1.1 & 12.2 & 3.9 & 22.9 & {\scriptsize W182, Y259-B, U069, Z245} \\
ZF-COS-20133 & 3.9 & 7.2 & 0.38 & 0.29 & 0.4 & 2.0 & 3.2 & 19.7 & {\scriptsize W182} \\
3D-EGS-18996 & 0.8 & 4.8 & 0.72 & 0.54 & 6.7 & 10.1 & 11.3 & 21.1 & {\scriptsize W057} \\
3D-EGS-26047 & 0.8 & 4.8 & 0.70 & 0.83 & 2.1 & 2.8 & 6.3 & 5.9 & {\scriptsize W057} \\
3D-EGS-27584 & 0.8 & 4.8 & -- & 0.86 & -- & 3.5 & -- & 6.0 & {\scriptsize W057} \\
3D-EGS-31322 & 0.8 & 4.8 & 0.50 & 0.51 & 3.9 & 7.4 & 10.1 & 15.4 & {\scriptsize W057} \\
3D-EGS-34322 & 0.8 & 4.8 & 0.63 & 0.72 & 1.1 & 1.4 & 3.8 & 5.7 & {\scriptsize W057} \\
3D-EGS-40032 & 0.8 & 4.8 & 0.50 & 0.52 & 7.1 & 11.0 & 16.0 & 19.3 & {\scriptsize W057} \\
ZF-UDS-3651 & 0.3 & 7.3 & 0.87 & 0.39 & 1.5 & 3.6 & 4.3 & 8.9 & {\scriptsize W182, Y259-A} \\
ZF-UDS-4347 & 0.3 & 2.4 & 0.89 & 0.52 & 1.3 & 2.6 & 4.2 & 6.8 & {\scriptsize W182} \\
ZF-UDS-6496 & -- & 4.9 & -- & 0.52 & -- & 3.6 & -- & 8.6 & {\scriptsize Y259-A} \\
ZF-UDS-7329 & -- & 9.6 & -- & 0.35 & -- & 7.2 & -- & 17.0 & {\scriptsize Y259-A, U069} \\
ZF-UDS-7542 & 0.3 & 7.3 & 0.93 & 0.34 & 1.4 & 4.2 & 3.8 & 9.1 & {\scriptsize W182, Y259-A} \\
ZF-UDS-8197 & 0.3 & 7.3 & 0.90 & 0.38 & 0.7 & 2.4 & 3.6 & 16.0 & {\scriptsize W182, Y259-A} \\
3D-UDS-27939 & -- & 4.0 & -- & 0.71 & -- & 3.2 & -- & 15.8 & {\scriptsize Y259-B} \\
3D-UDS-35168 & -- & 4.0 & -- & 0.50 & -- & 1.0 & -- & 4.9 & {\scriptsize Y259-B} \\
3D-UDS-39102 & -- & 4.0 & -- & 0.71 & -- & 0.8 & -- & 4.3 & {\scriptsize Y259-B} \\
3D-UDS-41232 & -- & 4.0 & -- & 0.60 & -- & 7.9 & -- & 17.3 & {\scriptsize Y259-B} \\
\hline
\end{tabular}
\end{center}
{\footnotesize $^a$ Median uncertainty of the MOSFIRE spectrum ($1\sigma$), accounting for slit loss. $^b$ For a binning of $70\,\AA$.}
\end{table*}

\clearpage

\begin{table*}
\begin{center}
\caption{Spectroscopic identification of our targets.\label{TAB:galaxies_zspecs}}
\begin{tabular}{lcccrrllc}
\hline\hline \\[-0.3cm]
ID       & $\zphot$ & $\zspec$ & \multirow{2}{*}{$\displaystyle \frac{\zphot-\zspec}{\Delta z}$} & $p$ $^a$ & $\chi^2_{\rm red}$ & Balmer   & Emission & $\sigma_{v}$ $^{d}$ \\
         &          &          &                            &          &                    & abs. $^b$ & lines $^c$ & $\kms$ \\
\hline \\[-0.3cm]
\multicolumn{2}{l}{Robust $\zspec$} \\
ZF-COS-20032 & $3.55^{+0.69}_{-0.68}$ & $2.4736^{+0.0008}_{-0.0010}$ & 1.6 & 96\% & 1.0 & -- & $\oone$ $\halpha$ $\nii_{\lambda\lambda}$  & $217\pm161$ \\[0.1cm]
ZF-COS-20115 & $3.64^{+0.08}_{-0.09}$ & $3.7145^{+0.0015}_{-0.0015}$ & -0.9 & 100\% & 1.0 & yes & -- & -- \\[0.1cm]
ZF-COS-20133 & $3.51^{+0.04}_{-0.04}$ & $3.4806^{+0.0002}_{-0.0002}$ & 0.6 & 100\% & 1.0 & -- & $\hbeta$ $\oiii_{\lambda\lambda}$  & $60\pm26$ \\[0.1cm]
3D-EGS-18996 & $2.99^{+0.03}_{-0.03}$ & $3.2390^{+0.0007}_{-0.0009}$ & -8.3 & 100\% & 1.0 & yes & $\oiii_{\lambda\lambda}$  & $530\pm199$ \\[0.1cm]
3D-EGS-26047 & $3.24^{+0.08}_{-0.08}$ & $3.2337^{+0.0020}_{-0.0016}$ & 0.0 & 99\% & 1.0 & -- & $\oii_{\lambda\lambda}$ $\hbeta$ $\oiii_{\lambda\lambda}$  & $530\pm263$ \\[0.1cm]
3D-EGS-40032 & $3.22^{+0.08}_{-0.09}$ & $3.2187^{+0.0011}_{-0.0013}$ & -0.0 & 100\% & 1.0 & yes & $\oii_{\lambda\lambda}$  & $582\pm236$ \\[0.1cm]
ZF-UDS-8197 & $3.47^{+0.06}_{-0.06}$ & $3.5431^{+0.0007}_{-0.0010}$ & -1.2 & 99\% & 1.0 & -- & $\oiii_{\lambda\lambda}$  & $530\pm53$ \\[0.1cm]
3D-UDS-27939 & $3.22^{+0.20}_{-0.22}$ & $2.2104^{+0.0004}_{-0.0004}$ & 4.5 & 100\% & 1.0 & -- & $\oone$ $\halpha$ $\nii_{\lambda\lambda}$ $\sii_{\lambda\lambda}$  & $154\pm7$ \\[0.1cm]
\hline \\[-0.3cm]
\multicolumn{2}{l}{Uncertain $\zspec$} \\
ZF-COS-17779 & $3.91^{+0.35}_{-0.33}$ & $3.4150^{+0.1320}_{-0.0003}$ & 1.4 & 77\% & 1.0 & -- & $\hbeta$ $\oiii_{\lambda\lambda}$  & $60\pm26$ \\[0.1cm]
ZF-COS-18842 & $3.47^{+0.07}_{-0.07}$ & $3.7823^{+0.0023}_{-0.0031}$ & -4.5 & 75\% & 1.1 & -- & $\oiii_{\lambda\lambda}$  & $60\pm279$ \\[0.1cm]
ZF-COS-19589 & $3.73^{+0.16}_{-0.15}$ & $3.7152^{+0.0094}_{-0.1589}$ & 0.1 & 32\% & 0.9 & -- & -- & -- \\[0.1cm]
3D-EGS-31322 & $3.47^{+0.07}_{-0.07}$ & $3.4344^{+0.0104}_{-0.0029}$ & 0.5 & 84\% & 1.0 & yes & -- & -- \\[0.1cm]
\hline \\[-0.3cm]
\multicolumn{2}{l}{Rejected $\zspec$} \\
ZF-COS-10559 & $3.34^{+0.30}_{-1.04}$ & $2.6376^{+1.6758}_{-0.2266}$ & 0.4 & 2\% & 0.9 & -- & -- & -- \\[0.1cm]
ZF-COS-14907 & $2.89^{+0.06}_{-0.06}$ & $4.1935^{+0.1241}_{-0.0030}$ & -23.1 & 63\% & 1.2 & -- & -- & -- \\[0.1cm]
3D-EGS-27584 & $3.60^{+0.18}_{-0.23}$ & $2.4620^{+0.1355}_{-0.3406}$ & 4.3 & 9\% & 1.0 & -- & -- & -- \\[0.1cm]
3D-EGS-34322 & $3.59^{+0.33}_{-0.32}$ & $3.5460^{+0.1150}_{-0.3692}$ & 0.1 & 18\% & 0.9 & -- & -- & -- \\[0.1cm]
ZF-UDS-3651 & $3.87^{+0.12}_{-0.12}$ & $3.4214^{+0.1193}_{-0.0600}$ & 2.7 & 33\% & 1.1 & -- & -- & -- \\[0.1cm]
ZF-UDS-4347 & $3.58^{+0.04}_{-0.05}$ & $3.4731^{+0.2220}_{-0.0186}$ & 0.5 & 39\% & 0.9 & -- & -- & -- \\[0.1cm]
ZF-UDS-6496 & $3.50^{+0.04}_{-0.04}$ & $2.0332^{+1.1110}_{-0.0003}$ & 1.3 & 33\% & 1.0 & -- & -- & -- \\[0.1cm]
ZF-UDS-7329 & $3.04^{+0.17}_{-0.17}$ & $3.1857^{+0.0188}_{-1.0035}$ & -0.1 & 25\% & 1.0 & -- & -- & -- \\[0.1cm]
ZF-UDS-7542 & $3.15^{+0.06}_{-0.06}$ & $3.1885^{+0.2709}_{-0.0029}$ & -0.6 & 35\% & 1.0 & -- & -- & -- \\[0.1cm]
3D-UDS-35168 & $3.46^{+0.32}_{-0.29}$ & $4.2240^{+0.2913}_{-1.5226}$ & -0.5 & 4\% & 1.0 & -- & -- & -- \\[0.1cm]
3D-UDS-39102 & $3.51^{+0.42}_{-0.36}$ & $3.3982^{+0.3368}_{-1.1912}$ & 0.2 & 4\% & 1.0 & -- & -- & -- \\[0.1cm]
3D-UDS-41232 & $3.01^{+0.07}_{-0.08}$ & $2.9810^{+0.1448}_{-0.5831}$ & 0.2 & 25\% & 1.0 & -- & -- & -- \\[0.1cm]
\hline
\end{tabular}
\end{center}
{\footnotesize This table is available in electronic format at the CDS. $^a$ Fraction of the redshift probability distribution enclosed within $\zspec\pm0.01$. $^b$ Indicates galaxies for which the redshift could be determined with $p>50\%$ from the continuum emission alone, using Balmer absorption lines. $^c$ List of emission lines detected with a significance of at least $2\sigma$. Line doublets names are shortened as in \rtab{TAB:galaxies_em}. $^d$ Best-fit velocity dispersion of the emission lines (assumed identical for all the lines in a spectrum).}
\end{table*}

\clearpage

\begin{table*}
\begin{center}
\caption{Measured emission line properties $^a$.\label{TAB:galaxies_em}}
\begin{tabular}{lccccccc}
\hline\hline \\[-0.3cm]
ID       & $\oii_{\lambda\lambda}$ & $\hbeta$ & $\oiii_{\lambda\lambda}$ & $\oone$ & $\halpha$ & $\nii_{\lambda\lambda}$ & $\sii_{\lambda\lambda}$ \\
\hline \\[-0.3cm]
\multicolumn{2}{l}{Luminosities ($10^{8}\,\lsun$)} \\
ZF-COS-17779 & $0.47\pm0.27$ & $0.62\pm0.14$ & $1.51\pm0.21$ & -- & -- & -- & -- \\
ZF-COS-18842 & -- & $0.66\pm1.63$ & $6.19\pm4.14$ & -- & -- & -- & -- \\
ZF-COS-19589 & -- & $0.70\pm1.02$ & -- & -- & -- & -- & -- \\
ZF-COS-20032 & -- & $0.19\pm0.50$ & $2.06\pm1.36$ & $0.52\pm0.23$ & $2.05\pm0.77$ & $0.94\pm0.45$ & $0.12\pm0.27$ \\
ZF-COS-20115 & -- & $0.00\pm0.33$ & $1.86\pm1.43$ & -- & -- & -- & -- \\
ZF-COS-20133 & -- & $2.05\pm0.21$ & $14.31\pm0.31$ & -- & -- & -- & -- \\
3D-EGS-18996 & $1.61\pm0.88$ & $0.00\pm0.35$ & $7.14\pm1.46$ & -- & -- & -- & -- \\
3D-EGS-26047 & $3.04\pm1.16$ & $6.61\pm3.12$ & $8.44\pm3.97$ & -- & -- & -- & -- \\
3D-EGS-31322 & $1.55\pm0.99$ & $2.18\pm1.60$ & $0.82\pm1.14$ & -- & -- & -- & -- \\
3D-EGS-40032 & $4.59\pm1.04$ & $1.02\pm0.99$ & $1.13\pm1.29$ & -- & -- & -- & -- \\
ZF-UDS-8197 & -- & $0.74\pm0.74$ & $18.26\pm1.49$ & -- & -- & -- & -- \\
3D-UDS-27939 & -- & -- & -- & $0.84\pm0.31$ & $6.69\pm0.46$ & $2.68\pm0.33$ & $3.96\pm0.47$ \\
\hline \\[-0.3cm]
\multicolumn{2}{l}{Rest-frame equivalent widths (\AA)} \\
ZF-COS-17779 & $13.4\pm7.9$ & $7.5\pm1.8$ & $21.0\pm3.1$ & -- & -- & -- & -- \\
ZF-COS-18842 & -- & $3.0\pm7.6$ & $32.8\pm23.2$ & -- & -- & -- & -- \\
ZF-COS-19589 & -- & $5.2\pm7.6$ & -- & -- & -- & -- & -- \\
ZF-COS-20032 & -- & $12.9\pm34.1$ & $136.6\pm87.3$ & $22.3\pm10.1$ & $84.1\pm31.2$ & $40.2\pm19.2$ & $5.2\pm11.3$ \\
ZF-COS-20115 & -- & $0.0\pm0.9$ & $5.8\pm4.5$ & -- & -- & -- & -- \\
ZF-COS-20133 & -- & $34.3\pm4.2$ & $282.2\pm19.4$ & -- & -- & -- & -- \\
3D-EGS-18996 & $6.7\pm3.7$ & $0.0\pm0.6$ & $14.2\pm3.0$ & -- & -- & -- & -- \\
3D-EGS-26047 & $42.8\pm16.5$ & $33.9\pm16.7$ & $49.2\pm24.2$ & -- & -- & -- & -- \\
3D-EGS-31322 & $9.6\pm6.1$ & $5.0\pm3.7$ & $2.2\pm3.1$ & -- & -- & -- & -- \\
3D-EGS-40032 & $23.0\pm5.3$ & $1.8\pm1.8$ & $2.3\pm2.7$ & -- & -- & -- & -- \\
ZF-UDS-8197 & -- & $5.7\pm5.8$ & $165.7\pm16.4$ & -- & -- & -- & -- \\
3D-UDS-27939 & -- & -- & -- & $13.8\pm5.3$ & $108.5\pm9.3$ & $43.1\pm6.0$ & $62.2\pm8.4$ \\
\hline
\end{tabular}
\end{center}
{\footnotesize $^a$ For doublets, luminosities and EW are the sum of the two lines in the doublet. $\oii_{\lambda\lambda}$ is the sum of $\oii_{3729}$ and $\oii_{3726}$. $\oiii_{\lambda\lambda}$ is the sum of $\oiii_{5007}$ and $\oii_{4959}$. $\nii_{\lambda\lambda}$ is the sum of $\nii_{6583}$ and $\nii_{6548}$. $\sii_{\lambda\lambda}$ is the sum of $\sii_{6731}$ and $\sii_{6716}$.}
\end{table*}

\clearpage

\begin{table*}
\begin{center}
\caption{Summary of $\sfr$ estimates from SED modeling or emission lines.\label{TAB:sfr}}
\begin{tabular}{lclcc}
\hline\hline \\[-0.3cm]
ID       & $\sfr_{10}$ $^a$ & line & $\sfr_{\rm line}^{\rm nodust}$ $^b$ & $\sfr_{\rm line}^{\rm corr}$ $^c$ \\
         & $\msun/\yr$ &      & $\msun/\yr$ & $\msun/\yr$ \\
\hline \\[-0.3cm]
\multicolumn{2}{l}{Confirmed $z>3$ galaxies} \\
ZF-COS-17779 & $0.66^{+100.77}_{-0.66}$ & $\hbeta$ & $3.4\pm0.8$ & $15.5^{+21.3}_{-10.6}$ \\[0.1cm]
\ldots       & \ldots                 & $\oii$ & $1.1\pm0.6$ & $8.7^{+24.8}_{-7.6}$ \\[0.1cm]
ZF-COS-18842 & $2.53^{+2.36}_{-1.21}$ & $\hbeta$ & $3.6\pm8.9$ & $3.6^{+12.0}_{-8.9}$ \\[0.1cm]
\ldots       & \ldots                 & $\oii$ & $0.0\pm1.9$ & $0.0^{+2.6}_{-1.9}$ \\[0.1cm]
ZF-COS-19589 & $0.00^{+4.41}_{-0.00}$ & $\hbeta$ & $3.8\pm5.6$ & $10.2^{+28.5}_{-12.9}$ \\[0.1cm]
ZF-COS-20115 & $0.00^{+0.69}_{-0.00}$ & $\hbeta$ & $0.0\pm1.8$ & $0.0^{+2.8}_{-2.5}$ \\[0.1cm]
\ldots       & \ldots                 & $\oii$ & $0.0\pm1.6$ & $0.0^{+2.9}_{-2.5}$ \\[0.1cm]
ZF-COS-20133 & $4.57^{+0.09}_{-1.30}$ & $\hbeta$ & $11.2\pm1.1$ & $12.5^{+1.3}_{-2.4}$ \\[0.1cm]
\ldots       & \ldots                 & $\oii$ & $0.0\pm0.8$ & $0.0^{+0.9}_{-0.8}$ \\[0.1cm]
3D-EGS-18996 & $0.88^{+0.91}_{-0.86}$ & $\hbeta$ & $0.0\pm1.9$ & $0.0^{+1.9}_{-1.9}$ \\[0.1cm]
\ldots       & \ldots                 & $\oii$ & $3.6\pm2.0$ & $3.6^{+2.0}_{-2.0}$ \\[0.1cm]
3D-EGS-26047 & $0.07^{+1.95}_{-0.04}$ & $\hbeta$ & $36.0\pm17.0$ & $44.8^{+37.3}_{-25.8}$ \\[0.1cm]
\ldots       & \ldots                 & $\oii$ & $6.8\pm2.6$ & $9.2^{+7.9}_{-5.0}$ \\[0.1cm]
3D-EGS-31322 & $0.04^{+2.26}_{-0.04}$ & $\hbeta$ & $11.9\pm8.8$ & $16.5^{+19.1}_{-13.0}$ \\[0.1cm]
\ldots       & \ldots                 & $\oii$ & $3.6\pm2.3$ & $5.7^{+6.8}_{-4.1}$ \\[0.1cm]
3D-EGS-40032 & $5.50^{+3.38}_{-3.07}$ & $\hbeta$ & $5.6\pm5.4$ & $8.6^{+10.3}_{-8.4}$ \\[0.1cm]
\ldots       & \ldots                 & $\oii$ & $10.3\pm2.3$ & $18.6^{+7.9}_{-6.2}$ \\[0.1cm]
ZF-UDS-8197 & $1.68^{+3.33}_{-0.73}$ & $\hbeta$ & $4.0\pm4.1$ & $4.0^{+7.2}_{-4.1}$ \\[0.1cm]
\ldots       & \ldots                 & $\oii$ & $0.0\pm0.2$ & $0.0^{+0.4}_{-0.2}$ \\[0.1cm]
\hline \\[-0.3cm]
\multicolumn{2}{l}{Interlopers at $z<3$} \\
ZF-COS-20032 & $139.80^{+75.78}_{-139.80}$ & $\halpha$ & $3.9\pm1.5$ & $63.6^{+46.0}_{-29.4}$ \\[0.1cm]
\ldots       & \ldots                 & $\hbeta$ & $1.0\pm2.7$ & $58.1^{+238.9}_{-136.8}$ \\[0.1cm]
3D-UDS-27939 & $1.81^{+7.97}_{-1.16}$ & $\halpha$ & $12.8\pm0.9$ & $27.1^{+18.4}_{-5.4}$ \\[0.1cm]
\hline
\end{tabular}
\end{center}
{\footnotesize $^a$ $\sfr$ estimated from the SED modeling, averaged over the last $10\,\Myr$ preceeding observation. $^b$ $\sfr$ estimated from the observed emission line luminosity, without correcting for dust attenuation. $^c$ Same, but corrected for dust attenuation using the best-fit $A_{\rm V}$ from the SED modeling.}
\end{table*}

\clearpage

\section{Reduction of MOSFIRE spectra \label{APP:reduction}}

\subsection{Transmission correction}

As in \cite{nanayakkara2016}, a standard star was observed at the beginning and end of the observing runs to monitor telluric absorption. Since this star was not observed at the same time as the science targets, it cannot capture variations of the effective transmission within the run, and it is affected by different systematic errors (e.g., in the background subtraction).

Each of our masks contained a ``slit star'' of moderate brightness, which was used mostly to monitor the seeing. These are typically M or K stars, which are not optimal flux calibrators because their continuum emission contains a variety of features which need to be properly modeled. However these features are relatively weak in $H$ and $K$, and we can model them using theoretical stellar spectra. We therefore used these ``slit stars'' to perform an independent telluric correction. Our approach, described below, generates an effective ``transmission correction'' which accounts for telluric absorption, filter transmission, slit losses for a point source, and absolute flux calibration (electron/s to cgs).

For each mask, we first fit the broadband photometry of the slit star with the PHOENIX theoretical star models \citep{husser2013} to estimate its intrinsic spectrum. We performed this fit only using the NIR photometry ($0.8\,\um < \lambda < 3\,\um$) to maximize the fidelity of the fit in the $H$ and $K$ bands. Because the stars are not extremely bright, this photometry is not saturated and can be used reliably. The best-fitting PHOENIX spectrum, normalized to fit the photometry, was used as the intrinsic spectrum of the star. We then used the MOSFIRE pipeline to reduce each pair of ``AB'' exposures into 2D spectra for all our targets, including the slit star (see next section for detail). At this stage, no transmission correction was applied, and the spectra were still in raw units. We collapsed the slit star's 2D spectrum along the wavelength axis to form the spatial profile of the star, which we fit with a Gaussian model to determine the location of the peak emission, as well as the seeing. We extracted the 1D spectrum of the star using this Gaussian model, normalizing it to the emission in the slit at each wavelength element while keeping the position and width of the Gaussian fixed. We finally estimated the transmission correction by computing the ratio of the intrinsic spectrum of the star to this 1D spectrum. Because the intrinsic spectrum was rescaled to match the broadband photometry, this correction also includes the conversion from raw units to physical flux, including slit loss correction for a point source.

\begin{figure*}
\begin{center}
\includegraphics[width=0.9\textwidth]{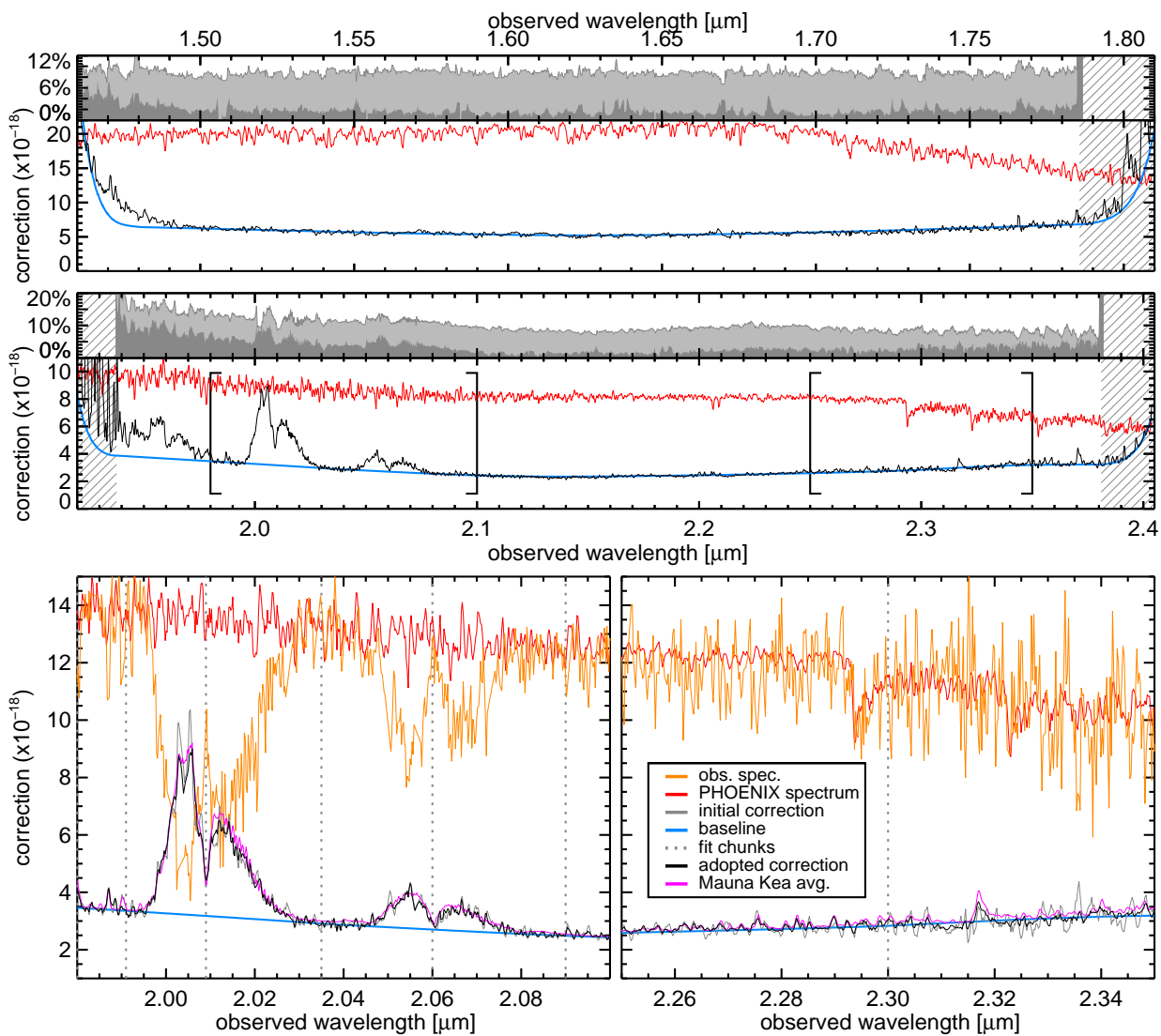}
\end{center}
\caption{Example transmission correction curve for one exposure of the COSMOS run using the slit star (an M5 star). As discussed in the text, this correction accounts for the telluric absorption, the filter transmission, slit loss (for a point source), and absolute calibration from electron/s to flux. {\bf Top:} adopted correction curve in $H$ (top) and $K$ (bottom) for this exposure (black line), assumed intrinsic spectrum of the slit star (red line) and adopted transmission ``baseline'' (blue line, see text). The hashed regions indicate the portion of the spectrum for which we used the Mauna Kea average transmission. The inset at the top of each plot shows the relative variation of the transmission among all exposures, showing either the full variation between exposures (light gray) or what remains after factoring out variations of global transmission (dark gray). {\bf Bottom:} zoom-ins on the $K$-band correction curve (areas bracketed in the plot above). In addition to the lines shown above, we show the raw observed spectrum multiplied by the baseline correction (orange line), the initial correction derived from this spectrum (light gray line), the individual wavelength chunks of the template curve (dotted vertical lines) and the expected correction based on the average Mauna Kea transmission (see text). \label{FIG:telluric}}
\end{figure*}

A downside of using the slit star is that its $S/N$ is not as high a that of the standard star. The derived transmission corrections are more noisy, and this can degrade the final $S/N$ of the science targets. To mitigate this, we first applied a 3 pixel boxcar filter to the star spectra, then stacked the transmission correction of all exposures to build a ``template'' curve with much higher $S/N$. We then fit this template to the transmission correction of each exposure. To allow the strength of the atmospheric absorption features to vary from one exposure to the other, we decomposed the template into multiple wavelength ``chunks'' -- each containing one major atmospheric feature -- and let their amplitude vary independently over a fixed baseline. For each exposure, we adopted the best-fitting combination of these chunks as the final transmission correction. An example fit is shown in \rfig{FIG:telluric}, and more detail on this procedure are provided in \rapp{APP:telluric}. We then extended each curve to cover wavelengths beyond that covered by the slit star's spectrum using the average atmospheric transmission at Mauna Kea\footnote{\url{https://www.gemini.edu/sciops/telescopes-and-sites/observing-condition-constraints/ir-transmission-spectra}}, and thus built the final transmission correction $T_i(\lambda)$ of each exposure $i$.

We find that the baseline of these corrections (which reflects the overall transmission across the band) shows variations of order $10$ to $20\%$ between exposures. Factoring out these global fluctuations, the residual wavelength-dependent transmission variations are smaller: of the order of $1\%$ and $3\%$ in $H$ and $K$, respectively, and reaching a maximum of $4\%$ and $10\%$ at the position of strong telluric absorption or toward the edges of the spectrum (see insets in \rfig{FIG:telluric}).

\subsection{Extracting spectra}

Before extracting 1D spectra, we removed hot pixels and cosmic rays by flagging strong pixel outliers from the 2D spectra produced by the pipeline. While the MOSFIRE pipeline does include a cosmic ray rejection algorithm, this feature can only be enabled when reducing all the exposures at once. In our case, where we reduced each ``AB'' exposure separately, we used a different approach which is described and illustrated in \rapp{APP:cosmicray}. We also systematically masked strong OH residuals.

For each galaxy, we then stacked the 2D spectra with uniform weighting and used it to identify the position within the slit of the target, either from the continuum emission (as for the slit star) or using clearly detected emission lines. Compared to the position predicted by the pipeline, we find systematic offsets within each masks of up to two pixels (a pixel is $0.18\,\arcsec$ wide), and residual per-target offsets of up to three pixels (the RMS ranges from $0.5$ to $1.5$ pixels, depending on the mask). For sources without detected continuum or lines, we used the average shift within their respective mask.

Using these positions, we extracted 1D spectra for our targets on each pair of ``AB'' exposures by fitting the amplitude of a fixed Gaussian profile along the slit axis, for each resolution element. We set the width of this Gaussian to that measured for the slit star for each exposure, which is equivalent to assuming that the targets are point-like. This is a reasonable assumption given that our galaxies are both distant and intrinsically compact objects \citep{straatman2015}, and potential flux loss resulting from this choice are accounted for at a later stage when the spectra are rescaled to match the broadband photometry.

For each target $t$ and exposure $i$, this produced a counts spectrum ${\rm ADU}_{t,i}(\lambda)$. We then applied the transmission corrections (described in the previous section) to all these 1D spectra to obtain the total flux, $F_{t,i}(\lambda) = {\rm ADU}_{t,i}(\lambda) \times T_i(\lambda)$. We also derived the formal uncertainty $\sigma^{\rm th}_{t,i}(\lambda)$ on this flux by propagating the uncertainties from the 2D error spectrum produced by the pipeline (see next section for more detail on the uncertainty estimates).

Then, for each target $t$ the final spectrum $F_{t}(\lambda)$ was obtained by stacking the 1D spectra of all exposures using inverse variance weighting, $w_{t,i}(\lambda) = 1/\sigma^{\rm th}_{t,i}(\lambda)^2$, and sigma clipping to remove any remaining strong outlier (exposures differing from the median by more than $5\,\sigma$ were given a weight of zero). This weighting scheme optimally penalizes exposures with poor observing conditions, in particular in case of bad seeing: the width of the Gaussian profile used for the flux extraction is larger, which results in larger formal uncertainties in the extraction. Furthermore, the transmission correction is also larger for these exposures because of larger slit losses, which contributes in weighting them down.

We also stacked the 2D spectra for visualization purposes, using the same weighting, flagging and telluric correction as for the 1D spectra. In case of small offsets along the slit between exposures, the spectra were shifted to a common grid before stacking.

\subsection{Determining uncertainties \label{SEC:uncertainty}}

\begin{figure}
\begin{center}
\includegraphics[width=0.5\textwidth]{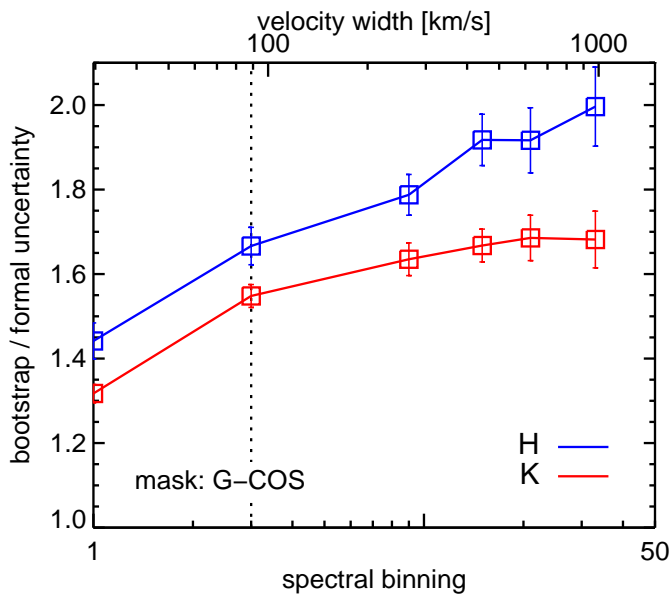}
\end{center}
\caption{Ratio of uncertainties derived by bootstrapping the 1D spectra against that obtained by formally propagating the uncertainties from the 2D spectra and assuming uncorrelated Gaussian noise. The values shown are the median of the ratios among all the targets within the mask (COS-W182) and error bars show the error on the median. We show this ratio for the two bands ($H$ in blue, $K$ in red) as a function of spectral binning. The dashed line indicates the binning adopted in this work. \label{FIG:rms}}
\end{figure}

We determined the uncertainties on the final stacked spectra in two ways. A first ``theoretical'' value, $\sigma_{{\rm th},t}(\lambda)$, was obtained for each target $t$ by propagating the formal uncertainties of each exposure in the stack, as derived originally from the pipeline's 2D error spectrum, namely, combining Poisson statistics on the measured electron counts with the detector read noise. As an alternative to this first method, we empirically estimated the uncertainties from the variance between exposures, $\sigma_{{\rm var},t}$, using:
\begin{align}
\sigma_{{\rm var},t}(\lambda) &= \frac{1}{\sum_{i} w_{t,i}(\lambda)} \sqrt{\frac{N_{\rm exp}}{N_{\rm exp} - 1} \sum_{i} w_{t,i}(\lambda)^2\, \big[F_{t,i}(\lambda) - F_{t}(\lambda)\big]^2}\,,
\end{align}
where $N_{\rm exp}$ is the number of exposures; the other quantities were defined in the previous section. This formula produces results identical to bootstrapping (as demonstrated in \citealt{gatz1995}, and we double checked that it was indeed the case on our data) but it requires much less computing time.

Using the native pixel scale of the spectra (i.e., without binning), we find that $\sigma_{\rm var}$ is systematically larger than $\sigma_{\rm th}$ by $30$ to $45\%$, depending on the mask and band as shown in \rfig{FIG:rms} (left). This suggests the uncertainties produced by the pipeline are substantially underestimated. In fact, the pipeline version we used was indeed reported to employ an incorrect treatment of the error spectra, which could be the source of this discrepancy\footnote{See \url{https://github.com/Keck-DataReductionPipelines/MosfireDRP/issues/40}.}. Using their own MOSFIRE pipeline (and a slightly different observing strategy), \cite{kriek2015} found no such issue when performing a similar test on their spectra. However, their formula for $\sigma_{\rm var}$ (their Eqs.~5 and 6) also differs from ours, making any direct comparison difficult. Using MC simulations of a weighted mean, we find that their formula actually over-predicts the uncertainty by 20 to 50\%, depending on the choice of weights, while our formula is accurate to 1\%.

We also find the discrepancy between $\sigma_{\rm var}$ and $\sigma_{\rm th}$ grows even larger for binned spectra: using a binning of three (resp.~nine) wavelength elements, the bootstrapped uncertainties are an additional $12$ to $18\%$ (resp.~$18$ to $25\%$) larger than the formal uncertainties. This suggests that, at the native pixel scale produced by the pipeline ($\lambda/\Delta\lambda=9000$--$11000$), the noise is spectrally correlated. Since we did not require high spectral resolution for this work, we chose to avoid the native pixel scale in the following, and considered instead a binning of at least three spectral elements corresponding to $\lambda/\Delta\lambda\sim3000$ (or resolution elements of $\ge80\,\kms$), which matched the spectral resolution of MOSFIRE with $0.7\arcsec$ slits, and ensured that our uncertainties were accurate within $20\%$ at worse. Consequently, the line spread function was reduced to little more than a pixel, and was thus ignored.

While $\sigma_{\rm var}$ is larger than $\sigma_{\rm th}$ on average, it is not always the case for every spectral element of a spectrum. In particular for masks with few exposures, the noise fluctuations can make the data of each exposure coincide (by pure chance) to similar flux values, leading to an underestimated $\sigma_{\rm var}$. To avoid this, we defined a ``rescaled'' formal uncertainty
\begin{equation}
\bar{\sigma}_{{\rm th},t}(\lambda) = \sigma_{{\rm th},t}(\lambda)\, \mean{\frac{\sigma_{{\rm var},t}}{\sigma_{{\rm th},t}}}_{\lambda}\,,
\end{equation}
where the second term is the median of $\sigma_{\rm var}/\sigma_{\rm th}$ across the spectrum of the target $t$. We then adopted as final uncertainty the largest of $\sigma_{\rm var}$ and $\bar{\sigma}_{{\rm th}}$.

As we have just shown, bootstrapping uncertainties are more conservative than the formal uncertainties in general. However, if some source of error affects all the exposures in a similar way, for example owing to an imperfect background subtraction, they will not be accounted for by bootstrapping by construction. To verify that our data were globally unaffected by such issues, we reduced $14$ ``sky'' spectra in the COSMOS mask, extracted at offset positions in the slits, avoiding known sources. If our reduction procedure was unbiased, these spectra should only contain noise with a zero flux average. We therefore stacked these spectra, aligned on the same wavelength grid, and find reduced $\chi^2=1.2$ and $1.7$ in $H$ and $K$, respectively, for a binning of $15$ resolution elements ($500\,\kms$, the expected full line width for our objects). These residuals were not perfect, as would be indicated by a reduced $\chi^2$ of unity, but they were still small: at this resolution the largest residuals were of the order of $3\times10^{-20}\,{\rm erg/s/cm^2/\AA}$, which is lower than the error bar of our deepest spectra. We therefore concluded that our spectra were not affected by background subtraction issues, and that the bootstrapping uncertainties derived above accounted for the main sources of error in the reduction.

\subsection{Rescaling to total flux}

\begin{figure}
\begin{center}
\includegraphics[width=0.5\textwidth]{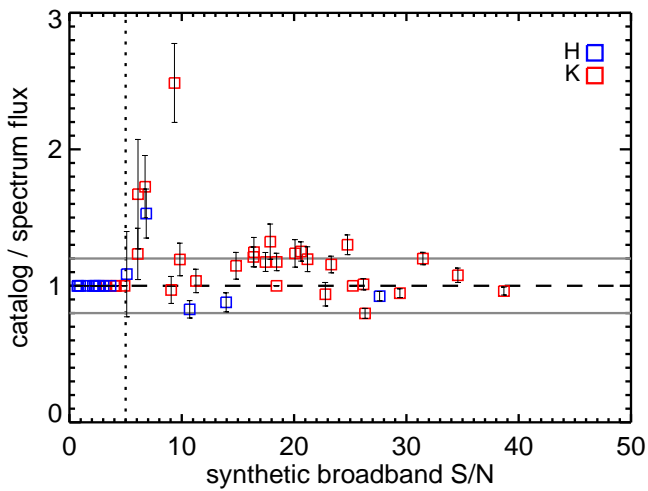}
\end{center}
\caption{Ratio of the flux listed in the ZFOURGE or 3DHST catalog to the flux measured in the MOSFIRE spectrum for each of our targets, as a function of the $S/N$ of the spectrum flux. The flux in the spectrum was integrated in the same passband as the catalog fluxes. Fluxes in $H$ band are shown in blue, and fluxes in $K$ band are shown in red. The vertical dotted line shows the $S/N$ limit below which we did not perform the flux rescaling to avoid introducing noise. \label{FIG:ftot}}
\end{figure}

\rfig{FIG:ftot} illustrates the final rescaling of the spectra using the catalog fluxes for each source.

\subsection{Detail of telluric correction \label{APP:telluric}}

Here we describe in more detail the procedure through which the telluric correction of each exposure was fit with a template curve to improve the $S/N$.

We first defined the ``initial'' observed correction (i.e., inverse transmission) for the $i$th exposure as
\begin{align}
T_i^{\rm init}(\lambda) = \frac{S_{\lambda}(\lambda)\,[{\rm cgs}]}{{\rm ADU}(\lambda)\,[e^-\!/{\rm s}]}\,,
\end{align}
where $S_\lambda$ is the physical flux of the star (taken from the PHOENIX synthetic spectrum and renormalized to the broadband fluxes of the star) and ${\rm ADU}$ is the raw electron flux measured by the MOSFIRE detector. This ratio could be used directly to perform the telluric correction of the science targets for the corresponding exposure, but since the slit star is only of moderate brightness, it can be undesirably noisy.

We therefore attempted to reduce the noise. To do so, we assumed the above correction can be empirically decomposed as
\begin{align}
T_i(\lambda) = B_i(\lambda) \times \Big[1 + \sum_k \alpha_{ik} \times A_k(\lambda)\Big]\,,
\end{align}
where $B_i$ is a strictly positive and smooth baseline function that captures the overall instrument transmission, $A_k$ is a function capturing the $k$th ``chunk'' of atmospheric absorption features, and $\alpha_{ik}$ is a strictly positive number that defines the strength of the $k$th set of features in the $i$th exposure. We obtained high $S/N$ versions of the $A_k$ functions by stacking all exposures, as described below.

For each exposure, we started by determining the baseline $B_i$. We first multiplied $T_i^{\rm init}$ by the response curve of the MOSFIRE filter to account for the sharp drop of transmission at the edges, and then isolated several disjoint wavelength regions within the band that are free of strong atmospheric feature. We computed the average correction within each of these regions, and interpolated these values using a cubic spline with natural boundary conditions (vanishing second derivatives). We eventually defined the baseline as the product of this spline with the inverse of the filter response curve.

We then stacked $T_i^{\rm init}/B_i$ among all exposures to produce the high $S/N$ template curve $\bar{T}$. We used the inverse of the Mauna Kea average transmission to fill the gaps in $\bar{T}$ which are not covered by the star's spectrum (at the edges of the $H$ and $K$ bands). We then used this template to define $A_k$ as
\begin{align}
A_k(\lambda) = \left\{\begin{array}{ll}
    \bar{T}(\lambda) - 1 & \text{for $\lambda^{\rm cut}_k \le \lambda < \lambda^{\rm cut}_{k+1}$} \\
    0 & \text{otherwise}\,. \\
\end{array}\right.
\end{align}
The values of $\lambda^{\rm cut}$ in both $H$ and $K$ are given in \rtab{TAB:lambdacut} and are illustrated in \rfig{FIG:telluric} (bottom).

\begin{table}
\begin{center}
\caption{Wavelength chunks used to define the $A_k$ functions in the telluric correction. \label{TAB:lambdacut}}
\begin{tabular}{cc}
\hline \\[-0.3cm]
\multicolumn{2}{c}{$\lambda^{\rm cut}_k$ ($\um$)} \\
$H$ & $K$ \\ \hline\hline\\[-0.3cm]
1.450 & 1.900 \\
1.565 & 1.951 \\
1.585 & 1.961 \\
1.615 & 1.980 \\
1.660 & 1.991 \\
1.670 & 2.009 \\
1.850 & 2.035 \\
      & 2.060 \\
      & 2.090 \\
      & 2.200 \\
      & 2.300 \\
      & 2.450 \\
\hline
\end{tabular}
\end{center}
\end{table}

As a last step, for each exposure $i$ we determined the values of $\alpha_{ik}$ by performing a linear fit of the $A_k$ functions to $(T_i^{\rm init}/B_i - 1)$. If some of the $\alpha_{ik}$ came out negative from the fit, we considered that the $S/N$ was too low and fixed them to $\alpha_{ik} = 1$ (i.e., assume the average transmission from the stack).

\subsection{Cosmic ray and hot pixel rejection \label{APP:cosmicray}}

\begin{figure}
\begin{center}
\includegraphics[width=0.5\textwidth]{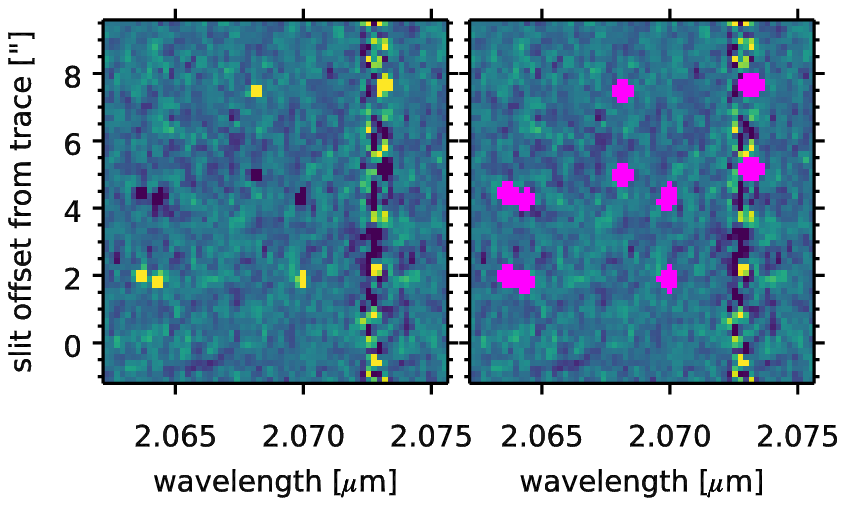} \\
\includegraphics[width=0.5\textwidth]{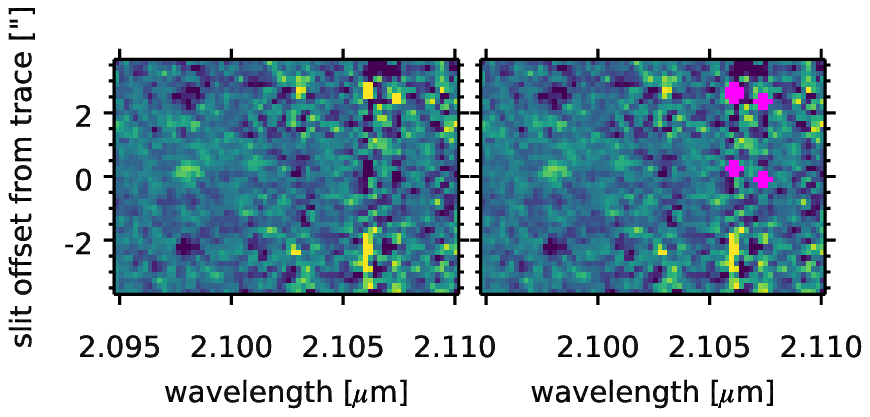}
\end{center}
\caption{Examples of the output of the cosmic ray and hot pixel rejection algorithm. The original unmasked 2D spectra are shown in the left, and the masked spectra are shown on the right with masked regions displayed in pink. {\bf Top:} excerpt from the brightest $z\sim4$ quiescent galaxy, {\bf bottom:} excerpt from a $z\sim2$ filler SFG, which has an emission line at $\lambda=2.098\,\um$. Both spectra are taken from the same $K$-band exposure in the COSMOS mask, with excellent seeing ($0.42\,\arcsec$). \label{FIG:crej}}
\end{figure}

In this section we describe the algorithm used for cosmic ray and hot pixel rejection in the 2D spectra produced by the MOSFIRE pipeline.

For each pixel $p$ in the 2D spectrum, we built two lists of neighboring pixels: a first list containing the $12$ neighboring pixels along the wavelength axis (corresponding to $\Delta \lambda \simeq 20\,\AA$), and a second list containing all the neighboring pixels along the slit axis. We then computed the median value $\bar{p}_{\rm wave}$ among the first list, and computed the scatter of the values in the two lists $\sigma_{\rm wave}$ and $\sigma_{\rm slit}$ (respectively) using the median absolute deviation, multiplied by 1.48 to get a robust standard deviation.

The median $\bar{p}_{\rm wave}$ measures the emission in the spectrum on scales larger or equal to $20\,\AA$ (i.e., continuum emission or relatively wide emission lines). The two scatter values, $\sigma_{\rm wave}$ and $\sigma_{\rm slit}$, are used to estimate the expected noise amplitude for the pixel $p$ without relying on the uncertainty spectrum produced by the pipeline (which, as we discuss in \rsec{SEC:uncertainty}, can be significantly underestimated and would cause many spurious hot pixel or cosmic ray rejections). The scatter $\sigma_{\rm slit}$ will capture increased noise at the wavelengths affected by OH lines, while $\sigma_{\rm wave}$ will be larger if the target shows detectable emission (which will result in increased Poisson noise). We therefore kept the highest value of the two as the best noise estimate, $\sigma$, and finally flagged out the pixel if $(p - \bar{p}_{\rm wave})/\sigma >4$.

The result of this filtering is shown in \rfig{FIG:crej} for the $K$-band exposure in COSMOS with the best seeing (i.e., where the risk of filtering out features of the science target is the highest). We show two examples: the brightest $z\sim4$ quiescent galaxy and a $z\sim2$ low-mass filler with an emission line. In the first case, the algorithm correctly identified a number of hot pixels, even within the residuals of an OH line. The rest of the spectrum appears visually well-behaved. In the second case, a few hot pixels were also identified, but the emission line of the target galaxy at $\lambda = 2.098\,\um$ was not inadvertently masked.

\end{document}